\documentclass[twocolumn]{jpsj3}
\usepackage{txfonts}
\usepackage{color}
\usepackage{bm}

\title{Review of U-based Ferromagnetic Superconductors:
Comparison between UGe$_2$, URhGe, and UCoGe}

\author{
Dai Aoki$^{1,2}$\thanks{E-mail: aoki@imr.tohoku.ac.jp},
Kenji Ishida$^3$\thanks{E-mail: kishida@scphys.kyoto-u.ac.jp},
and 
Jacques Flouquet$^1$\thanks{E-mail: jflouc@aol.com}
}

\inst{
$^1$Universit\'{e} Grenoble Alpes, CEA, INAC-PHELIQS, F-38000 Grenoble, France\\
$^2$IMR, Tohoku University, Oarai, Ibaraki 311-1313, Japan \\
$^3$Department of Physics, Graduate School of Science, Kyoto University, Kyoto 606-8502, Japan \\}

\abst{
The discovery in 2000 that the ferromagnetic (FM) compound UGe$_2$ ($T_{\rm Curie}$ = 52 K at ambient pressure) becomes superconducting under a pressure of $P = 1.1\,{\rm GPa}$ until it enters the paramagnetic (PM) phase above $P_c$ = 1.6 GPa was a surprise, despite the fact that such a possibility was emphasized in theory four decades ago. 
Successive searches for new materials (URhGe and UCoGe) led to the discovery of the coexistence of superconductivity (SC) and ferromagnetism at ambient pressure. 
Furthermore in UCoGe, it was found that SC survives in the PM regime from $P_{\rm c} = 1.1$ to $4\,{\rm GPa}$. 
The novelty is that SC also emerges deep inside the FM regime but with strong FM fluctuations.
Focus has been on low-temperature experiments under extreme conditions of magnetic field ($H$), pressure, and uniaxial stress. 
NQR and NMR experiments are unique tools to understand the interplay between the spin dynamics and the Cooper pairing. 
We choose to present the SC properties from the knowledge of quasiparticle dressing in the normal phase (renormalized band mass, $m_{\rm B}$ plus the extra dressing originating from FM fluctuations, $m^{**}$).
In UGe$_2$, strong interplay exists between Fermi surface (FS) reconstructions in the cascade of different FM and PM ground states and their magnetic fluctuations. 
Similar phenomena occur in URhGe and UCoGe but, at first glance, the SC seems to be driven by the FM fluctuations. 
The weakness of the FM interaction in these two compounds gives the opportunity to observe singular features in magnetic field scans depending on their field orientation with respect to the FM sublattice magnetization ($M_0$). 
We will show that for UCoGe, which has the smallest ordered moment, a longitudinal field scan ($H \parallel M_0$) leads to a drastic decrease in the FM fluctuations with direct consequences on SC properties such as the upward curvature of the upper critical field. 
A transverse field scan ($H \perp M_0$) leads to suppression of the Curie temperature, $T_{\rm Curie}$; 
the consequence is a boost in FM fluctuations, which leads to a reinforcement of SC. 
Contrary to the two examples of Ising FM UGe$_2$ and UCoGe, the singularity in URhGe is the weakness of the magnetocrystalline term between the choice of ferromagnetism along the $c$- or $b$-axis; 
the most noteworthy feature is the detection of reentrant SC on each side of the $H$ switch at $H_{\rm R}$ from the $c$ easy axis of magnetization to the $b$-axis.  
All the experimental results give evidence that the SC in these three materials originates from the  FM fluctuations, which are amplitude modes of magnetic excitations in the FM state. 
Spin-triplet pairing has been anticipated in the FM superconductors and was actually observed by Knight-shift measurements in the SC state of UCoGe. 
Their fascinating ($p$, $T$, $H$) phase diagrams are now well established.   
Of course, a new generation of experiments will elucidate subtle effects by obtaining their SC order parameters.
While the FSs of UGe$_2$ have been experimentally well determined, those of URhGe and UCoGe have been poorly determined, and thus a quantitative comparison with band structure calculations cannot be achieved.
Up to now, Angle-resolved photoemission spectroscopy (ARPES) measurements have only given the flavor of the electronic bands at the Fermi level. 
Discussion is presented on how different theoretical approaches can describe the various phenomena discovered by experimentalists. 
Following the new hot subject of topological superconductors, proposals have been made for UCoGe, 
which is a great challenge for ambitious researchers! 
}

\begin{document}

\maketitle
\section{Introduction: Experimental Probes, Crystal Structure}
\subsection{Introduction}
The interplay between ferromagnetism and superconductivity (SC) is a challenging problem in the coupling between the two major ground states of condensed matter. 
This problem was theoretically posed six decades ago in 1957,~\cite{GinzburgJETP1957}. 
Early experiments in 1958 showed that ferromagnetic (FM) impurities, such as Gd, dissolved in La ($T_{\rm SC}$ = 5.7 K) destroy SC with 2\% doping\cite{MatthiasPRL1958}. 
Exchange interactions put stringent limits on the occurrence of SC.
However, SC can easily coexist with antiferromagnetic (AF) sublattices of localized rare earth (RE) atoms. 
The first discovered cases were the Chevrel phases REMo$_6$S$_8$~\cite{IshikawaSSC1977}, and soon after, in 1977, another example of RERh$_4$B$_4$ was discovered~\cite{FertigPRL1977}. 
Basically, on the scale of the SC coherence length, which is larger than the magnetic one, the Cooper pairs go through zero exchange interaction.

Two singular cases were ErRh$_4$B$_4$ and HoMo$_6$S$_8$, where SC and ferromagnetism are in competition. 
Despite the SC temperature ($T_{\rm SC}$) being higher than the Curie temperature ($T_{\rm Curie}$ ) of ferromagnetism, the ground state ends up in the FM ground state with the collapse of the singlet SC. 
For example, ErRh$_4$B$_4$ is a superconductor below $T_{\rm SC}=8.7\,{\rm K}$~\cite{SinhaPRL1982}; up on cooling to $T = 1\,{\rm K}$, 
a compromise between ferromagnetism and SC is realized by forming a modulated structure with a domain of alternating magnetic moments.
The period of $100\,{\rm \AA}$ is smaller than the SC coherence length. 
Upon further cooling below 0.8 K, ferromagnetism becomes the ground state and SC disappears. 
Here the energy gained by the FM atoms exceeds that of the Cooper pair formation at $T_{\rm SC}$ as the number of quasiparticles involved, $k_{\rm B}T_{\rm SC} \times \rho (E_{\rm F})$, is much lower than 1, where $\rho(E_{\rm F})$ is the electronic density of states~\cite{FlouquetBuzdin}. 
The occurrence of the modulated structure is discussed theoretically in Refs.~\cite{AndersonPR1959} and  \citen{BlountPRL1979}. 
An exotic observation was the detection of magnetic-field-induced SC in 1984 in  Eu$_{0.75}$Sn$_{0.25}$Mo$_6$S$_{7.2}$Se$_{0.8}$.~\cite{MeulPRL1984}.
This is in agreement with a theoretical prediction given by Jaccarino and Peter in 1962~\cite{JaccarinoPRL1962}, which stresses that the compensation of the exchange internal field by the opposite external magnetic field can overcome the bare Pauli limit of the upper critical field $H_{\rm c2}$.
In these localized magnetic SC compounds, $T_{\rm SC}$ is higher than $T_{\rm Curie}$ or $T_{\rm N}$ (for antiferromagnets). 
There are two types of electrons: localized ones that carries the magnetic moment and itinerant ones that are paired via electron-phonon coupling (see Refs.~\citen{Fis90_FM,Mul01,CanfieldPhysToday1998} for a review). 

In the reported U compounds UGe$_2$, URhGe, and UCoGe, the $5f$ electrons participate both in the magnetic coupling and in the formation of heavy quasiparticles; quasiparticles with high effective mass $m^*$ ($\sim 20\,m_0$, $m_0$: bare mass of an electron) are detected on the orbit at the Fermi surface (FS).  
The suggestion of unconventional SC and itinerant FM was given many decades ago on the basis of a Cooper pairing generated by FM spin fluctuation\cite{FayPRB1980}, in which the expression for $T_{\rm SC}$ vanishes at the FM critical pressure because of the poor approximation of the theory.~\cite{MonthouxPRB2001}
For a Fermi liquid, a well-known example is liquid $^3$He~\cite{OsheroffPRL1972,LeggettRMP1975}; however, the system is very far from FM instability\cite{Flo06_review}. 
In bulk electronic materials, the first observation was made on UGe$_2$\cite{SaxenaNature2000},
SC emerges under pressure ($P$) near the switch at $P_x \sim 1.2\,{\rm GPa}$ between two FM phases (FM2, FM1)~\cite{HuxleyPRB2001},
upon entering the paramagnetic (PM) phase at a higher pressure, no SC is detected. 
A breakthrough in research on the domain was realized with the discovery of two new cases, URhGe\cite{AokiNature2001} and UCoGe\cite{HuyPRL2007}, which show FM and SC transitions at ambient pressure. 
In both examples, the singular feature is that a transverse magnetic field with respect to the easy FM magnetization axis ($H\perp M_0$) leads to the suppression of $T_{\rm Curie}$, enhancing the SC pairing via the enhancement of FM fluctuations.
We will see that in the particular case of UCoGe, the weakness of the FM interaction is associated with a strong decrease in $H$ of the SC pairing in a longitudinal field scan ($H \parallel M_0$).
 
An important approach is to restore the normal phase by $H$ above the upper critical field $H_{\rm c2}$. 
The magnetic field can destroy singlet-pairing SC in two ways. 
The first one, called the orbital limit, is a manifestation of the Lorentz force on electrons, 
and the second one, called the Pauli limit, occurs when $H$ breaks the spin antialignment 
and orients the spin along the field direction due to the Zeeman effect.
The critical field of the Pauli limit, $H_{\rm P}$, is expressed as $\mu_0 H_{\rm P}/T_{\rm SC} \sim 1.86\,{\rm T/K}$ on the assumption that the electronic $g$ factor is 2 and $2\Delta(0) = 3.53 k_{\rm B}T_{\rm SC}$.     
In contrast to most unconventional singlet superconductors, where AF fluctuations are the main glue, 
in UGe$_2$, URhGe, and UCoGe\cite{SheikinPRB2001, Har05, HuyPRL2007}, the upper critical field $H_{\rm c2}$(0) in the three main directions  ($a$, $b$, $c$) of these orthorhombic structures exceeds the Pauli limiting field $H_{\rm P}$, as shown in Fig.~\ref{fig1} (see Sec.~\ref{sec:UCoGe} for UCoGe), (and the necessity to take into account the strong $H$ dependence of the pairing).
Thus, it seems established that a triplet pairing with equal spin pairing (ESP) is realized.
In the three cases, the coexistence of SC with FM has been directly verified by neutron scattering, $\mu$SR, and NQR experiments.~\cite{AokiJPSJ2012Rev,AokiJPSJ2014Rev,Aok14_actinide,HattoriJPSJ2014Rev,HuxleyPhysicaC2015}
Previous experimental reviews can be found in Refs.~\citen{AokiJPSJ2012Rev, AokiJPSJ2014Rev, HattoriJPSJ2014Rev}, and \citen{HuxleyPhysicaC2015}.
Reviews concerning the theory are given in Refs.~\citen{MineevPUsr2017} and \citen{Min18}. 

Basically, ESP between the same effective spin components seems to be established here. 
For the PM phase of $^3$He,
a beautiful clear case is the A phase of superfluid $^3$He~\cite{OsheroffPRL1972,LeggettRMP1975}; 
in the B phase of $^3$He, mixing occurs between spin-down and spin-up components.
In our case of FM superconductors, the general form of the
order parameter can be more complex with mixing of the magnetic component (see Ref.~\citen{MineevPUsr2017}). 
However, owing to the strength of FM exchange coupling in the FM phase,
the realization of ESP seems to be achieved.
The situation may change on entering the PM phase.

The aim of this review paper is to show our new experimental progress with the strong interplay between macroscopic and microscopic properties revealed recently via various NMR experiments (see Ref.~\citen{Asayamatextbook} for the strengths of NMR experiments in the study of SC). 
Special focus will be made on combined magnetic field, pressure, and uniaxial stress ($\sigma$) scans to cross the FM instability and observe their effect on SC. 
Our report mainly concerns results achieved in Grenoble, Oarai, Kyoto, and Tokai. 
The discovery of SC in UGe$_2$ was made through a collaboration between Grenoble and Cambridge~\cite{SaxenaNature2000}.
The SC of URhGe was discovered in Grenoble~\cite{AokiNature2001} and that of UCoGe was discovered in Amsterdam~\cite{HuyPRL2007}.
Special attention on the duality between the local and itinerant character of the 5$f$ electron notably in UGe$_2$, was stressed in the work of Wroclaw\cite{TrocPRB2012}, 
and the effect of doping with Ru on UCoGe was studied in Prague\cite{Val15}, as well as analyses of SC in both FM and PM phases\cite{VejpravovaPRB2010}.
\begin{fullfigure}[!tbh]
\begin{center}
\includegraphics[width=0.8\hsize]{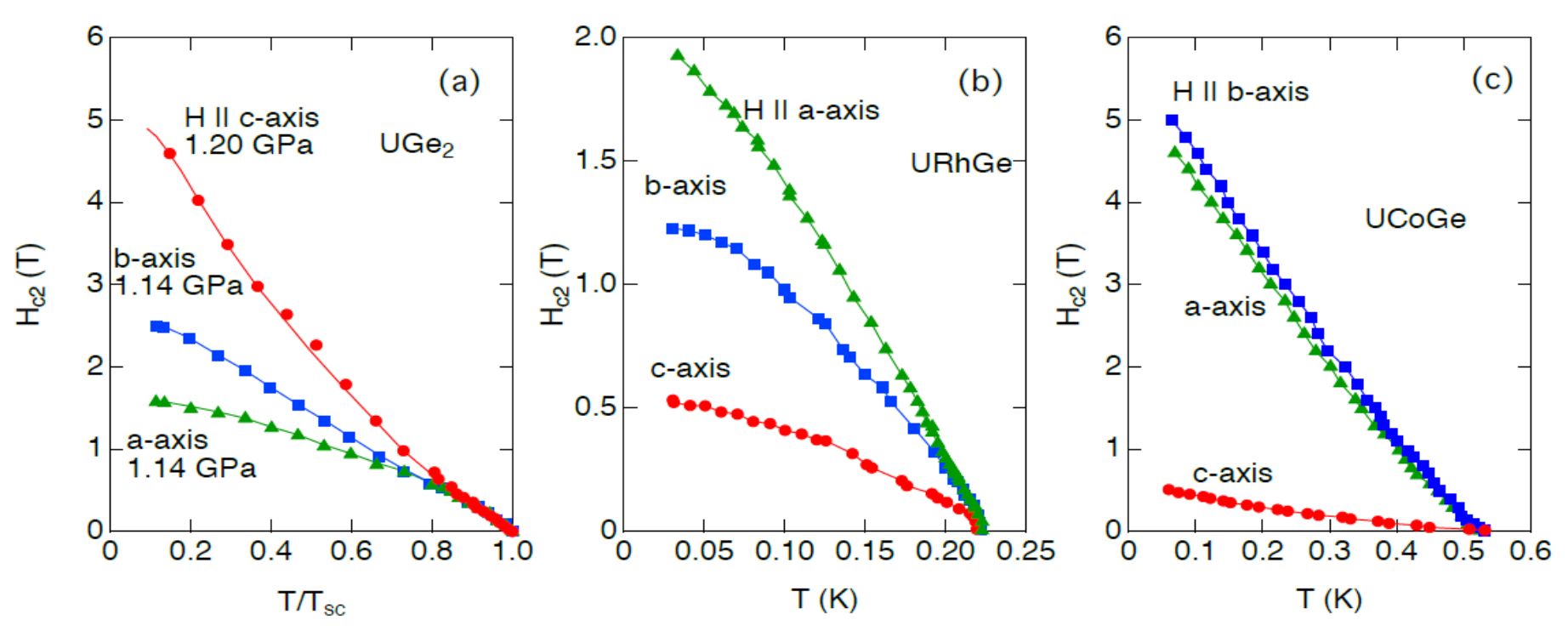}
\end{center}
\caption{(Color online) Temperature dependence of $H_{\rm c2}$ for UGe$_2$~\cite{SheikinPRB2001}, URhGe~\cite{Har05}, and UCoGe~\cite{HuyPRL2008}. 
The large upper critical field $H_{\rm c2}$ exceeding the Pauli limit is a mark of triplet SC. For UCoGe, the pairing is strongly field-dependent for $H\parallel c$. 
$T_{\rm SC}$ in a high field is not $T_{\rm SC}$ at $H = 0$. 
Note that the $H_{\rm c2}$ curves in (c) are the results in early studies. 
After fine-tuning of the field direction with high-quality single crystals, 
the $H_{\rm c2}$ curves for $H\parallel b$ and $a$ were found to have a more pronounced upturn, as shown later.}
\label{fig1}
\end{fullfigure}

\subsection{Experimental probes}
Of course, thermodynamic measurements [specific heat ($C$), thermal expansion ($\alpha$), magnetization ($M$)] as well as transport measurements [resistivity ($\rho$), Hall effect ($R_{\rm H}$), thermoelectric power (TEP, $S$), thermal conductivity ($\kappa$)] are basic experiments to establish the normal and SC properties. 
In the crude frame chosen to relate to the physical parameters in the normal and SC phases, 
we take the view that the SC is driven by the effective mass enhancement due to magnetic fluctuations over the renormalized band mass $m_{\rm B}$ induced by local fluctuations connected with Kondo phenomena. 
The basic experiments on the normal phase allow the evaluation of the band mass $m_{\rm B}$ and the additional contribution $m^{**}$ given by FM fluctuation to the effective mass 
$m^\ast = m_{\rm B} + m^{\ast\ast} = (1+\lambda) m_{\rm B}$, where $\lambda \equiv m^{\ast\ast}/m_{\rm B}$ is the so-called mass enhancement factor due to the many-body effect. 
The renormalized band mass ($m_{\rm B}$) in heavy fermions is the consequence of complex electronic couplings which cannot be restricted to a single impurity Kondo effect.~\cite{Flo06_review} 
Short-range interactions modify the Kondo temperature of a single U atom;
The estimation of the Kondo temperature must be performed in the intermediate-temperature domain, for example, just above $T_{\rm Curie}$ or by decoupling the interaction by the magnetic field.
For a large number of heavy-fermion compounds, the relation $\gamma^2 \propto A$ between the Sommerfeld coefficient $\gamma$ of the linear $T$ term of the specific heat $C \sim \gamma T$ and the $A$ coefficient of the $T^2$ term of the resistivity~\cite{Kad86} is used to follow the $P$ and $H$ variations of $m^*$, despite the fact that such a relation is not valid near FM instabilities.~\cite{Mor95,Mor03}
In our studies, the $\gamma^2 \propto A$ relation is roughly obeyed even close to $P_c$; 
the ``hidden'' responsible for this cutoff of FM quantum criticality may indicate a sign of the strong first-order nature of $P_c$ and/or the associated change in the FS.   
Thermal conductivity experiments are a supplementary tool for revealing spin fluctuation phenomena and for trying to derive the anisotropic gap structure in the SC phase. 
TEP is a very sensitive probe for detecting topological changes in the FS in these complex multiband systems, where classical quantum oscillation techniques as well as photoemission often fail to resolve the full FS structure.

Elastic and inelastic neutron scattering are powerful probes, mainly for clarifying the FM transition. 
Here, we will stress the strengths of NMR experiments. 
NMR has already been quite successful in the study of conventional~\cite{Slichtertextbook} and unconventional~\cite{Asayamatextbook} superconductors. 
A major interest is the anisotropy of the spin-lattice ($T_1$) and spin-spin ($T_2$) relaxation times as well as its field dependence. 
We will see how unique responses are obtained between longitudinal and transverse fluctuations with respect to the FM magnetization axes. 
Along the three $x$, $y$, and $z$ crystallographic axes, $1/T_1$ is related to the transverse fluctuation via the dynamical susceptibility $\chi''({\bm q},\omega_0)$ as~\cite{Slichtertextbook},
\begin{eqnarray*}
\lefteqn{\left(\frac{1}{T_1T}\right)_{x}}\\
 & = & \frac{\gamma_n^2 k_{\rm B}}{(\gamma_e\hbar)^2} \sum_{{\bm q}} \left[
|A^{y}_{\rm hf}|^2\frac{\chi''_{y}({\bm q},\omega_0)}{\omega_0} + 
|A^{z}_{\rm hf}|^2\frac{\chi''_{z}({\bm q},\omega_0)}{\omega_0}\right], 
\label{eq:T1}
\end{eqnarray*} 
while $1/T_2$ (spin-spin relaxation rate) is mainly sensitive to the longitudinal magnetic fluctuation with an extra contribution originating from $1/T_1$ and is expressed as
\begin{eqnarray*}
\lefteqn{\left(\frac{1}{T_2}\right)_{x} =\frac{\gamma_n^2 k_{\rm B}T}{(\gamma_e\hbar)^2} \lim_{\omega \rightarrow 0} \sum_{{\bm q}} \left[ |A^{x}_{\rm hf}|^2\frac{\chi''_{x}({\bm q},\omega_0)}{\omega_0}\right]} \\
& & +\left[I(I+1)-m(m+1)-1/2\right]\left(\frac{1}{T_1}\right)_x.
\end{eqnarray*}

In an uncorrelated condition, 
the product of $T_1T$ and the spin part of the Knight shift $K_s$ is a constant, $T_1TK_s^2 = R_0$.
This product $R$ will differ strongly in correlated systems ($R > R_0$ for FM, $R < R_0$ for AF), and the characteristic of magnetic correlations can be established from the value of $R/R_0$.

In the SC phase, pairing in a spin singlet or spin triplet is related to a decrease or invariance of the Knight shift across $T_{\rm SC}$. 
The temperature dependence of $1/T_1$ below $T_{\rm SC}$ reflects the gap structure. 
$1/T_1$ varies 
as $\exp{( -\Delta/T)}$ for a finite gap, 
as $T^3$ for a line-node gap, and
as $T^5$ for a point-node gap.~\cite{Asayamatextbook}
In the presence of the residual density of states induced by impurities and the inhomogeneity of samples, $1/T_1$ deviates from the above temperature dependences and approaches the Korringa relation.

\subsection{Crystal structure}
\begin{figure}[!tbh]
\begin{center}
\includegraphics[width=\hsize,pagebox=cropbox,clip]{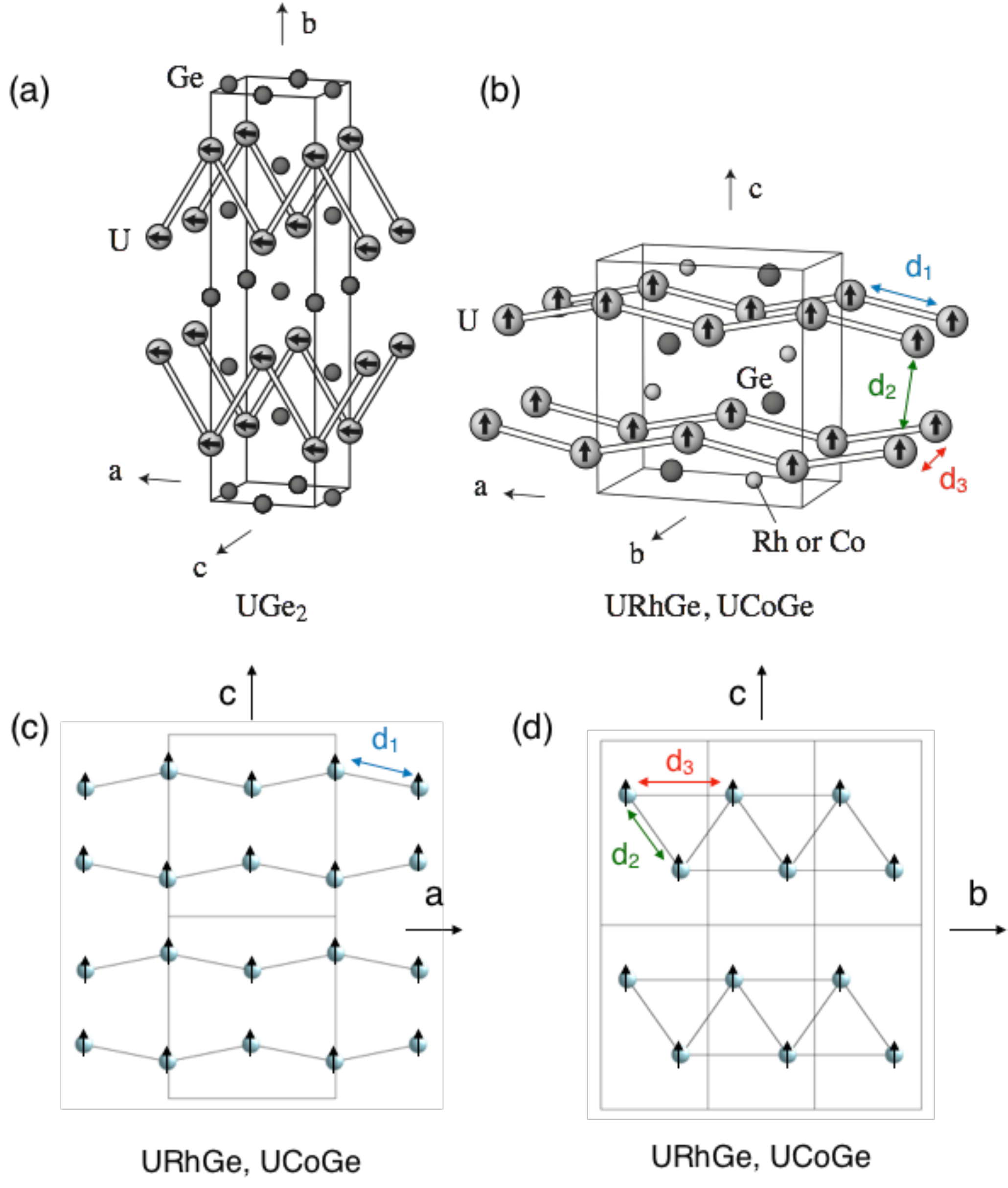}
\end{center}
\caption{(Color online) Crystal structures of (a) UGe$_2$ and (b) URhGe/UCoGe. 
(c), (d) Projections of URhGe/UCoGe from $b$- and $a$-axes, respectively.}
\label{fig:structure}
\end{figure}

\begin{figure}[!tbh]
\begin{center}
\includegraphics[width=0.8\hsize,pagebox=cropbox,clip]{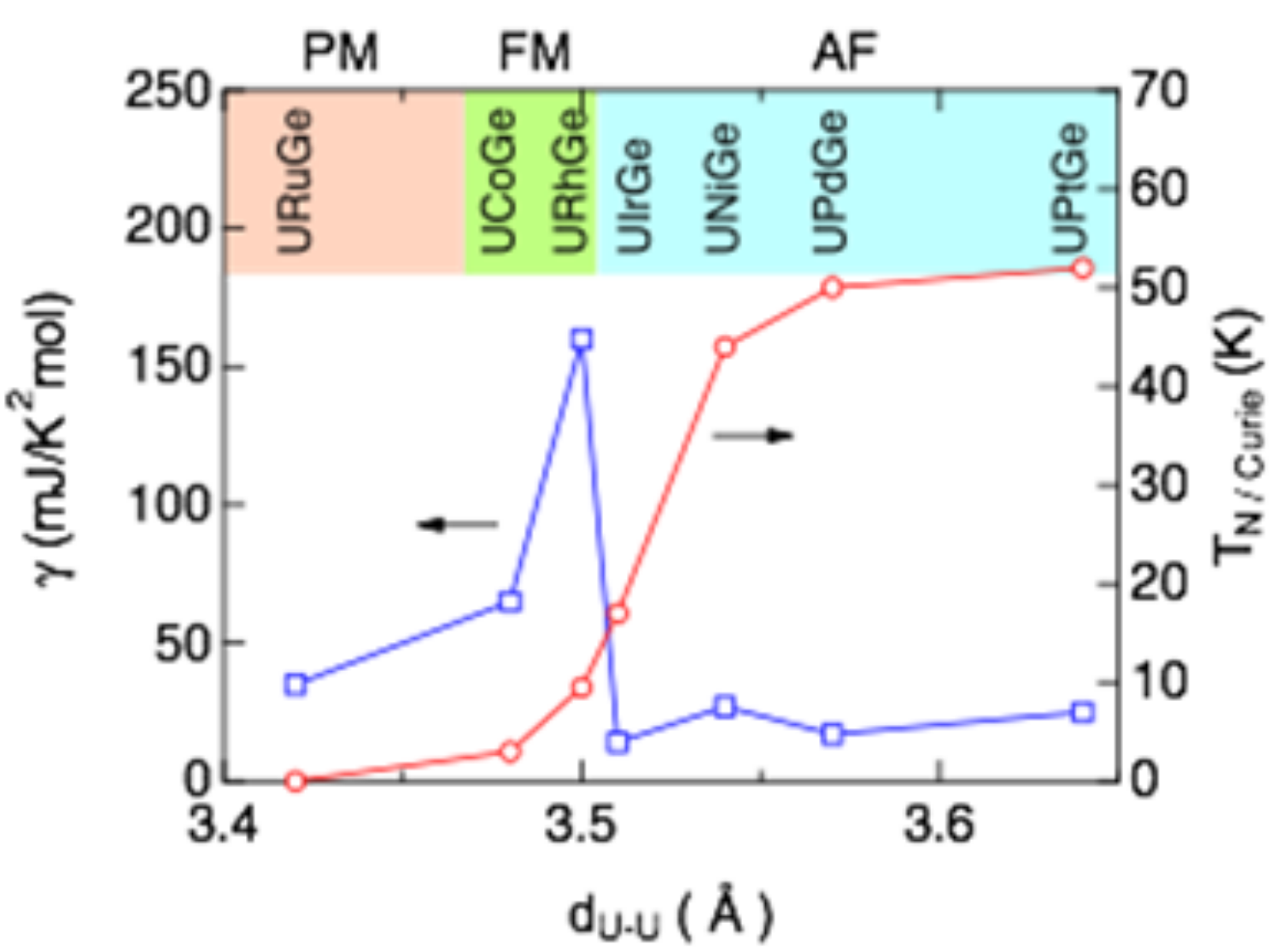}
\end{center}
\caption{(Color online) Sommerfeld coefficient and magnetic ordering temperature as a function of the distance of uranium atoms from the first nearest neighbors in U$T$Ge ($T$: transition metal). UCoGe and URhGe are located at the border between paramagnetism and antiferromagnetism.}
\label{fig2A}
\end{figure}

UGe$_2$ crystallizes in the ZrGa$_2$-type orthorhombic structure with the space group $Cmmm$ ({\#}65, $D_{2h}^{19}$), which belongs to the symmorphic space group, as shown in Fig.~\ref{fig:structure}(a). 
In early studies~\cite{Mak59}, the crystal structure was determined to be the orthorhombic ZrSi$_2$-type with the space group $Cmcm$, but after the refinement of the structure, 
the ZrGa$_2$-type structure was confirmed~\cite{Oik96}. 
The U atoms form zigzag chains along the $a$-axis and the FM moment is directed along the $a$-axis. 
The distance of the first nearest neighbor of the U atom is 3.854 \AA, which is larger than the so-called Hill limit.
URhGe and UCoGe have the TiNiSi-type orthorhombic structure with the space group $Pnma$ ({\#}62, $D_{2h}^{16}$), which belongs to the nonsymmorphic space group, as shown in Fig.~\ref{fig:structure}(b). 
The U atoms form the zigzag chains along the $a$-axis, and the distances between two U atoms are $d_1 = 3.497$ and $3.481\,{\rm \AA}$ for URhGe and UCoGe, respectively, which are close to Hill limit. 
Most U$T$Ge ($T$: transition metal) compounds crystallize in the TiNiSi-type structure or its non-ordered variant CeCu$_2$-type structure. 
As shown in Fig.~\ref{fig2A}, URhGe and UCoGe are located between the PM ground state and the AF ground state, as a function of the next-nearest-neighbor distance, and have relatively large $\gamma$-values.

The global inversion symmetry is preserved in the TiNiSi-type structure; however, the local inversion symmetry at the sites of the U atoms is broken because of the zigzag chain. 
Theoretically, it has been proposed that a spatially inhomogeneous antisymmetric spin-orbit interaction and peculiar physical properties that depend on the energy scale of the band structure might appear. 
In URhGe, a small AF component along the $a$-axis as a result of the small relativistic Dzyaloshinskii--Moriya interaction has been theoretically proposed.~\cite{Min06} 
However, it has not been observed experimentally in a high-quality single crystal, 
although the early studies using polycrystalline powder samples revealed a small AF component~\cite{Tra98}.

\begin{figure}[!tbh]
\begin{center}
\includegraphics[width=0.8\hsize,pagebox=cropbox,clip]{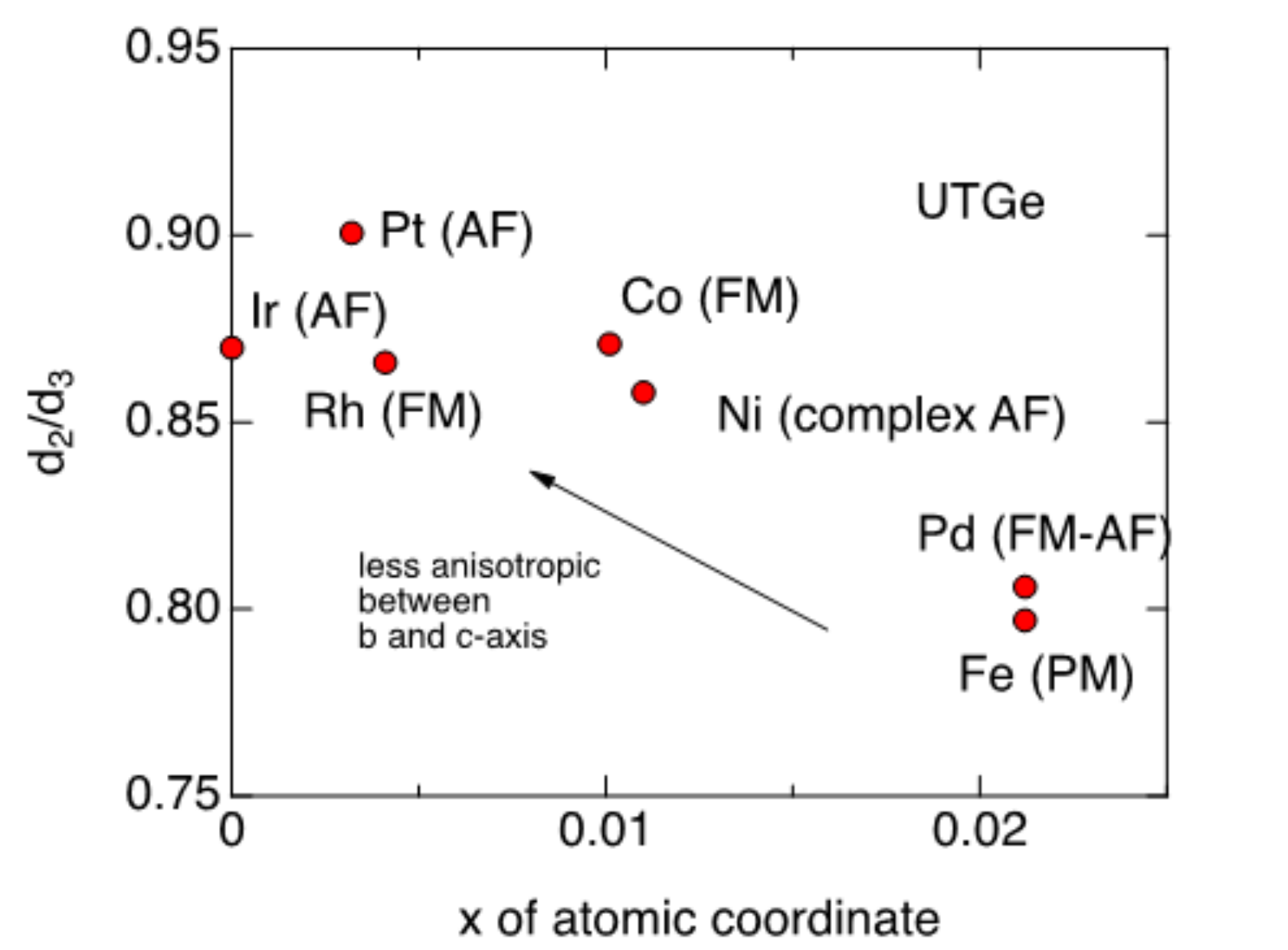}
\end{center}
\caption{(Color online) $d_2/d_3$ vs atomic coordinate $x$ of uranium atom. A large value of $d_2/d_3$ close to 1 and small $x$ are preferable for spin reorientation.}
\label{fig2B}
\end{figure}

Another important point is that the TiNiSi-type orthorhombic structure is derived from the distorted AlB$_2$-type hexagonal structure. 
In fact, the U atoms of URhGe are almost located in the $bc$-plane ($a$-plane), but their alignment is slightly corrugated owing to the $x$ parameter of the atomic coordinate ($x$ = 0.0041), which corresponds to a quite small displacement of $0.028\,{\rm \AA}$ from the $bc$-plane. 
If we neglect this small corrugation, the U atoms form a distorted hexagon or successive triangles, as shown in Fig. 2(d). 
The distances $d_2$ and $d_3$ between the U atoms are $3.746$ and $4.327\,{\rm \AA}$, respectively, and the ratio $d_2/d_3$ is 0.866 in URhGe. 
If $d_2/d_3$ is 1, the U atoms form equilateral triangles, and the magnetic anisotropy between the $b$- and $c$-axes will be very small because the exchange interactions due to $d_2$ and $d_3$ are almost equivalent. 
As described later, URhGe shows spin reorientation from the $c$- to $b$-axis at low temperatures when a field is applied along the $b$-axis, indicating that the magnetic anisotropy between the $b$- and $c$-axes is relatively small in spite of the Ising magnetic character. 
The small anisotropy between the $b$- and $c$-axes and the hard-magnetization $a$-axis are a general trend in the U$T$Ge system. 
The key parameters are the $x$-value of the atomic coordinate of the U atom and the ratio $d_2/d_3$. 
The small $x$-value and the large $d_2/d_3$ close to 1 are preferable conditions for small anisotropy between the $b$- and $c$-axes, leading to the spin reorientation at low fields. 
Figure \ref{fig2B} shows $d_2/d_3$ plotted against the $x$-value for different U$T$Ge compounds. 
URhGe satisfies the preferable conditions for spin reorientation, 
while UCoGe has a larger $x$-value, increasing the difficulty of spin reorientation. 
In Fig. \ref{fig2B}, one can recognize that spin reorientation is more likely to occur in UIrGe and UPtGe. 
In fact, the antiferromagnet UIrGe shows metamagnetic transitions from an AF state to a polarized PM state at 21 and 14 T for $H \parallel  b$ and $H \parallel c$, respectively, 
while the hard-magnetization axis ($a$-axis) shows no metamagnetism up to $50\,{\rm T}$.~\cite{Yos06} 
In UPtGe, an incommensurate cycloidal magnetic structure is found in the $bc$-plane for the TiNiSi-type structure~\cite{Man00}, although the structure is refined to the EuAuGe type from the TiNiSi type with small modifications. 
Upon 10 \% Ir doping of URhGe, the spin reorientation field is significantly reduced to $\mu_0 H_{\rm R}$ = 9.4 T with $T_{\rm Curie}$ = 9.3 K. 
A similar trend is also observed for 2\% Pt-doped URhGe with $\mu_0 H_{\rm R}$ = 10 T and $T_{\rm Curie}$ = 8.6 K. 

\section{Common Features and Particularities}
The goal is to present the domain of existence of the different phases in $(T,P,H)$ space,
to see the consequences of the self-induced vortex (SIV) created in the FM phase, and to stress the unique opportunity for the modification of the SC pairing by $H$ acting on the FM interaction.
\subsection{($P$, $T$) phase diagram}
\begin{fullfigure}[!tbh]
\begin{center}
\includegraphics[width=0.8\textwidth,pagebox=cropbox,clip]{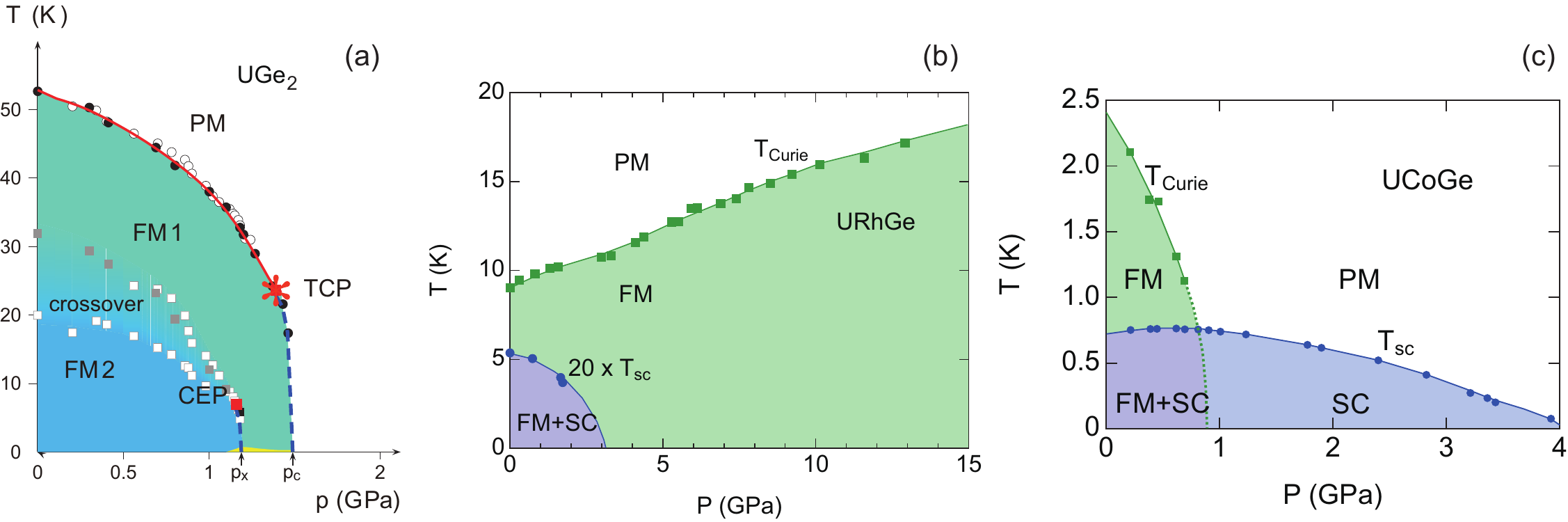}
\end{center}
\caption{(Color online) (Color online) ($T, P$) phase diagram of (a) UGe$_2$, (b) URhGe, and (c) UCoGe at $H=0$. 
For UGe$_2$, 
two FM phases (FM1: weakly polarized phase, FM2: strongly polarized phase) are clearly separated below the CEP ($P_{\rm CEP}$ = 1 GPa, $T_{\rm CEP}$ = 7 K) by the first-order transition.
The TCP is marked by a star ($T_{\rm TCP}$ = 24 K, $P_{\rm TCP}$ = 1.42 GPa). 
For URhGe, $T_{\rm Curie}$ increases with $P$ up to $13\,{\rm GPa}$; SC will collapse at $P_{\rm S} \sim\,{\rm GPa}$. 
For UCoGe, $P_{\rm C} \sim 1\,{\rm GPa}$ and $P_{\rm S} = 4\,{\rm GPa}$; SC survives deep inside the PM domain.  }
\label{fig3}
\end{fullfigure}
Figure~\ref{fig3} shows the ($P, T$) phase diagram of the three compounds. 
As shown in Table~\ref{tab:1}, the FM moments at ambient pressure
are $1.5$, $0.4$, and $0.06\,\mu_{\rm B}$ in UGe$_2$, URhGe, and UCoGe, respectively.
Thus, the duality between the localized and itinerant character of the 5$f$ electron is strong in UGe$_2$,
while an itinerant description of 5$f$ electrons seems to be justified in UCoGe.
UGe$_2$  has a rather high Curie temperature $T_{\rm Curie} \sim 52\,{\rm K}$ with a large magnetic moment $M_0  \sim 1.5\,\mu_{\rm B}$ at $P = 0$ in the FM2 ground state. 
On cooling below $T_{\rm Curie}$, the competition between the low-magnetic-moment ($M_0 \sim 0.9 \mu_{\rm B}$) phase FM1  and the FM2 ground state is marked by a large $T$ crossover.~\cite{Har09_UGe2} 
At $P \sim 1$ GPa, a first-order transition between FM2  and FM1 appears at $P_{\rm CEP}$ and $T_{\rm CEP}$, which will end up at $P_x \sim$ 1.2 GPa. 
Specific attention is given to the FM-PM change from a second order to first-order transition at the tricritical point (TCP) at $T_{\rm TCP}$ and $P_{\rm TCP}$; 
this change is directly associated with the observation of the FM wings created by the restoration of ferromagnetism in magnetic fields\cite{TaufourPRL2010}. 
The particularities of SC in UGe$_2$ are that
(1) it exists only in the FM domain with the maximum $T_{\rm sc}$ at $P_x$, and
(2) $P_x$ is coupled with a drastic change of the FS on switching from FM2 to FM1. 

In URhGe, $T_{\rm Curie}$ ($=9.5\,{\rm K}$) and $M_0$ ($=0.4\,\mu_{\rm B}$) at ambient pressure 
are much lower than those of UGe$_2$ at $P_x$ where SC emerges.
Under pressure, the FM becomes more stable and $T_{\rm Curie}$ still increases even at 13 GPa\cite{HardyPhysicaB2005}.  
As $T_{\rm Curie}$ increases under pressure, the magnetic fluctuations become weaker and $T_{\rm SC}$ decreases. 
The collapse of SC will occur at around $P_{\rm S} \sim$ 4 GPa.

UCoGe has a small FM moment, $M_0 \sim 0.06 \mu_{\rm B}$, with a small magnetic entropy release at $T_{\rm Curie}$,
indicating a clear example of itinerant FM. 
Here, pressure drives the system towards FM instabilities.
FM disappears at around 1 GPa\cite{SlootenPRL2009, HassingerJPSJ2008, BastienPRB2016}, 
while SC survives deep inside the PM phase up to $P_{\rm S} \sim 4\,{\rm GPa}$.~\cite{BastienPRB2016} 
The characteristic values for the three compounds are shown in Table~\ref{tab:1}.
The internal field $H_{\rm int}$ created by $M_0$ below $T_{\rm Curie}$ as well as the FM molecular field $H_{\rm mol}$ are shown in Table~\ref{tab:2}.
\begin{fulltable}[!tbh]
\caption{
Easy axis, Curie temperature $T_{\rm Curie}$, ordered moment along the easy axis $M_0$,
Sommerfeld coefficient $\gamma$ at $P=0$, 
critical pressure $P_{\rm C}$ for switching from FM to PM phase, and
SC transition temperature $T_{\rm SC}$ in UGe$_2$ at $P_x$ and in URhGe, and UCoGe at $P=0$.}  
\begin{tabular}{ccccccc} \hline
             &Easy& $T_{\rm Curie}$ & $M_0$           & $\gamma$   & $P_{\rm C}$ & $T_{\rm SC}$  \\
             & axis&  (K)           & $(\mu_{\rm B})$  & (${\rm mJ\,mol^{-1} K^{-2}}$) &  (GPa)       &   (K)             \\    \hline 
UGe$_2$ & a   & 52                    & 1.48           &  34             &  1.6             &  0.8 at $P = P_x$ \\
URhGe   & c   & 9.5                   & 0.4             &  163            &  $> 13$          &   0.25 at $P$ = 0  \\
UCoGe   & c   & 2.7                   & 0.06           &   57             &  $\sim 1$         &   0.8 at $P$ = 0   \\
\hline
\end{tabular}
\label{tab:1}
\end{fulltable}
\begin{table}[!tbh]
\caption{Internal field $H_{\rm int}$ and molecular field $H_{\rm mol}$ for $T\to 0$.}
\begin{tabular}{ccc} \hline
             &$H_{\rm int}$ (T)      & $H_{\rm mol}$ (T)  \\ \hline 
UGe$_2$ & 0.28                   & 50                       \\
URhGe   & 0.08                   & 10                       \\
UCoGe   & 0.01                   & 2.5                       \\
\hline
\end{tabular}
\label{tab:2}
\end{table}

\subsection{SC depairing and self-induced vortex state}
\begin{figure}[!tbh]
\begin{center}
\includegraphics[width=0.6\hsize,pagebox=cropbox,clip]{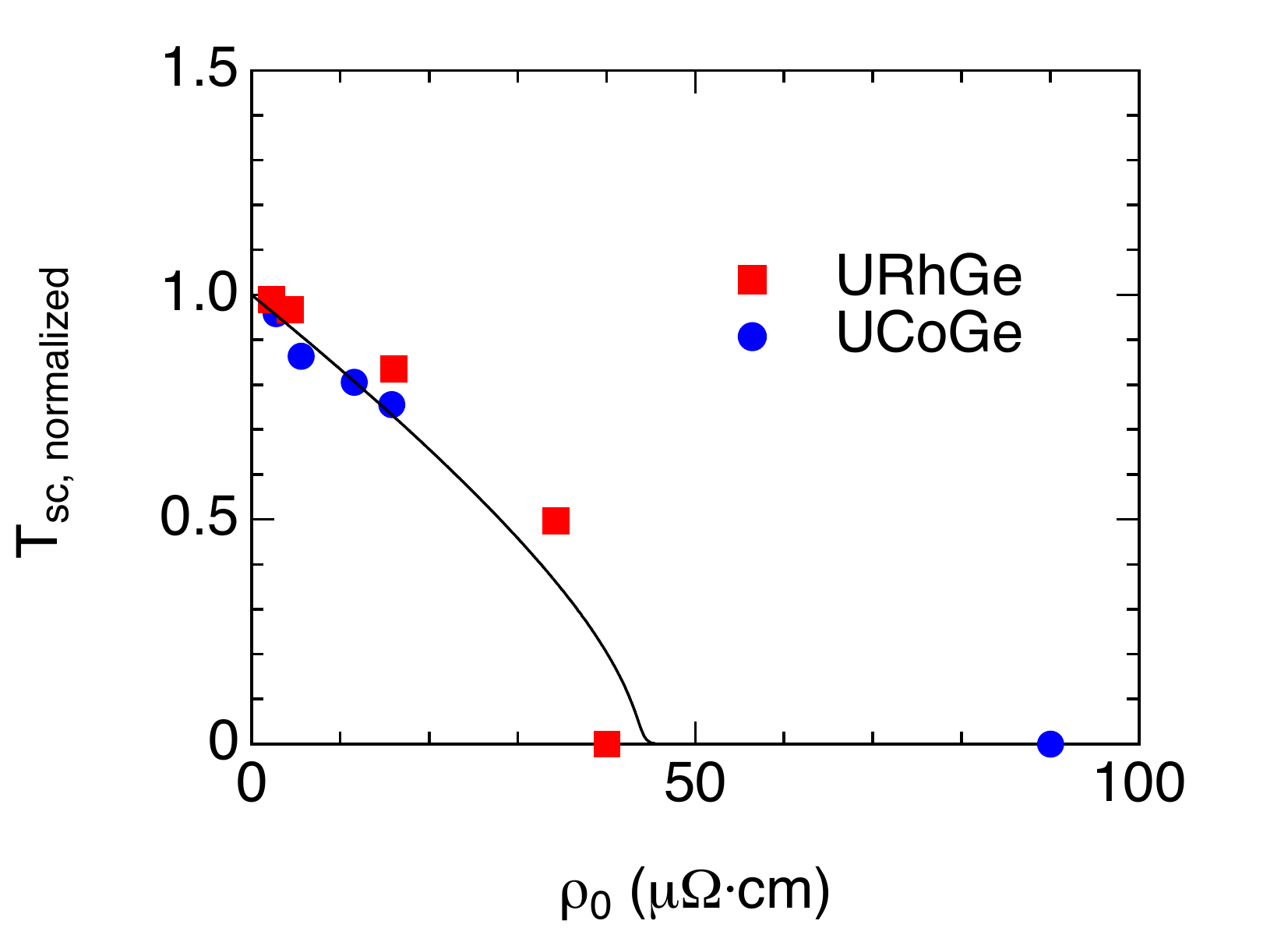}
\end{center}
\caption{(Color online) Relative dependence of $T_{\rm SC}$ as a function of residual resistivity $\rho_0$ in URhGe and UCoGe. The solid line is obtained from the Abrikosov--Gor'kov pair-breaking function.}
\label{fig4}
\end{figure}

The relative variation of $T_{\rm SC}$ as a function of the residual resistivity ($\rho_0$) (Fig.~\ref{fig4}) shows that, as expected in unconventional SC, $T_{\rm SC} / T^0_{\rm SC}$ depends strongly on $\rho_0$, which is inversely proportional to the electronic mean free path.

\begin{figure}[!tbh]
\begin{center}
\includegraphics[width=\hsize]{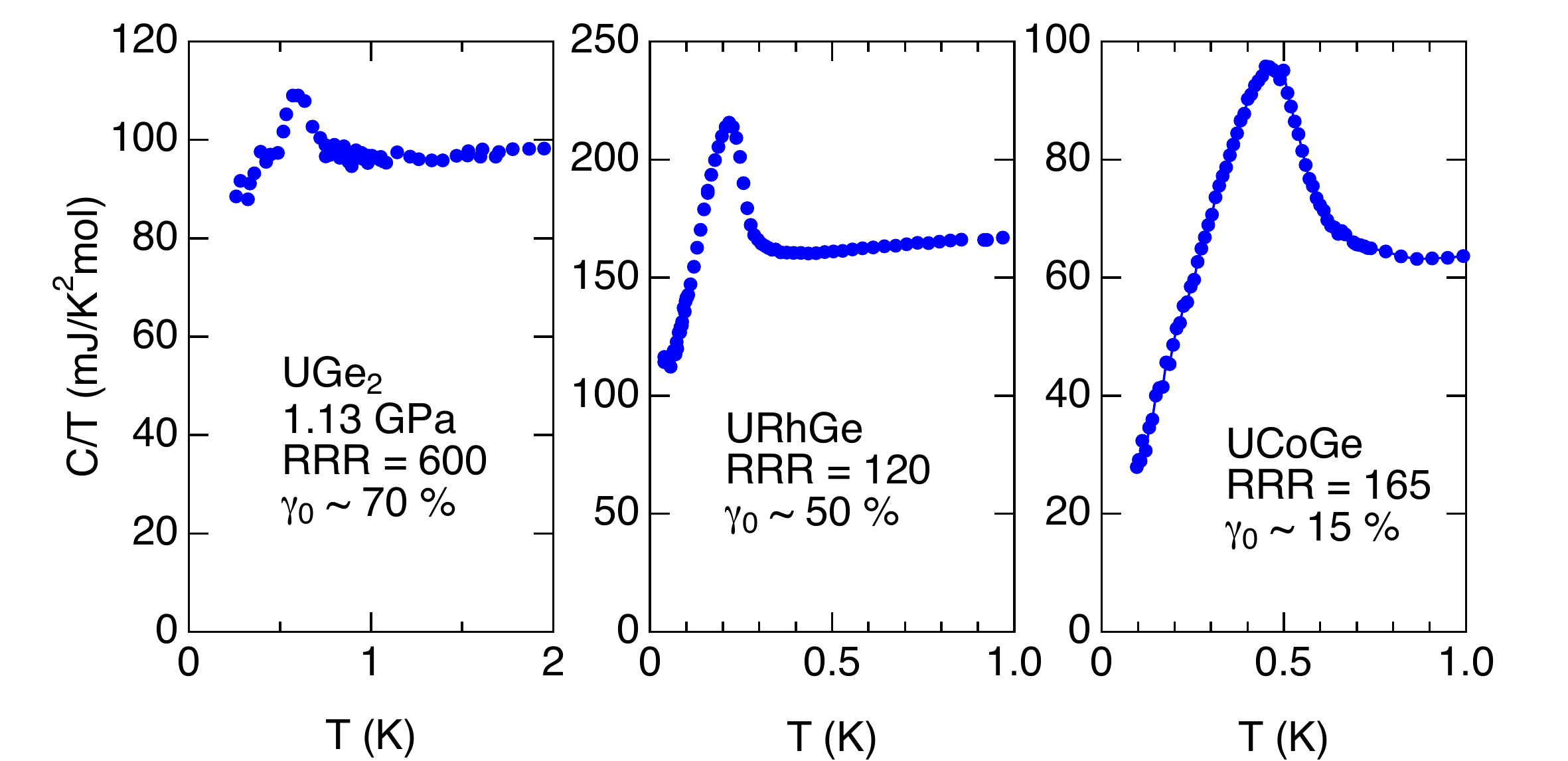}
\end{center}
\caption{(Color online) Specific heat in UGe$_2$, URhGe, and UCoGe.}
\label{fig:Cp}
\end{figure}

\begin{figure}[!tbh]
\begin{center}
\includegraphics[width=0.6\hsize,pagebox=cropbox,clip]{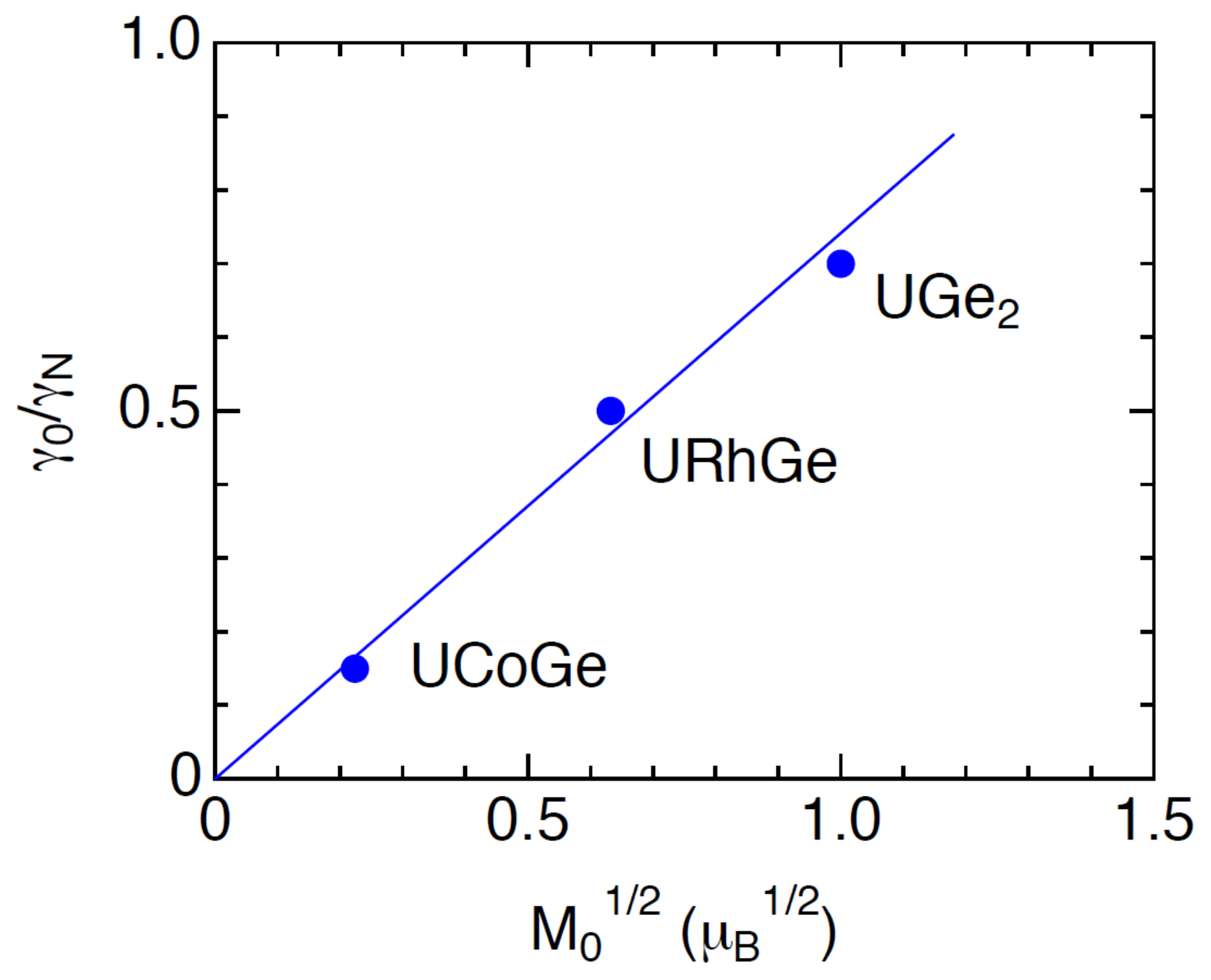}
\end{center}
\caption{(Color online) Relative variation of the residual value $\gamma_0$ normalized by $\gamma_{\rm N}^{\mathstrut}$ as a function of $\sqrt{M_0}$. This graph proves that strong Volovik vortices already contribute  at $H = 0$.}
\label{fig6}
\end{figure}

Another singularity is that as $H_{\rm int}$ is much higher than $H_{\rm c1}$ (as shown later in magnetization curves for UCoGe),
thus SIVs exist at $H = 0$. 
The creation of self induced vortex below $T_{\rm SC}$ will lead to the residual contribution $\gamma_1 \sim H_{\rm int} /H_{\rm c2}$ to the linear $T$ term. 
In addition, the phenomenon is enhanced by the additional contribution $\gamma_2$ given by the Volovik effect in the inter vortex phase\cite{MineevPUsr2017,VolovikJETPL1993}.
Figure \ref{fig:Cp} shows the SC anomaly of the three compounds; the SC jump at $T_{\rm SC}$ is directly related to the weakness of $M_0$.
Even the residual $\gamma$ term in the SC phase follows a $\sqrt{M_0}$ dependence as predicted for the main origin of the Volovik effect (Fig.~\ref{fig6})~\cite{MineevPUsr2017}.

\subsection{Transverse and longitudinal $H$ scan: consequences on SC pairing}
\begin{figure}[!tbh]
\begin{center}
\includegraphics[width=0.6\hsize,pagebox=cropbox,clip]{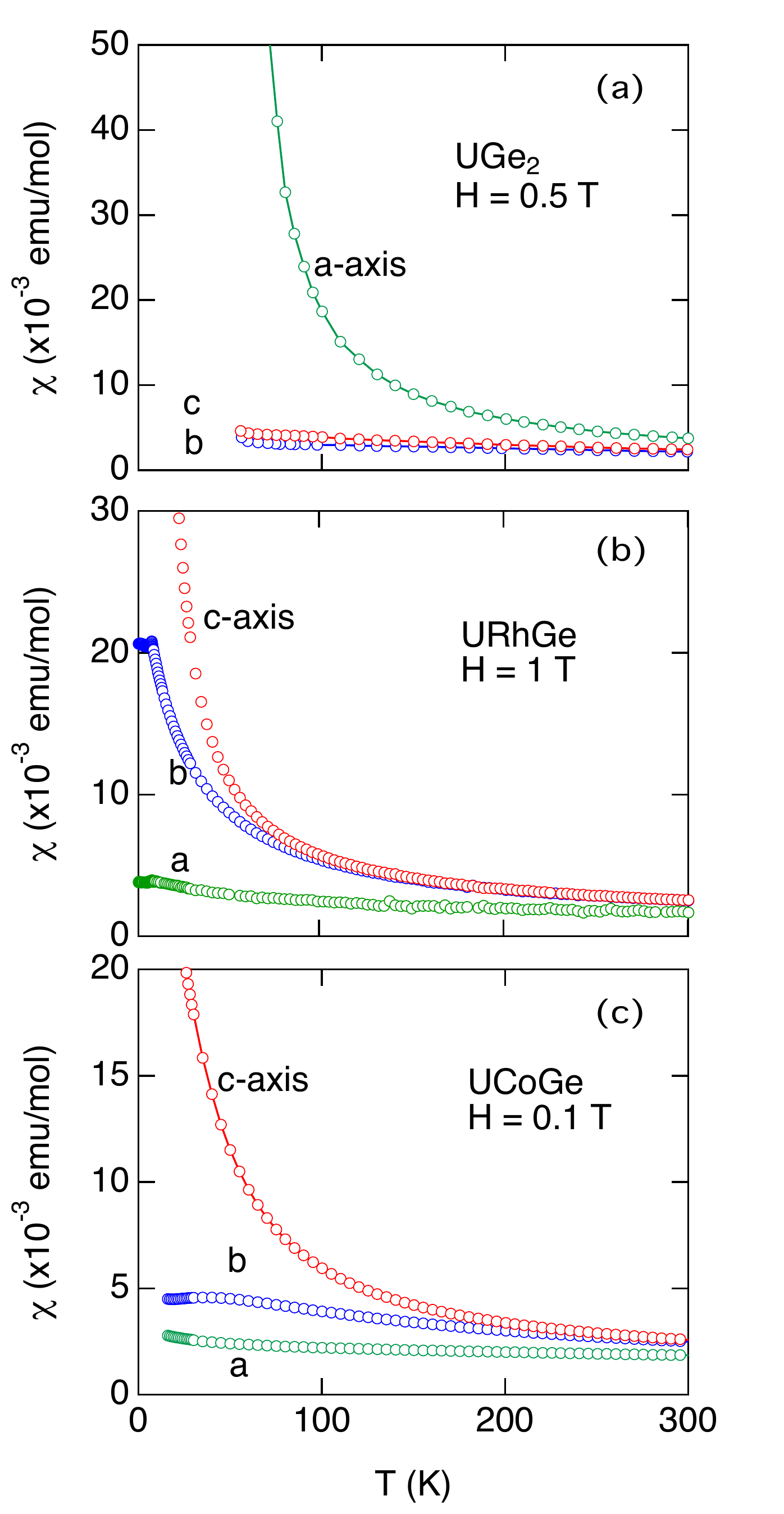}
\end{center}
\caption{(Color online) Susceptibilities of (a) UGe$_2$, (b) URhGe, and (c) UCoGe. In URhGe, owing to the weak magnetocrystalline term, $\chi_c$ clearly becomes the easy magnetization axis only below 50 K. The susceptibility in UGe$_2$ is replotted from Ref.~\citen{TrocPRB2012}.}
\label{fig7}
\end{figure}

Figure \ref{fig7} shows the low-field susceptibility $\chi$ of the three compounds measured along their principal axes.
At room temperature, the $a$-axis is already the easy magnetization axis in UGe$_2$,
and the $c$-axis becomes the easy axis in UCoGe.~\cite{HuyPRL2008}
However, in URhGe, almost no anisotropy appears between the $b$- and $c$-axes at room temperature, and
the $c$-axis can be differentiated from the $b$-axis only below $50\,{\rm K}$.
UGe$_2$ and UCoGe are considered to be Ising ferromagnets.
For URhGe, the duality between FM along the $c$- and $b$-axes is at the core of its extremely high field-reentrant superconductivity (RSC) for $H\parallel b$.

\begin{figure}[!tbh]
\begin{center}
\includegraphics[width=0.6\hsize,pagebox=cropbox,clip]{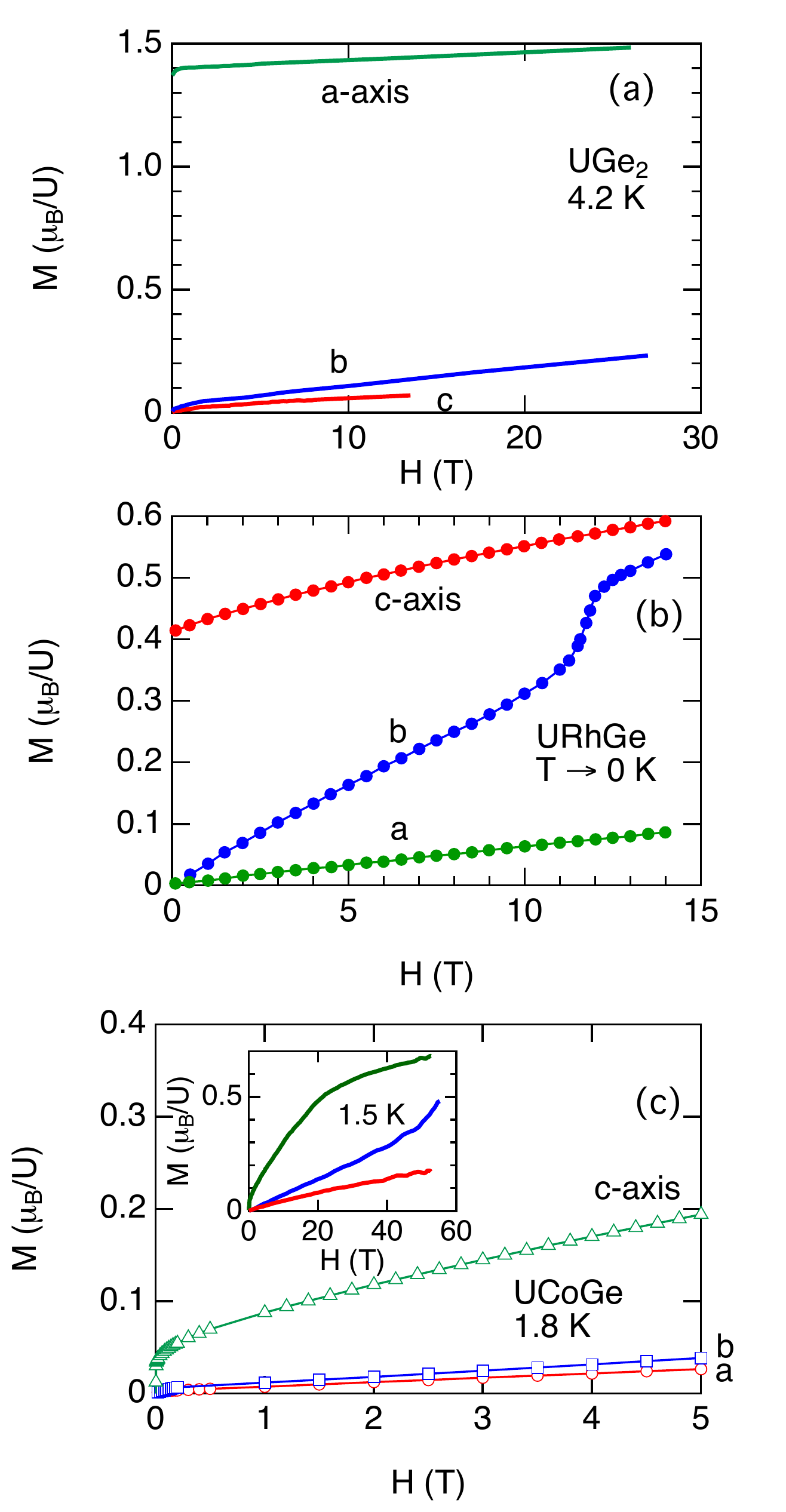}
\end{center}
\caption{(Color online) 
Magnetization of (a) UGe$_2$~\cite{Sakon07}, (b) URhGe, and (c) UCoGe in their normal phase.
In URhGe, the magnetization for $H \parallel b$ shows ``metamagnetic''-like transition at $H = H_{\rm R}$, indicating a switch of the easy magnetization axis from the $c$-axis to the $b$-axis.~\cite{Har11} 
Note the high value of $\chi_b = dM_b /dH$ up to $H_{\rm R}$ and the weak curvature of $M(H)$ for $H \parallel c$.  In UCoGe, the strong curvature of $M(H)$ for $H \parallel c$ is directly linked with the strong decrease in $m^{**}$ with increasing $H$ and that in $\lambda$,  the relation, $\chi_c > \chi_b \sim \chi_a$, is preserved  regardless of the magnetic field.~\cite{HuyPRL2008,KnafoPRB2012}}
\label{fig8}
\end{figure}

Figure \ref{fig8} shows the magnetizations of UGe$_2$, URhGe, and UCoGe at low temperatures along their $a$-, $b$-, and $c$-axes.~\cite{Sakon07,Har11,HuyPRL2008}
In URhGe, the large value of the initial slope of magnetization, $dM/dH\equiv \chi$, for the $b$-axis compared with that for the $c$-axis indicates that 
under a magnetic field of $\mu_0 H_{\rm R} \sim 12\,{\rm T}$, 
the $b$-axis will become the easy magnetization axis. 
On the other hand, $\chi$ for the $c$-axis always exceeds that for the $b$-axis in UCoGe; thus
the Ising character will be preserved up to a very high magnetic field. 
The strong curvature of the magnetization curve for the $c$-axis in UCoGe implies that 
the contribution of the spin fluctuation will drastically decrease with increasing $H$;
this unusual $M(H)$ curve has major consequences on the SC properties for $H \parallel c$, particularly on the $H_{\rm c2}$ curvature (see below).

\begin{figure}[!tbh]
\begin{center}
\includegraphics[width=\hsize,pagebox=cropbox,clip]{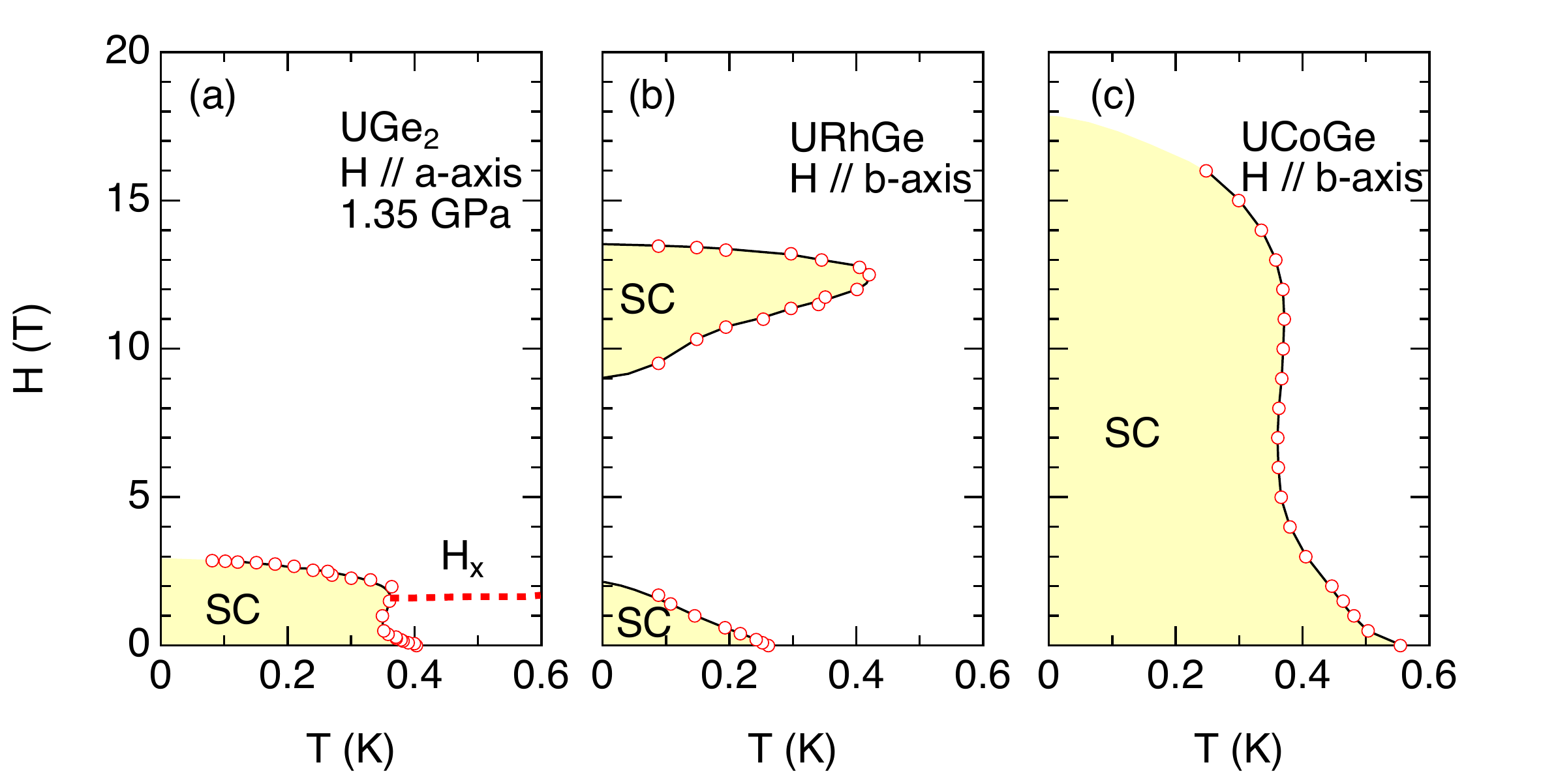}
\end{center}
\caption{(Color online) (a) $H_{\rm c2}$ versus $T$ for $H \parallel a$ (longitudinal field scan with respect to $H \parallel M_0$) in UGe$_2$~\cite{SheikinPRB2001}, providing evidence that the $T_{\rm SC}$ dependence close to $P_x$ is sharp.  
$H_{\rm c2}$ versus $T$ for $H \parallel b$ (transverse field scan $H \perp M_0$) in (b) URhGe~\cite{LevyScience2005} and (c) UCoGe~\cite{AokiJPSJ2009}. In these cases, the evidence of $H$ reinforcement of the pairing is connected with the collapse of $T_{\rm Curie}$.    }
\label{fig:Hc2}
\end{figure}

In UGe$_2$ for $H \parallel a$ [Fig.~\ref{fig:Hc2}(a)], the sudden enhancement of $H_{\rm c2}$ for $P$ just above $P_x$  has been taken to be a consequence of the $H$-switch from the FM1 to FM2 phases, 
since these two phases have different SC temperatures at $H$ = 0.~\cite{SheikinPRB2001}

\begin{figure}[!tbh]
\begin{center}
\includegraphics[width=\hsize]{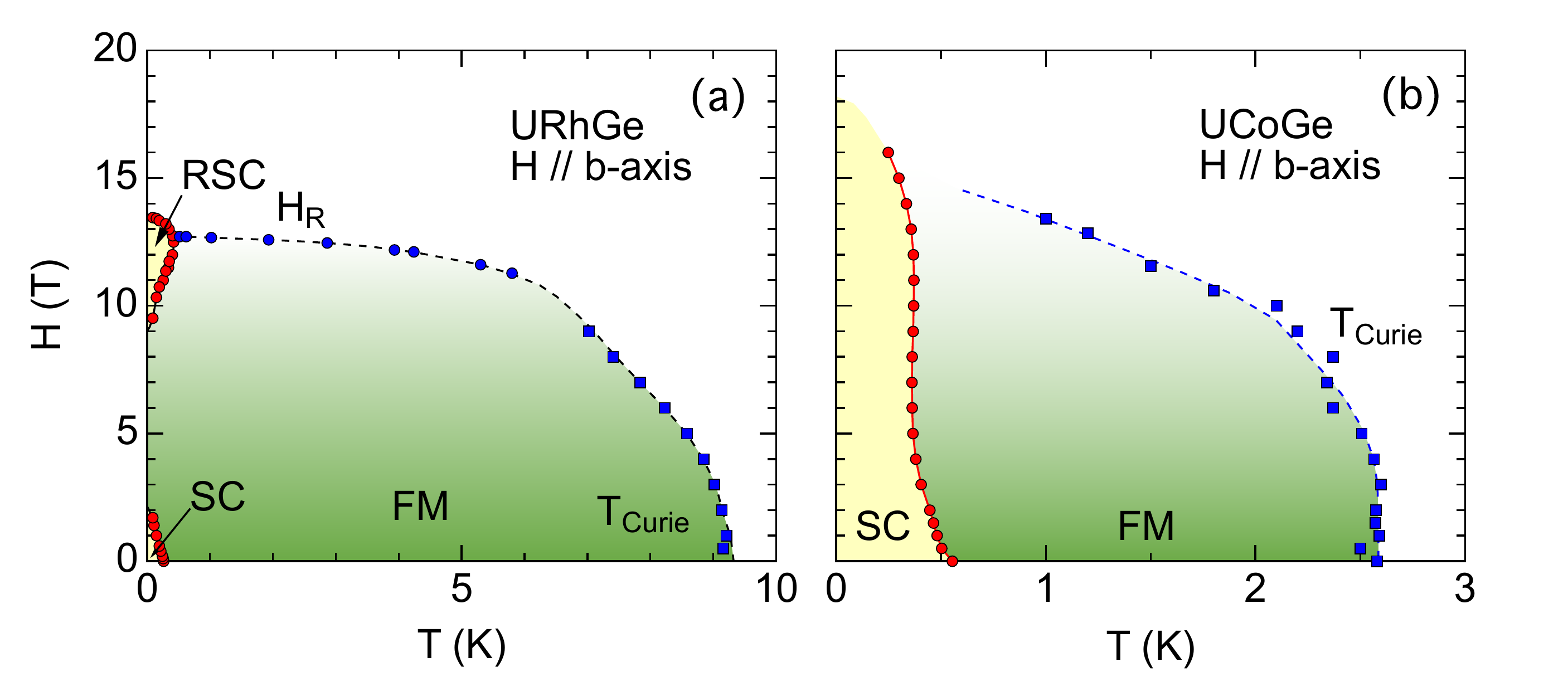}
\end{center}
\caption{(Color online) Overlap between the SC and FM domains in the ($H, T$) plane at $P = 0$ for $H \parallel b$. For URhGe (a), data are taken from transport, magnetization, and thermal expansion measurements and for UCoGe (b), data are taken from transport, thermal expansion, and NMR measurements.}
\label{fig10}
\end{figure}

Considerably more interesting in Figs.~\ref{fig:Hc2}(b) and \ref{fig:Hc2}(c) are the cases of URhGe and UCoGe,
where reentrant SC in URhGe and field-reinforced SC in UCoGe occur for a transverse magnetic field scan ($H\parallel b$) with respect to the initial FM direction ($M_0 \parallel c$). 
The key origin of this singular behavior is that a transverse $H$ scan leads to the collapse of the Ising FM along the $c$-axis and thus gives a unique elegant opportunity to cross the FM-PM instability with the enhancement of FM fluctuations. 
Proof of the collapse of ferromagnetism has been observed in thermodynamic and transport experiments as well as in NMR measurements. 
Figure~\ref{fig10} shows the FM and SC domains of URhGe and UCoGe.
Table~\ref{tab:mag}~\cite{Har09_UGe2} summarizes the different values of the low-temperature susceptibility at $H \rightarrow 0$ and the estimated critical magnetic fields $H^\ast_a$, $H^\ast_b$, and $H^\ast_c$, where each magnetization $M_a$, $M_b$, and $M_c$ reaches $M_0$.~\cite{Har11} 
Note that for URhGe, $\chi_b > \chi_c > \chi_a$ and that $\chi_b$ in URhGe is even larger than $\chi_c$ in UCoGe. 
The weakness of the magnetic anisotropy leads to FM instabilities with $M_0 \parallel c$ and $M_0 \parallel b$ for URhGe occuring at $H_{\rm R}$.
For UCoGe, $\chi_c$ is much greater than $\chi_b$; this property will reflect only the Ising FM proximity with $M_0 \parallel c$. 
These different behaviors will be demonstrated in NMR experiments. 

\begin{table}
\caption{Field derivative magnetizations at $H \to 0$ and the estimated critical field along each axis.~\cite{Har09_UGe2}} 
\begin{tabular}{ccccccc} \hline
             &$\chi_a$                      & $\chi_b$                     & $\chi_c$                     & $H^*_a$           & $H^*_b$             & $H^*_c$           \\
             & ($\mu_{\rm B}/T$)   &  ($\mu_{\rm B}/T$)                 &  ($\mu_{\rm B}/T$)    &  (T)      &   (T)               & (T)               \\    \hline 
UGe$_2$ & 0.006                         & 0.0055                        & 0.011                         &  230              &  250               &  122             \\
URhGe   & 0.006                         & 0.03                           & 0.01                           &  66                &  13                &  40             \\
UCoGe   & 0.006                         & 0.0055                        & 0.011                         &  29              &  12               &  2.5             \\
\hline
\end{tabular}
\label{tab:mag}
\end{table}

Now we will describe the singular features of each compound. \\
For UGe$_2$: 
\begin{itemize}
\item FM transition switches from second order to first order at $T_{\rm TCP}$, $P_{\rm TCP}$
\item Detection of FM wings in the $(T,P,H)$ phase diagram for $H \parallel a$ (easy magnetization axis) up to the quantum critical end point
\item Drastic change of the FSs on entering the three different phases FM2, FM1, and PM.
\item Consequences on the SC domain with the interplay between FS instability and FM spin fluctuations.
\end{itemize}
For URhGe:
\begin{itemize}
\item Appearance of FM wings upon tilting the field direction from the $b$- to $c$-axis around $H_{\rm R}$.
\item Duality between FM along the $c$- and $b$-axes: concomitant longitudinal and transversal fluctuations detected by NMR near $H_{\rm R}$.
\item Link between RSC and $H$ dependence of $m^{**}$.
\item Pressure, uniaxial stress dependences of RSC: evidence of scaling in $m^\ast (H_{\rm R})/m^\ast (0)$ supporting Lifshitz transition.
\end{itemize}
For UCoGe:
\begin{itemize}
\item Precise magnetization measurements at very low temperature: hierarchy between $H_{\rm c1}$, $H_{\rm int}$ /strong $H$ curvature of $M$($H$) for $H \parallel c$.
\item Observation by NMR of mainly longitudinal spin fluctuations along $c$-axis, regardless whether the field direction is along the $a$-, $b$-, or $c$-axis.
\item Huge decrease in $m^{\ast\ast}$ with increasing $H$ in longitudinal scan and strong increase in $m^{**}$ in transverse magnetic field on approaching $H^\ast_b$.
\item Description of $H_{\rm c2}$ curve via the field dependence of the parameter $\lambda$ defined by $m^{**} / m_{\rm B}$.
\item ($P$, $H$) phase diagram with collapse of FM at $P_C \sim 1\,{\rm GPa}$, persistence of SC up to $P_{\rm S} \sim 4\,{\rm GPa}$, link between FM collapse and $H_{\rm c2}$ singularities for $H \parallel M_0$ and $H \perp M_0$.
\end{itemize}
The different theoretical approaches will be presented with focus on the interplay between magnetism and unconventional SC, and on additional phenomena related to the Lifshitz transition.
Special attention will be given later to the present knowledges of FSs referring to band structure calculations. 

\section{Properties of UGe$_2$}\label{sec:UGe2}
Contrary to the canonical example of SC around $P_{\rm c}$ driven by spin fluctuations,
for UGe$_2$, the singularity is that SC appears close to $P_{\rm x}$
where the system switches from FM2 to FM1 phases.
The clear feature is that FS reconstruction at $P_x$ must be considered to evaluate the SC pairing.
\subsection{Two FM ground states, FM wing, and FS instability}
\begin{figure}[!tbh]
\begin{center}
\includegraphics[width=0.6\hsize,pagebox=cropbox,clip]{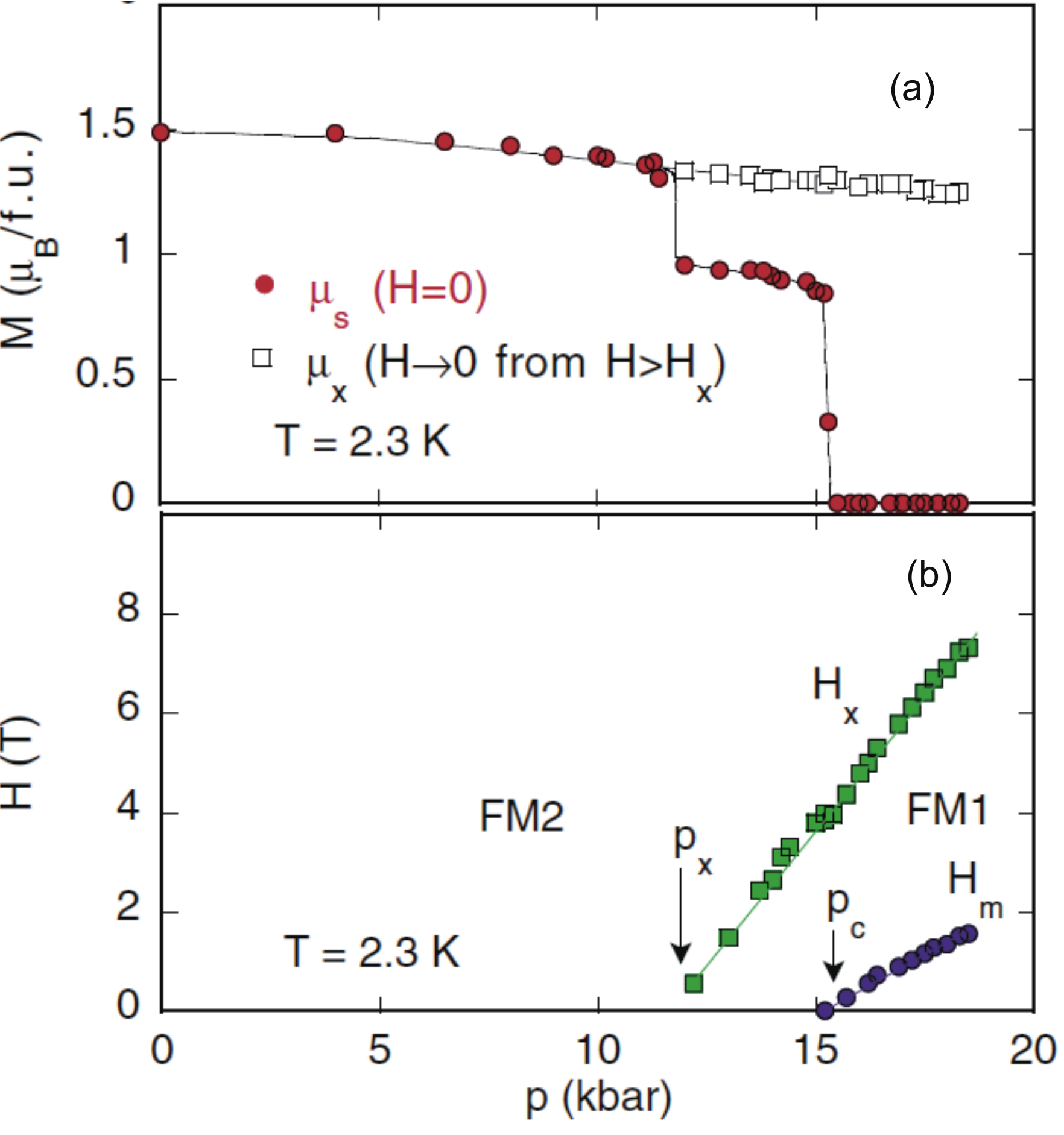}
\end{center}
\caption{(Color online) (a) Variation of the FM $M_0$ component at $H$ = 0 for the FM2 and FM1 states in UGe$_2$ at $2.3\,{\rm K}$.~\cite{PfleidererPRL2002} Extrapolation of $M_0$ above $H_{\rm X}$. 
(b) Field variation of $H_{\rm X}$ and $H_c$ as a function of pressure ($P$). }
\label{fig11}
\end{figure}

\begin{figure}[!tbh]
\begin{center}
\includegraphics[width=\hsize,pagebox=cropbox,clip]{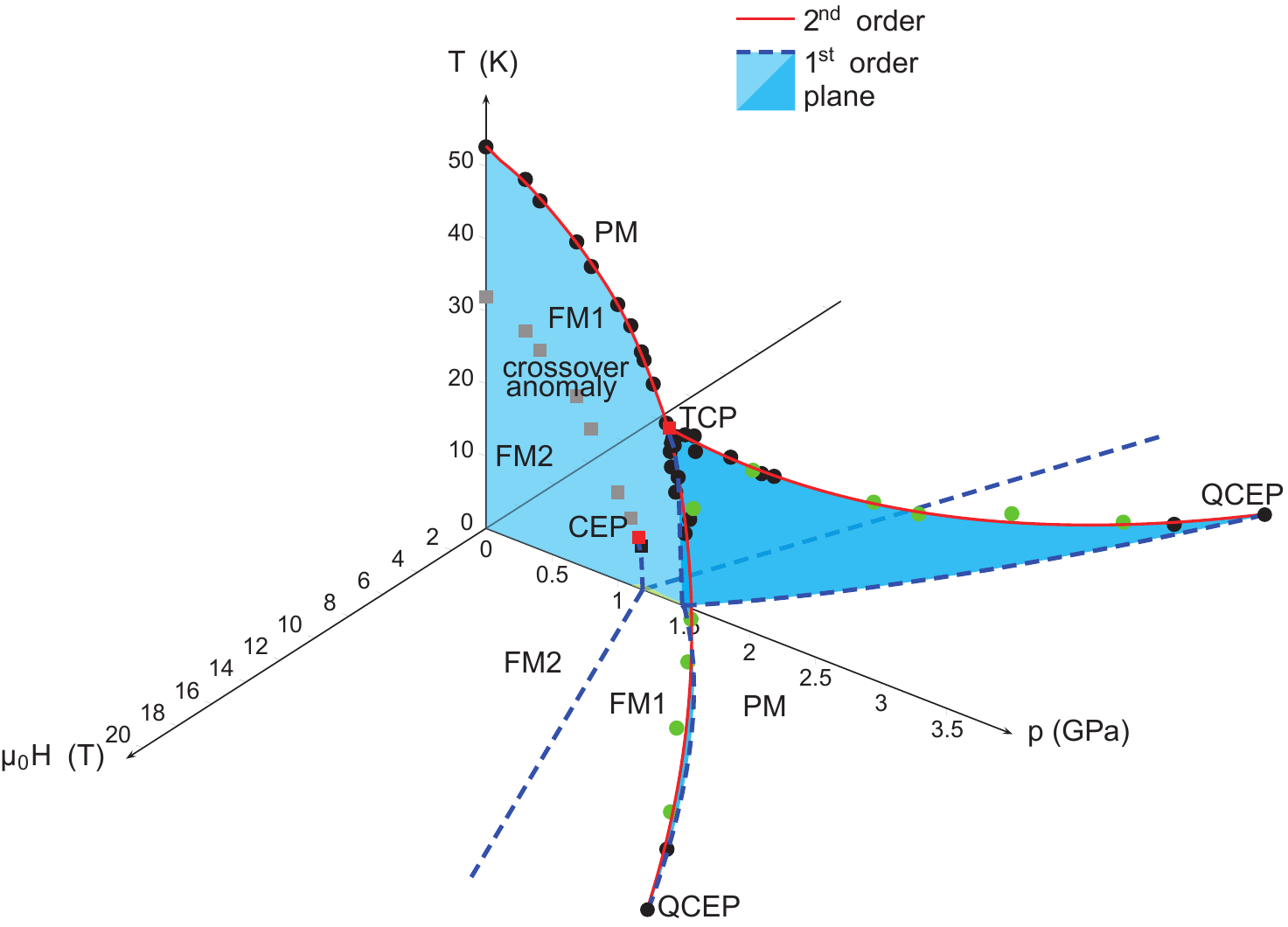}
\end{center}
\caption{(Color online) Three-dimensional ($T, H, P$) phase diagram of UGe$_2$, indicating evidence of the FM wings extending far above $P_{\rm C}$ up to the QCEPs ($H_{\rm QCEP}\sim 18\,{\rm T}$, $P_{\rm QCEP}\sim 3.5\,{\rm GPa}$).~\cite{KotegawaJPSJ2011,TaufourPhD}}
\label{fig12}
\end{figure}

\begin{figure}[!tbh]
\begin{center}
\includegraphics[width=\hsize]{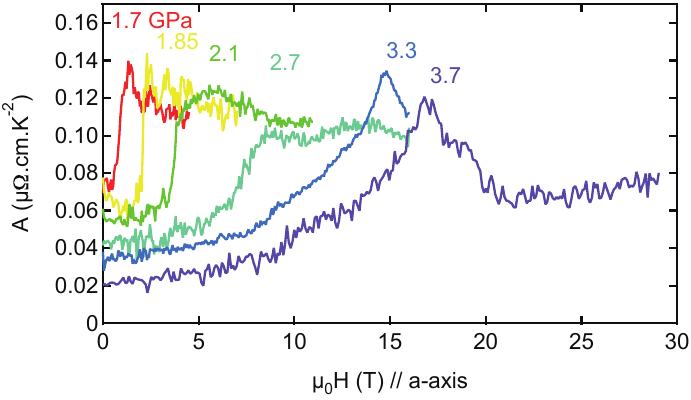}
\end{center}
\caption{(Color online) Variation of the $A$ coefficient of the $AT^2$ resistivity term upon crossing $H_{\rm X}$ at different pressures through the QCEP of UGe$_2$.~\cite{TaufourPhD}}
\label{fig13}
\end{figure}

The first determination of the ($P$, $T$) phase diagram of UGe$_2$ was realized in 1993, showing a collapse of ferromagnetism between 1.5 and 2 GPa\cite{TakahashiPhysicaB1993}. 
Evidence of an anomaly at $T_x$ (signature of the competition between FM2 and FM1) was reported in 1998\cite{OomiJAlloyComp1998}. 
The next breakthrough was the discovery of SC in the FM domain ($P < 1.5$ GPa) in 2000\cite{SaxenaNature2000}. 
The key role of the switch from FM2 to FM1 in the SC onset was pointed out in 2001~\cite{HuxleyPRB2001,BauerJPCM2001,TateiwaJPSJ2001,MotoyamaPRB2001}. 
Above $P_x$, $M_0 = 0.9\,\mu_{\rm B}$/U in the FM1 phase
and $M_0 = 1.5\,\mu_{\rm B}$/U in the FM2 phase 
(see Fig.~\ref{fig11})~\cite{TateiwaJPSJ2001,MotoyamaPRB2001,PfleidererPRL2002,HuxleyJPCM2003}. 
Above $P_x$, FM2 will be reached at a magnetic field $H_x$ for $H\parallel a$ (easy magnetization axis).  
At a field $H_c$ above $P_{\rm C}$, the PM phase will switch to FM1; increasing the field to $H_x$ leads to a transition to FM2.
Figure~\ref{fig11} shows the $P$ dependences of $M_0(P)$, $H_x$, and $H_c$~\cite{PfleidererPRL2002}. 
Complementary studies can be found in Refs.~\citen{TateiwaJPSJ2001,HuxleyJPCM2003}, and \citen{TerashimaPRB2002}.
The jump of $M_0$ at $P_x$ and $P_c$ clearly shows that both transitions at $P_x$ and $P_c$ are of the first order.
The transition line $T_{\rm Curie}(P)$ between FM and PM at zero field changes from second order to first order at a TCP of $T_{\rm TCP}\sim 24\,{\rm K}$ and $P_{\rm TCP} = 1.42\,{\rm GPa}$.
The range of the first-order transition is quite narrow $(P_c-P_{\rm TCP})/P_{\rm TCP} \sim 0.05$~\cite{TaufourPRL2010}.
However, for $H \parallel M_0$ along the easy axis, the first-order FM wing appears up to the quantum critical end point (QCEP) at $P_{\rm QCEP}\sim 3.5\,{\rm GPa}$ and $\mu_0 H_{\rm QCEP}\sim 18\,{\rm T}$, as shown in Fig.~\ref{fig12}.~\cite{KotegawaJPSJ2011,TaufourPhD}
Note the large separation between $P_{\rm QCEP}$ and $P_c$ directly linked to the large jump of $M_0$ at $P_c$. 
Figure~\ref{fig13} shows how the step like jump of the resistivity coefficient $A$ below $P_{\rm QCEP}$ is replaced by a maximum above $P_{\rm QCEP}$.~\cite{TaufourPhD} 
The phase transition at $T_x$ between FM1 and FM2 starts at the critical end point (CEP) equal to $T_{\rm CEP} = 7\,{\rm K}$ and $P_{\rm CEP}$ =1.16 GPa. 
At $T = 0$ K, it collapses at $P_x \sim 1.2\,{\rm GPa}$.

\begin{figure}[!tbh]
\begin{center}
\includegraphics[width=0.6\hsize,pagebox=cropbox,clip]{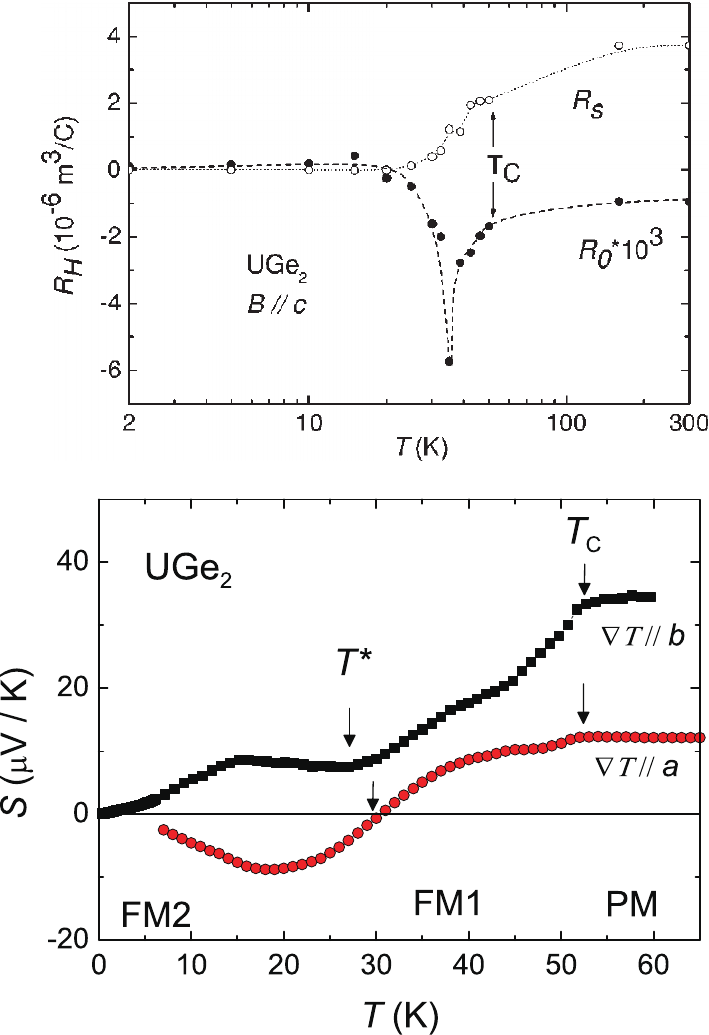}
\end{center}
\caption{(Color online) Variation of the Hall constant $R_{\rm H}$ and the thermoelectric power through the crossover temperature $T^*$, which is a characteristic of the choice of FM2 as the ground state of UGe$_2$ at low pressures.}
\label{fig14}
\end{figure}

At $P = 0$, the specific heat and thermal expansion show a crossover between FM1 and FM2 ground states.~\cite{Har09_UGe2}.                                
Note that drastic changes are observed in the normal component of the Hall effect\cite{TranPRB2004} as well as in the thermoelectric power (Fig.~\ref{fig14})\cite{MoralesPRB2016}.
These low-pressure effects are precursors of the drastic changes in the FS on entering the different ground states through $P_x$.

\begin{figure}[!tbh]
\begin{center}
\includegraphics[width=0.9\hsize,pagebox=cropbox,clip]{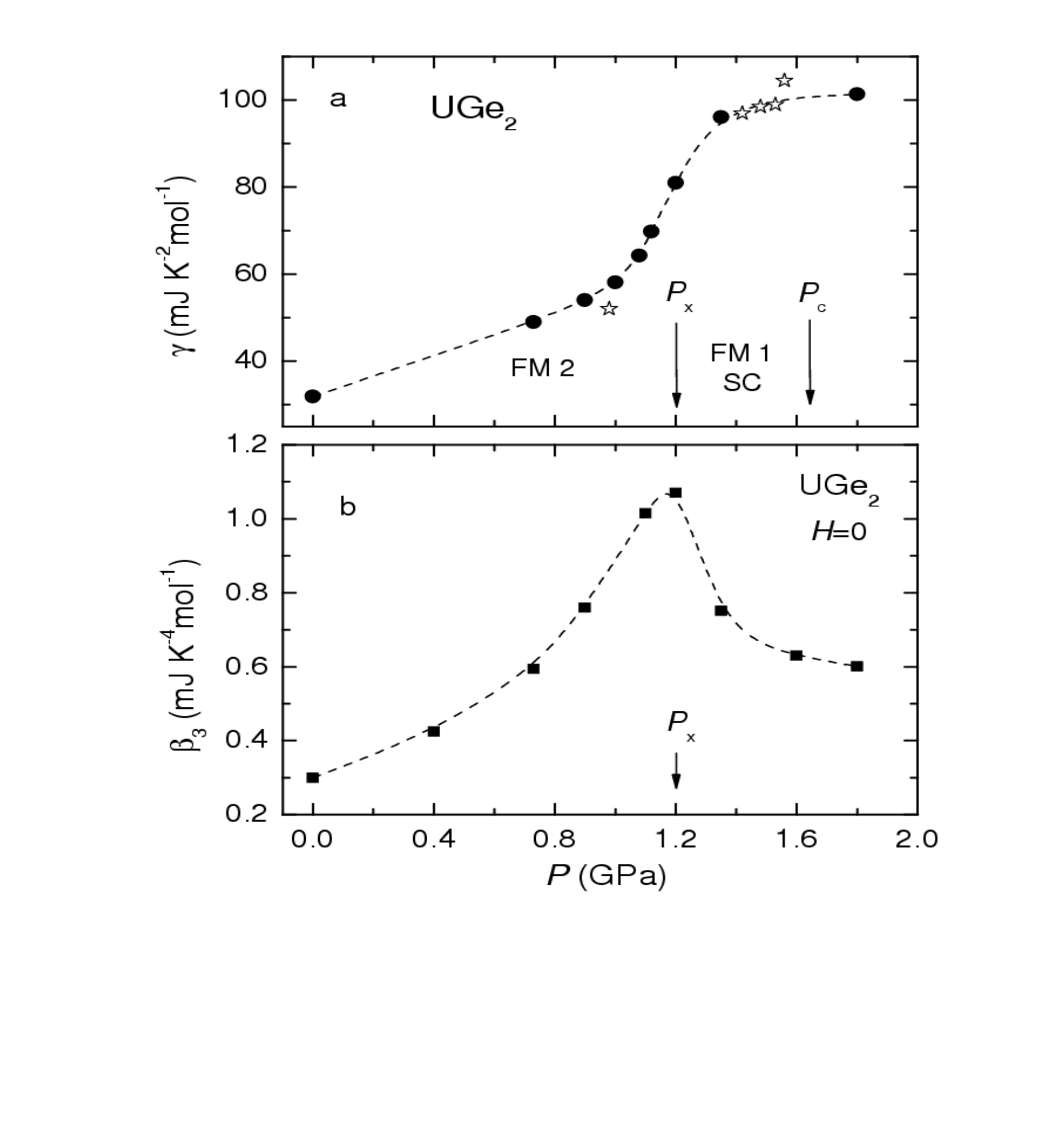}
\end{center}
\caption{(a) Pressure dependence of $\gamma$ in UGe$_2$. (b) Note that the jump of $\gamma$ at $P_x$ is associated with a maximum of $\beta$ for the $T^3$ term of the specific heat.~\cite{Flo06_review}}
\label{fig15}
\end{figure}

Measurements of the specific heat under pressure (Fig. \ref{fig15}) indicate that the $\gamma$ coefficient jumps at $P_x$ while the $\beta T^3$ term also unexpectedly has a maximum at $P_x$~\cite{PhilippsPrivateComm}.
There is no maximum of $\gamma$ at $P_x$, suggesting that the additional effect, rather than spin fluctuation, should occur for the establishment of SC. 
In addition, the $P$ dependence of the $A$ coefficient obeys $A \propto \gamma^2$ for the current $I \parallel b$, but has a pronounced maximum for $I \parallel a$~\cite{TerashimaPRB2006,KobayashiJPCM2002,SettaiJPCM2002}.

\begin{figure}[!tbh]
\begin{center}
\includegraphics[width=0.8\hsize,pagebox=cropbox,clip]{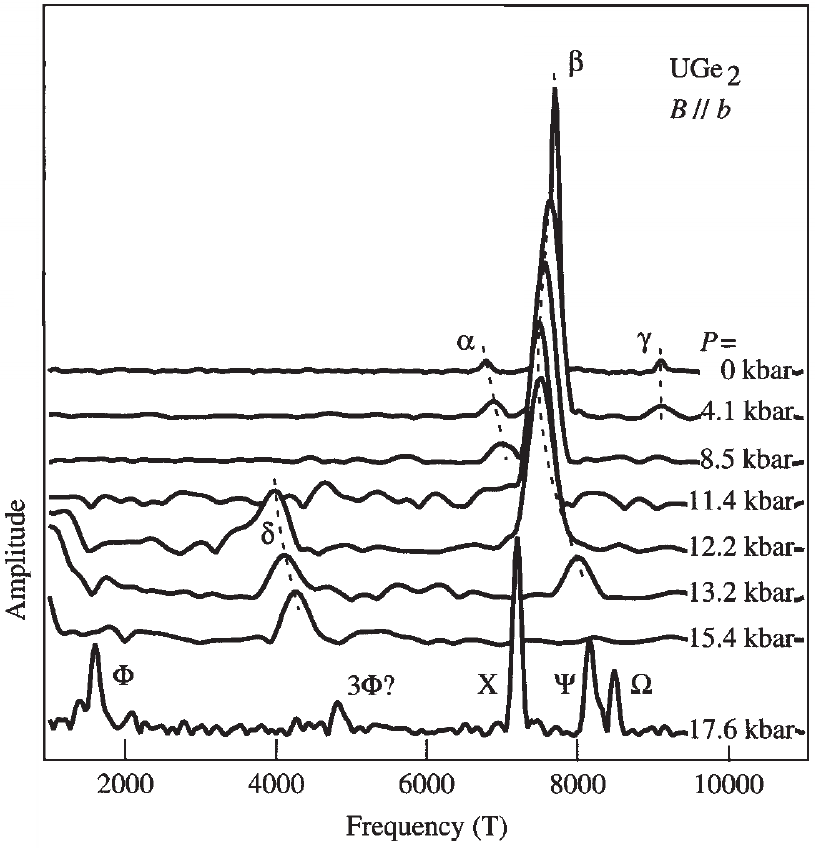}
\end{center}
\caption{Pressure dependence of FFT spectra obtained from dHvA experiments for $H\parallel b$ in UGe$_2$. The FSs are drastically changed above $P_{\rm c}$.~\cite{TerashimaPRL2001}}
\label{fig16}
\end{figure}

\begin{figure}[!tbh]
\begin{center}
\includegraphics[width=0.8\hsize,pagebox=cropbox,clip]{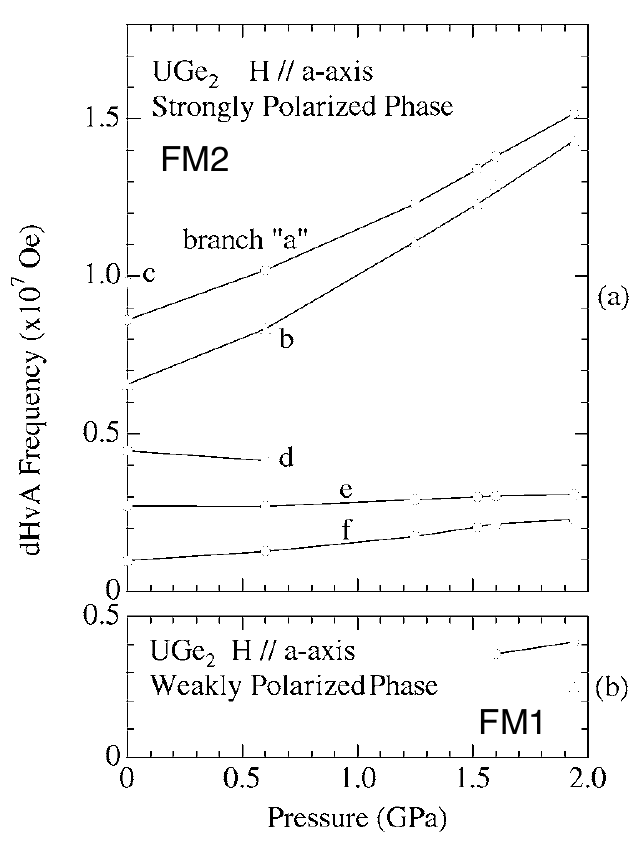}
\end{center}
\caption{Pressure dependences of the dHvA frequencies for $H\parallel a$ in UGe$_2$ in the (a) FM2 and (b) FM1 phases.~\cite{Hag02}}
\label{fig:UGe2_Freq_pressure}
\end{figure}

The new feature is that quantum-oscillation measurements demonstrate drastic changes in the FS through $P_{x}$ and $P_c$. 
For $H\parallel b$, corresponding to the hard-magnetization axis, the three phases, FM1, FM2, and PM, are not affected by the magnetic field.
The main dHvA branches in FM1, namely $\alpha$, $\beta$, and $\gamma$, which might be due to the nearly cylindrical FSs~\cite{SettaiJPCM2002}, disappear in the PM state, and new branches are detected~\cite{TerashimaPRL2001}, as shown in Fig.~\ref{fig16}.
In the PM state, a heavy electronic state is realized with large effective masses of up to $64\,m_0$, which is consistent with the large Sommerfeld coefficient ($100\,{\rm mJ\,K^{-2}mol^{-1}}$) measured under pressures above $P_{\rm c}$.
For $H\parallel a$ (easy-magnetization axis), FM1, FM2, and PM are separated by metamagnetic transitions.
Thus, the results are more complicated. 
Figure~\ref{fig:UGe2_Freq_pressure} shows the pressure dependence of the dHvA frequencies.~\cite{Hag02,TerashimaPRB2002}
The observed frequencies, which are less than approximately $1\times 10^7\,{\rm Oe}$, correspond to the small FSs, revealing the relatively strong pressure dependence in the FM2 phase.
It is also clear that the FS changes with the transition from FM2 to FM1.

One interesting theoretical proposal is that the transition at $T_x$ may be the signature of a charge density wave (CDW) onset linked to supplementary nesting of the FS\cite{WatanabeJPSJ2002}, but attempts to detect a CDW have failed\cite{AsoJPSJ2006}.
Band-structure calculation indicates that over a range of pressures, the two FM states are nearly degenerate with different orbital and spin moments on the U sites\cite{ShickPRB2004}.
The ($P$, $T$) phase diagram was qualitatively explained by a phenomenological model with two initial maxima in the density of states\cite{SandemanPRL2003}.
In contrast to the case of AF-PM instability, which is often of the second order, a first-order collapse in clean FM itinerant materials is observed; 
theoretical arguments to justify this were given through the nonanalytic term in the Landau expression for the free energy\cite{BelitzPRL1999,BelitzPRL2005} and through the feedback with magnetoelastic coupling.~\cite{GehringEPL2008,MineevPUsr2017,MineevCRP2011}.

To summarize the normal properties of UGe$_2$, the main features are as follows.
\begin{itemize}
\item $T_{\rm Curie}$ is suppressed with increasing $P$, and a switch to a first order transition occurs at the TCP.
\item FM wings exist far above $P_c$ up to $P_{\rm QCEP} \sim 3.5\,{\rm GPa}$.
\item There is a drastic change in the FS at $P_x$ and $P_c$
\end{itemize}

\subsection{SC phase: interplay of FS instability and FM fluctuations}
The SC domain was first determined by resistivity measurements; 
the two important points are that optimum of $T_{\rm SC}$ coincides with $P_x$ and that the collapse of SC occurs at $P_{\rm SC} \sim P_c$.
Unexpectedly, the specific heat jump at $T_{\rm SC}$ (Fig.~\ref{fig18}) can only be observed near $P_x$\cite{TateiwaPRB2004}. 
A series of supplementary results derived from the $ac$ susceptibility measurements\cite{NakaneJPSJ2005,BanJMMM2007,KabeyaPhysicaB2009} suggests that SC will exist only in the FM1 phase (Fig.~\ref{fig19}).
The different behavior of SC in FM2 and FM1 has been confirmed by the fast broadening in the SC transition below $P_x$ (see Fig.~\ref{fig20}). 
The behavior of the intrinsic SC region remains an open question.
One difficulty is the narrow $P$ width of SC (0.3 GPa) compared with the pressure inhomogeneity close to the first-order transition with a volume change of approximately $10^{-3}$.

The unusual field dependence of $H_{\rm c2}$ reported in Fig.~\ref{fig:Hc2} with pressure slightly above $P_x$ corresponds to the field switching between FM1 and FM2 at $H_x$. 
The initial claim that the maximum of $T_{\rm SC}$ occurs in the FM2 phase just below $P_x$ must be verified. 
A sharp structure of $T_{\rm SC}$ in FM1 with a maximum upon approaching $P_x$ may give an alternative explanation. 
\begin{figure}[!tbh]
\begin{center}
\includegraphics[width=0.7\hsize,pagebox=cropbox,clip]{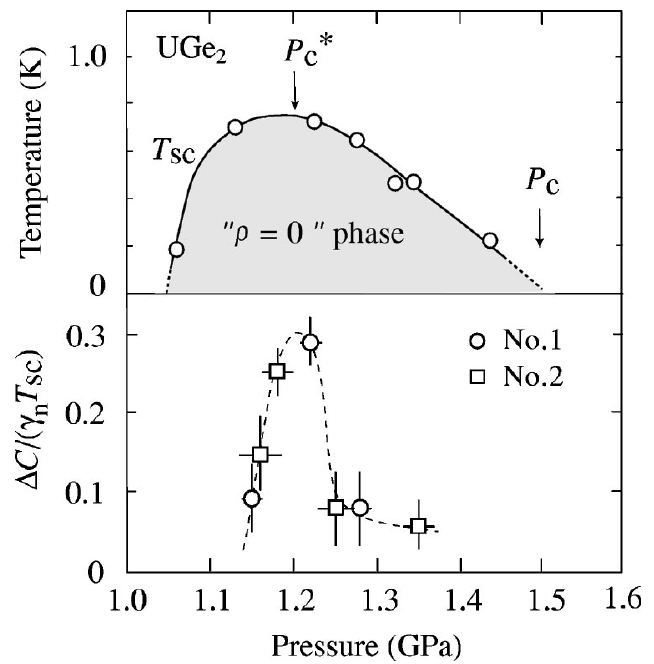}
\end{center}
\caption{Unexpectedly, a clear jump $\Delta C/\gamma_{\rm N}T_{\rm SC}$ can be detected in UGe$_2$ only at around $P_x$ but with a very weak value compared with the conventional BCS value (1.4) (see Fig.~\ref{fig:Cp}).}
\label{fig18}
\end{figure}
\begin{figure}[!tbh]
\begin{center}
\includegraphics[width=0.8\hsize,pagebox=cropbox,clip]{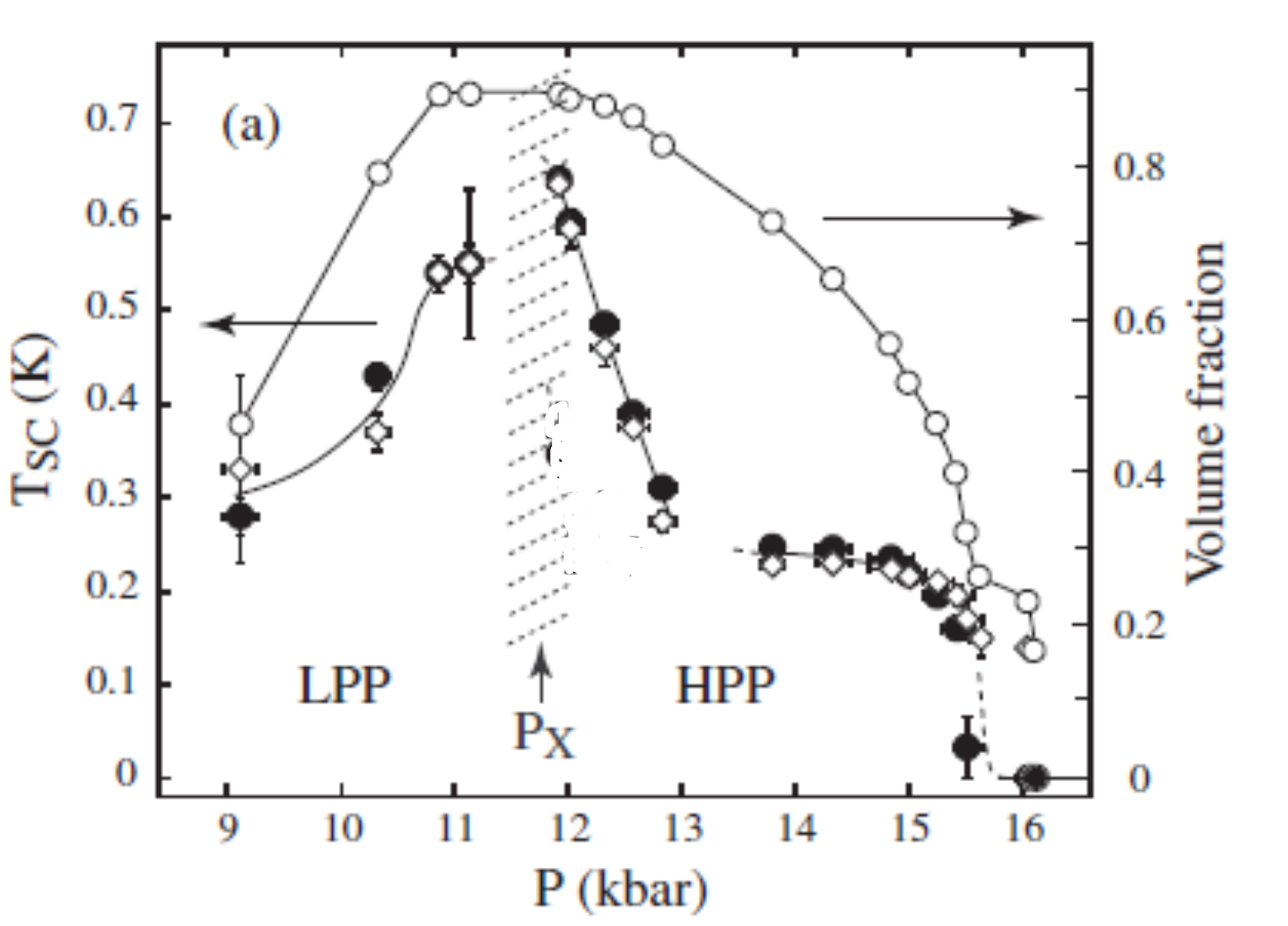}
\end{center}
\caption{$T_{\rm SC}$ of UGe$_2$ detected in susceptibility measurements from the peak in the $\chi''(T)$ curve. The volume fraction is obtained from $-4\pi\chi'(T)$ at 60 mK. A double structure in $\chi''(T)$ is detected for $P > P_x$.~\cite{NakaneJPSJ2005,BanJMMM2007,KabeyaPhysicaB2009}}
\label{fig19}
\end{figure}
\begin{figure}[!tbh]
\begin{center}
\includegraphics[width=\hsize]{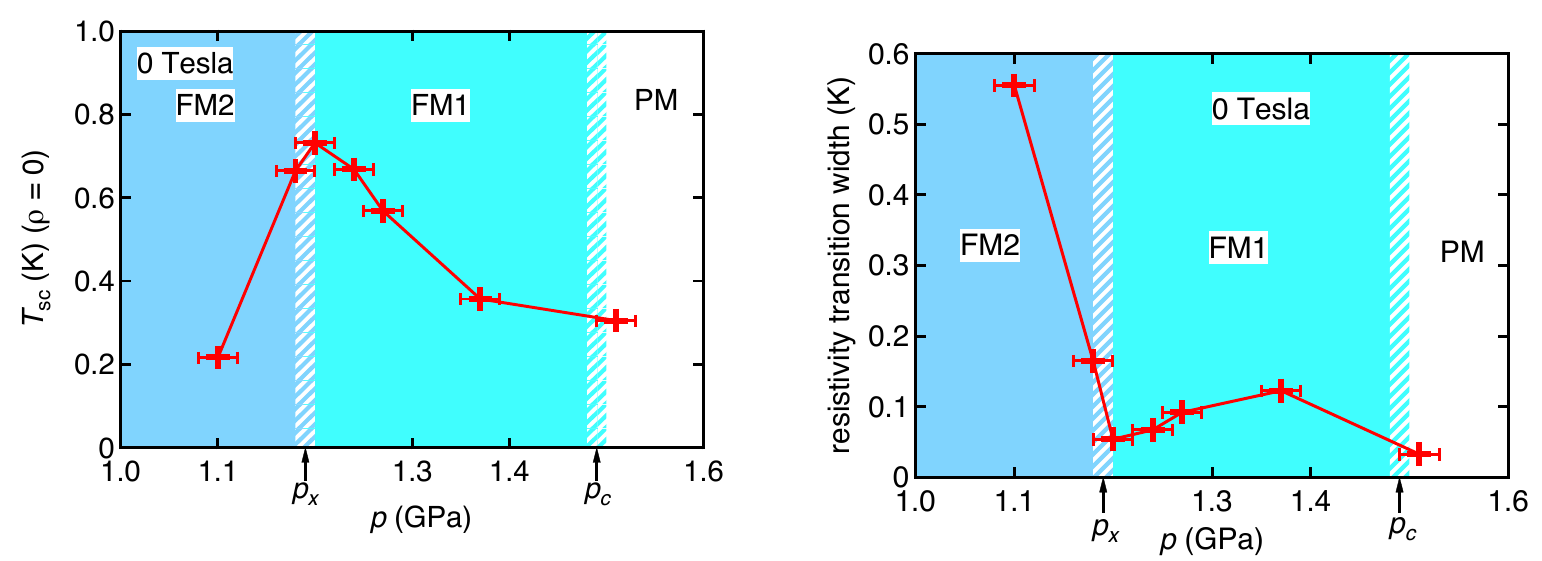}
\end{center}
\caption{(Color online) Determination of $T_{\rm SC}$ by the achievement of zero resistivity. Note the huge broadening of the $\rho$ anomaly on entering FM2.~\cite{TaufourPhD}}
\label{fig20}
\end{figure}

\begin{figure}[!tbh]
\begin{center}
\includegraphics[width=\hsize,pagebox=cropbox,clip]{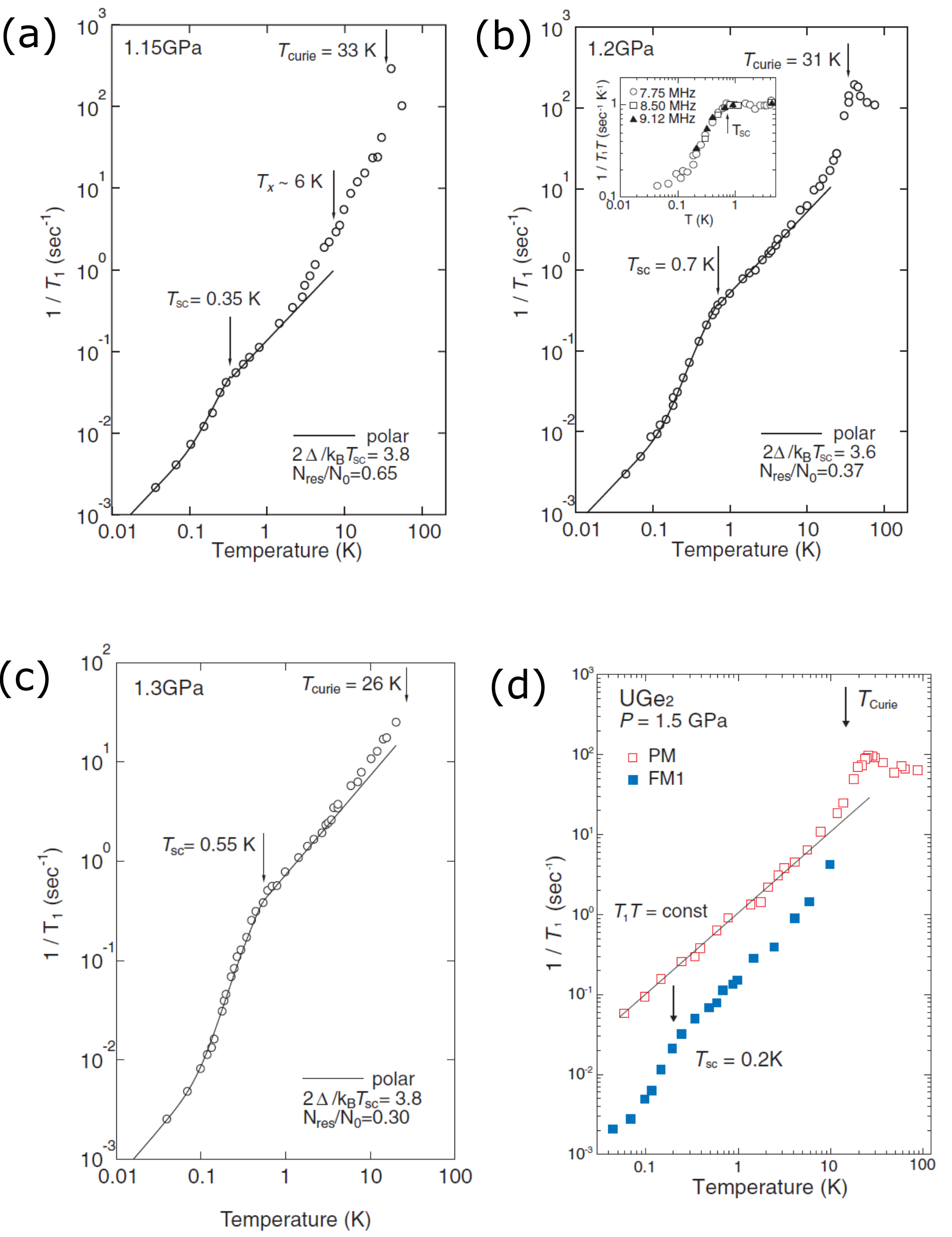}
\end{center}
\caption{(Color online) Temperature dependence of $1/T_1$ of $^{73}$Ge at (a) 1.15, (b) 1.2, (c) 1.3, and (d) 1.5 GPa. $1/T_1$ in panels (a), (b), and (c) was measured at the peak of the Ge1 (4$i$) site for FM1. The solid curves in (a), (b), and (c) are the results of calculations based on an unconventional superconducting model with a line-node gap. The identification of the phase transitions into both SC and FM ensures a phase with their uniform coexistence. $T_{\rm Curie}$ was determined by ac-$\chi$ measurement. The inset in (c) shows the frequency dependence of $1/T_1T$ at $P$ = 1.2 GPa and $f$ = 7.75, 8.5, and 9.12 MHz. The observation of a similar $T$ dependence of $1/T_1T$ ensures the onset of SC over the whole sample.  $1/T_1$ in (d) was measured at the peak of PM (open squares) and FM1 (solid squares). The long component in $1/T_1$ for FM1 indicates that SC sets in at $T_{\rm SC} \sim 0.2\,{\rm K}$, but the short components for PM do not.~\cite{KotegawaJPSJ2005,HaradaJPSJ2005}}
\label{fig:UGe2_T1}
\end{figure}

Unique information is given by NQR-NMR experiments using the $^{73}$Ge isotope. 
At low pressures, in agreement with the neutron scattering experiments\cite{HuxleyPRL2003}, the Ising FM character of the fluctuations is clearly observed\cite{NomaJPSJ2018}.  
Figure~\ref{fig:UGe2_T1} illustrates the temperature dependence of $1/T_1$~\cite{KotegawaJPSJ2005,HaradaJPSJ2005}.
In the FM phase (FM2 or FM1), $T_{\rm Curie}$ and $T_{\rm SC}$ are clearly detected; 
above $T_{\rm SC}$, $1/T_1$ follows a $T$-linear Korringa dependence linked with the value of $\gamma$,
and below $T_{\rm SC}$, a $T^3$ term is observed and regarded as an indication of a line-node SC gap.
At $P_c$, because of the first-order nature of the transition, both FM and PM signals are detected with the noteworthy feature that SC ($T_{\rm SC} \sim 0.2\,{\rm K}$) is only detected in the FM1 phase and not in the PM phase\cite{HaradaJPSJ2005}.

Note that the results of muon experiments~\cite{Sak10_UGe2} emphasize the duality between the localized and itinerant character of the 5$f$ electrons (see also Ref.~\citen{TrocPRB2012} and recent neutron data in Ref.~\citen{Has18}).
The common point with our previous consideration is that
the FM1 phase has an electronic density larger than the FM2 phases, and hence
multiband effects must be taken into account.
The conduction electron carries a quasi isotropic magnetic moment considerably lower than $M_0$.
The statement that $P_x$ collapses close to $P_c$ is in contradiction to all previous data;
this can be evidence that the muon signal is not directly linked with the switch of magnetism at $P_x$ or a indicator of nonhydrostaticity, which happens in the chamber of a pressure cell fixed at a constant volume but not at a constant pressure (the reaction of the pressure cell causing the deformation of the crystal depends on the specific arrangement).
This puzzle should be solved.

\begin{figure}[!tbh]
\begin{center}
\includegraphics[width=0.8\hsize,pagebox=cropbox,clip]{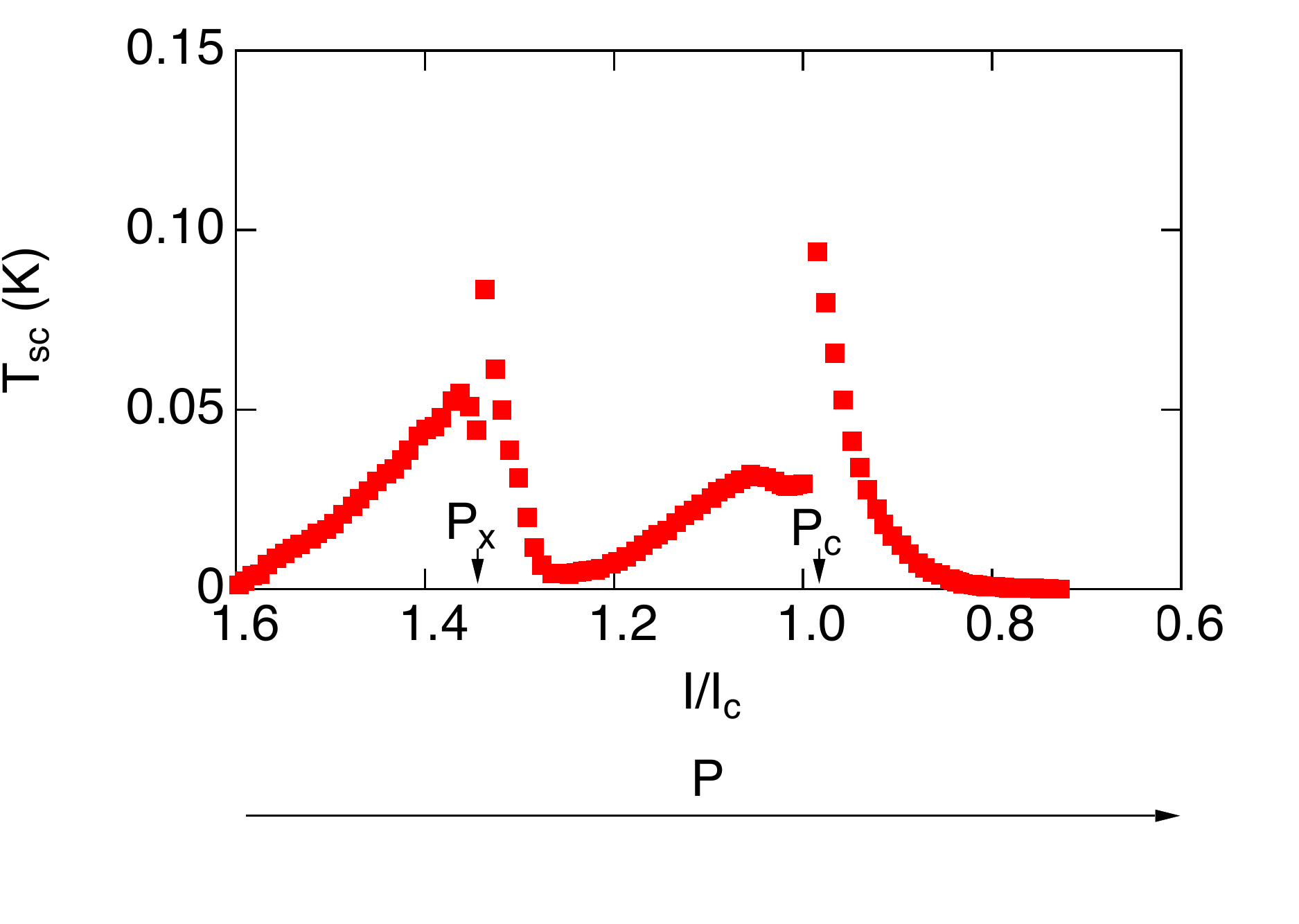}
\end{center}
\caption{(Color online) For UGe$_2$, $T_{\rm SC}$ versus $P$ or $I / I_c$ the Stoner factor using the parameter of the phenomenological model of Ref.~\citen{SandemanPRL2003}.}
\label{fig22}
\end{figure}

The main attempt to derive the SC properties from the normal ones was made by changing the FS topology in two critical peaks of density of states of the PM phase~\cite{SandemanPRL2003}.
By comparison with a later consideration on URhGe and UCoGe, the pressure dependences of $m_{\rm B}$ and $m^{**}$ were considered to have singularities at $P_x$ and $P_c$.
Using the parameter in Ref.~\citen{SandemanPRL2003}, the pressure variation of $T_{\rm SC}$ is shown in Fig.~\ref{fig22}. 
In agreement with Figs.~\ref{fig19}-\ref{fig:UGe2_T1}, a sharp SC singularity occurs at $P_x$. 
The prediction of SC in the PM region is not verified by the experiments; 
it is difficult to predict SC quantitatively in the complex case of UGe$_2$.

Recently, a mechanism of spin-triplet pairing was proposed~\cite{Kad18},
as an alternative to FM spin-fluctuation pairing. 
Robust SC is predicted to exist only in the FM1 phase 
by the combination of the FM exchange based on Hund's rule
and the interelectronic Coulomb interaction (see Fig.~4 in Ref.~\citen{Kad18}).
A sound proof will be microscopic evidence of orbital-selective Mott-type delocalization of the 5$f$ electrons in the transition from FM2 to FM1. 
Although the effect is small (see Fig.~5 in Ref.~\citen{Kad18}),
the effect on SC pairing is predicted to be large.

\section{Properties of URhGe}\label{sec:URhGe}
We focus on the RSC observed in the transverse field scan along the $b$ axis ($H_b$),
and its link with the FM wing structure detected by tilting the field angle $\theta$ from the $b$ to $c$-axis.
The possibility of crude modeling through the field dependence of $m^{\ast\ast}(H)$
and its collapse under pressure is shown.
The additional new possibility of boosting SC via uniaxial stress is also presented.
\subsection{Reentrant superconductivity, FM wing, QCEP, FS instabilities}
\begin{figure}[!tbh]
\begin{center}
\includegraphics[width=0.8\hsize,pagebox=cropbox,clip]{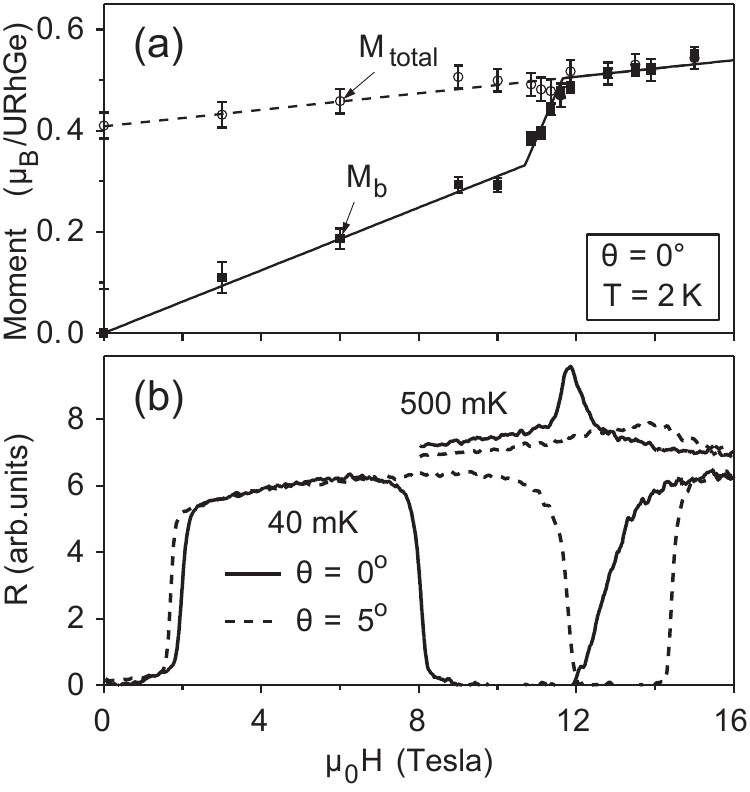}
\end{center}
\caption{(a) Magnetization of URhGe determined from neutron scattering experiments.
$M_{\rm total}$ is the total magnetization contributed from both the $M_b$ and $M_c$ components, where $M_b$ is the magnetization of the $b$-axis component for $H \parallel b$ with perfect alignment ($\theta$ = 0).
Spin reorientation occurs at $H_{\rm R}$ = 11.8 T. 
(b) Consequence of 5$^{\circ}$ misalignment on resistivity curve at 40 mK, for $\theta = 0^{\circ}$. SC ($\rho = 0$) extends from  8 to 13 T, 
while a sharp maximum in $\rho$ can be found in the normal phase ($T \geq$ 500 mK). 
A weak misalignment of 5$^{\circ}$ towards the $c$-axis leads to an increase in $\mu_0 H_{\rm R} to \sim$ 13 T, a broadening of the maximum $\rho$ in the normal phase, and a shrinking of the SC domain ($12 - 14$ T). For $\theta$ = 5$^{\circ}$, we are already close to the QCEP indicated in Fig.~26.~\cite{LevyScience2005,LevyThesis}}
\label{fig23}
\end{figure}

\begin{figure}[!tbh]
\begin{center}
\includegraphics[width=\hsize,pagebox=cropbox,clip]{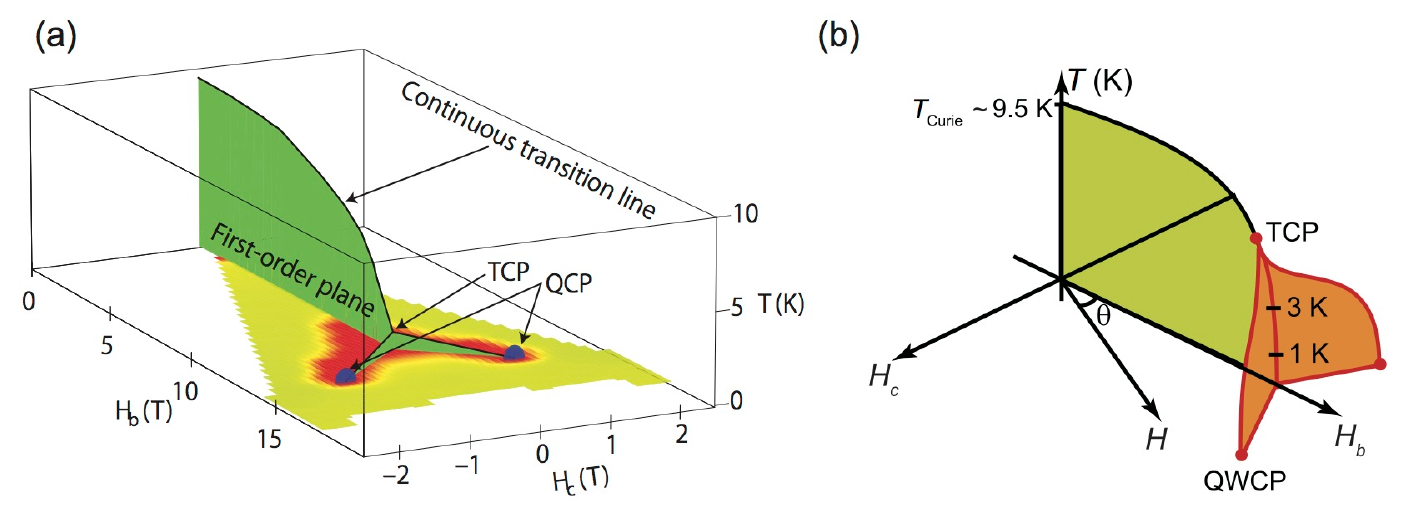}
\end{center}
\caption{(Color online) (a) Existence of FM wings in the ($T$, $H_b$, $H_c$) phase in URhGe reported in Refs.~\citen{LevyNatPhys2007,HuxleyJPSJ2007}. The QCEP is located at $H_c \sim$1.1 T and $H_{\rm R} \sim$13.5 T, corresponding to a misalignment of $\theta \sim5^{\circ}$. 
(b) Confirmation of the wings in URhGe by precise magnetization measurements.~\cite{NakamuraPRB2017}}
\label{fig:FM_wing}
\end{figure}

\begin{figure}[!tbh]
\begin{center}
\includegraphics[width=0.8\hsize,pagebox=cropbox,clip]{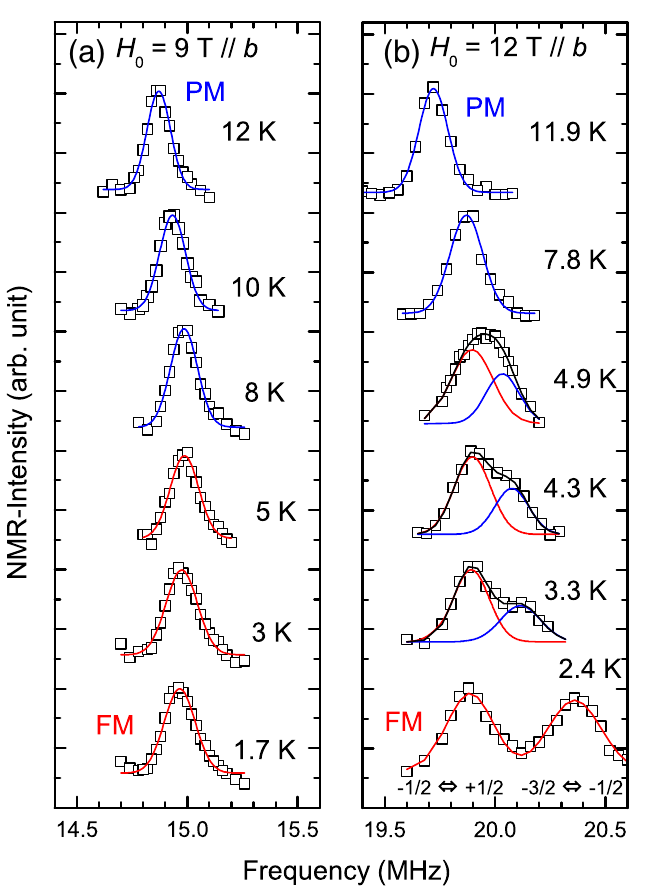}
\end{center}
\caption{(Color online) NMR spectra of the central transitions for $H \parallel b$ in URhGe. At 9 T, the peak position shifts to a higher frequency down to $\sim8$ K and then returns to a slightly lower frequency. This temperature of $\sim8$ K corresponds to $T_{\rm Curie}$ under $H \parallel b$. At 12 T, the spectrum is composed of two components at around $T_{\rm Curie}$, indicating that the transition is of the first order and accompanied by phase separation\cite{KotegawaJPSJ2015}.}
\label{fig:URhGe_NMR}
\end{figure}

In a transverse field scan ($H\parallel b$),
RSC appears in the field range from 8 to $13\,{\rm T}$, and 
the easy magnetization axis switches from the $c$- to $b$ axis at $H_{\rm R}=12\,{\rm T}$.~\cite{LevyScience2005,LevyThesis}.
Figure~\ref{fig23}(a) shows the evolution of the total magnetization $M_{\rm tot}$ with $H$ and its component along the $b$-axis in an $H$ scan along the $b$-axis at 2 K. 
Figure~\ref{fig23}(b) shows the resistivity measurements in this $H$ sweep revealing a sharp maximum at  $H_{\rm R}=12\,{\rm T}$ for $500\,{\rm mK}$,
and zero resistivity in the field range from 8 to $13\,{\rm T}$ at $40\,{\rm mK}$.~\cite{LevyScience2005}
Misalignment of $H$ by $5^{\circ}$ towards the $c$-axis leads to a weak maximum of resistivity at 500 mK and a narrowing of RSC from $12$ to $14\,{\rm T}$ at 40 mK. 
Based on the evidence that in the $H_b$-$H_c$ plane, FM wings appear upon adding an extra $H_c$ component, a QCEP at $14\,{\rm T}$ tilted $\theta \sim6^\circ$ to the $c$-axis was proposed (Fig.~\ref{fig:FM_wing})~\cite{LevyThesis,HuxleyJPSJ2007}.
The wing structure has been investigated extensively by angle-resolved magnetization measurements~\cite{NakamuraPRB2017}; the schematic $T$, $H_b$, and $H_c$ phase diagram is shown in Fig.~\ref{fig:FM_wing}(b)~\cite{NakamuraPRB2017}.
The QCEP is located at $H_{\rm QCEP}$ = 13.5 T and $H_c$ = 1.1 T, in good agreement with the previous estimation of $H_{\rm QCEP}$ = 14 T~\cite{LevyNatPhys2007}.   
Confirmation of the TCP emerges from Hall effect,\cite{AokiJPSJ2014} ac susceptibility\cite{AokiJPSJ2014}, TEP,\cite{GourgoutPRL2016,GourgoutThesis} and NMR\cite{KotegawaJPSJ2015} measurements. 
For example, as shown in Fig.~\ref{fig:URhGe_NMR}, NMR spectra show both FM and PM signals at around $4\,{\rm K}$ at $12\,{\rm T}$ close to $H_{\rm R}$.

\begin{figure}[!tbh]
\begin{center}
\includegraphics[width=0.8\hsize,pagebox=cropbox,clip]{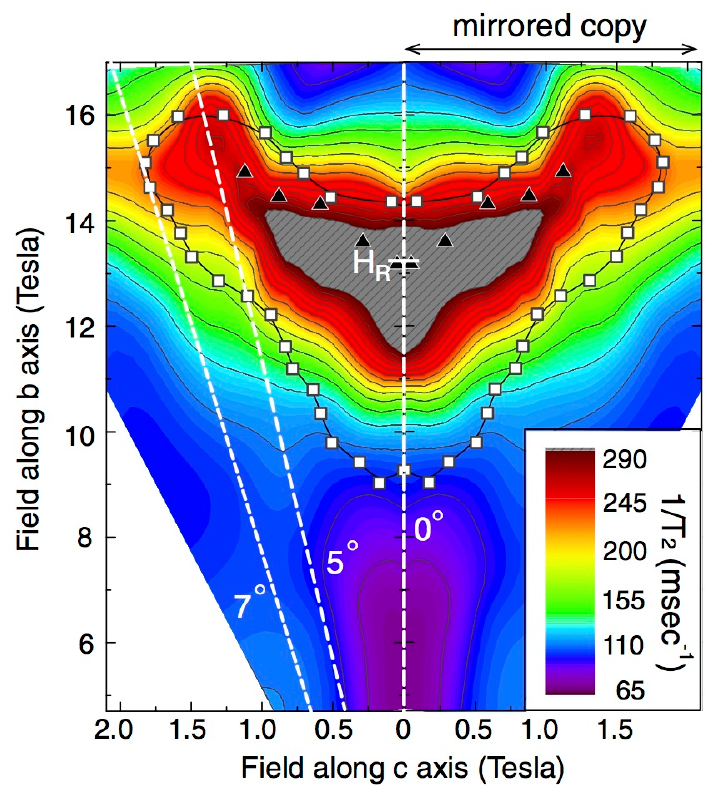}
\end{center}
\caption{(Color online) Map of the magnetic fluctuations detected via $1/T_2$ at 1.6 K for URh$_{0.9}$Co$_{0.1}$Ge with $H \parallel b$ and field misalignment along the $c$-axis ($H_c$).  $\theta$ is varied from 0 to 11$^{\circ}$. The open squares indicate where RSC occurs at low temperatures. The solid triangles show the variation of $H_{\rm R}$.~\cite{TokunagaPRL2015,TokunagaPRB2016}}
\label{fig27}
\end{figure}

$^{59}$Co-NMR experiments on URhGe doped with 10\% clearly show the strong increase in $1/T_2$ in a field scan along the $b$-axis towards $H_{\rm R}$\cite{TokunagaPRL2015,TokunagaPRB2016}.
The huge increase in the longitudinal fluctuations at $H_{\rm R}$ shown by the $1/T_2$ measurement coincides with a concomitant increase in $1/T_1$, which is sensitive to transverse fluctuations.
The transition towards an FM instability along the $b$-axis has been indicated already by the huge value of $\chi_b$ compared with $\chi_c$ in a low field.
Because of the weakness of the magnetocrystalline coupling, the specific feature in URhGe is the competition between FM along two axes ($c$ and $b$): the transverse fluctuations of one mode become longitudinal fluctuations for the other mode. 
Figure~\ref{fig27} shows the variation of $1/T_2$ in the $H_b$-$H_c$ plane.
In contrast to UCoGe\cite{IharaPRL2010}, where, in low fields, $(1/T_1)_b/(1/T_1)_c$ is larger than 20,~\cite{IharaPRL2010} the ratio in URhGe is small, $(1/T_1)_b/(1/T_1)_c \sim2.5$\cite{KotegawaJPSJ2015,TokunagaPRL2015}.

\begin{figure}[!tbh]
\begin{center}
\includegraphics[width=0.8\hsize,pagebox=cropbox,clip]{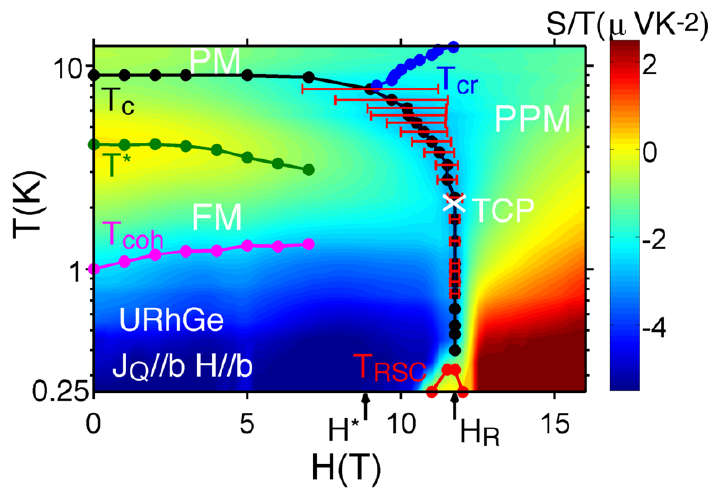}
\end{center}
\caption{(Color online) Results of thermoelectric power (TEP) experiments on URhGe down to $T$ = 0.25 K. $T_{\rm CR}$ represents the crossover line between the PM and polarized PM phase above $T_{\rm Curie}$. The transition widths observed in the TEP around $H_{\rm R}$ are shown by horizontal lines. The TCP is located close to 2 K.~\cite{GourgoutPRL2016,GourgoutThesis}}
\label{fig28}
\end{figure}

An additional effect detected from the TEP\cite{GourgoutPRL2016,GourgoutThesis} is that FS instability occurs at $H_{\rm R}$, as demonstrated by the change in the sign of $S$ (Fig.~\ref{fig28}).
The reconstruction of the FS has already been reported on the basis on Shubnikov$-$de Haas oscillations at $\theta \sim12^{\circ}$ from the $b$- to $c$-axis in order to escape from RSC.~\cite{YellandNatPhys2011}
However, the FM wing is never crossed in field at $\theta \sim12^\circ$ as it is beyond the QCEP.
We will discuss how a Lifshitz transition enhances the electronic correlation in Section 6. 
Contrary to the case of UGe$_2$ mentioned before, in URhGe as well as in UCoGe, at first glance, it seems that the SC pairing is driven mainly by the strength of the FM fluctuation (constant $m_{\rm B}$) without the necessity of invoking FS reconstruction.

\subsection{Modeling by considering field enhancement of $m^{\ast\ast}$ at $H_{\rm R}$}
\begin{figure}[!tbh]
\begin{center}
\includegraphics[width=0.8\hsize,pagebox=cropbox,clip]{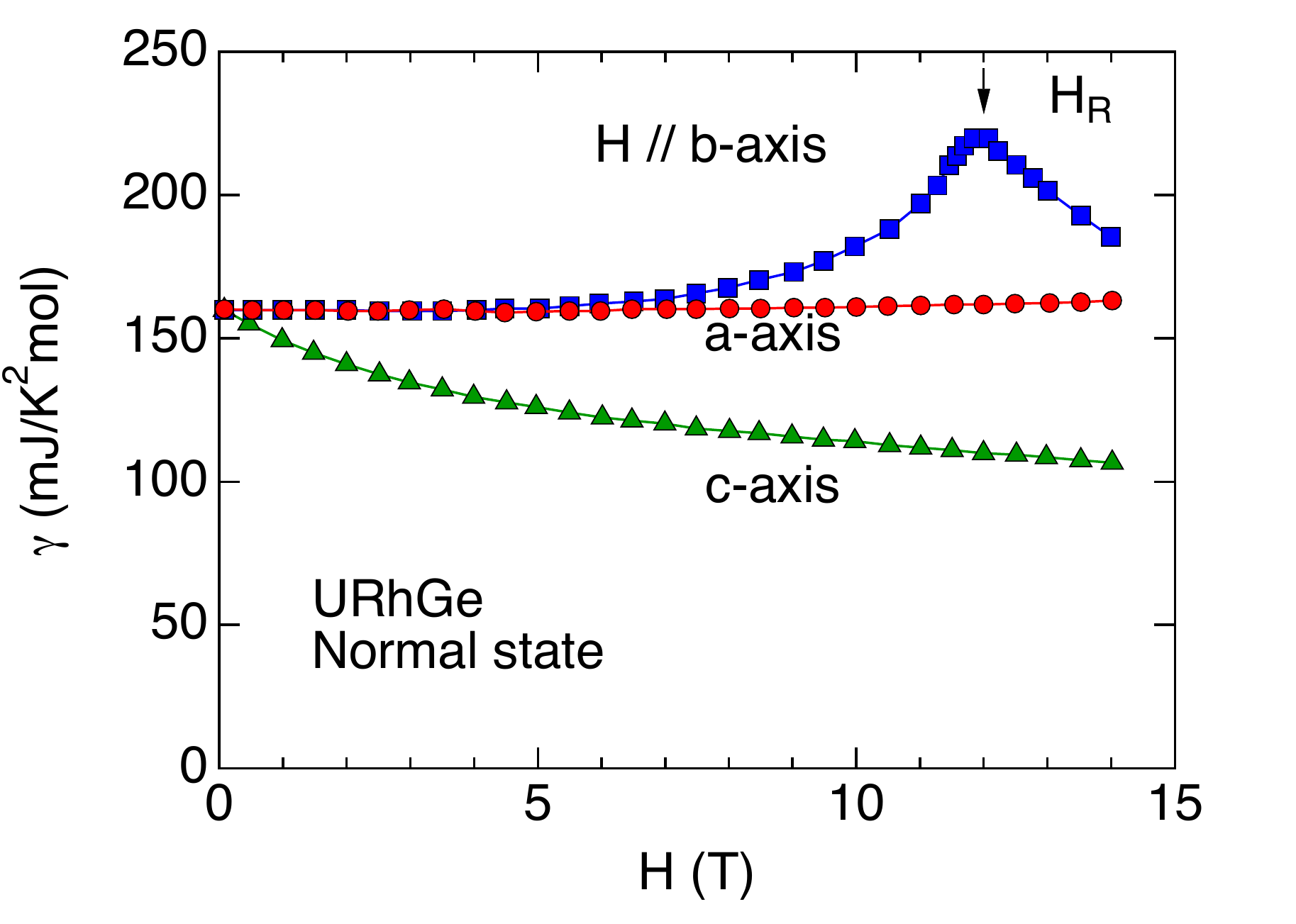}
\end{center}
\caption{(Color online) Field variation of the Sommerfeld coefficient $\gamma$ as a function of $H$.~\cite{Har11}}
\label{fig29}
\end{figure}

The proximity of the FM instabilities at $H_{\rm R}$ is indicated by the enhancement of the Sommerfeld coefficient (Fig.~\ref{fig29}),
the enhancement of the  $A$ coefficient in the $T^2$ dependence of resistivity,
and the concomitant increase in $1/T_1$ and $1/T_2$.
The $H_{\rm c2}$ curve for $H \parallel b$ can almost be quantitatively explained by a crude model, where sweeping $H$ drives an enhancement of $m^{**}$ linked to the approach of the FM instabilities\cite{MiyakeJPSJ2008, MiyakeJPSJ2009}.
In the normal FM phase, the $H$ and $P$ dependence of $m^{**}$ and $m_{\rm B}$ can be estimated from the temperature dependence of $C/T$ through $T_{\rm Curie}$\cite{Aok11_CR} or the $H$ dependence of the $A$ coefficient.

\begin{figure}[!tbh]
\begin{center}
\includegraphics[width=0.6\hsize,pagebox=cropbox,clip]{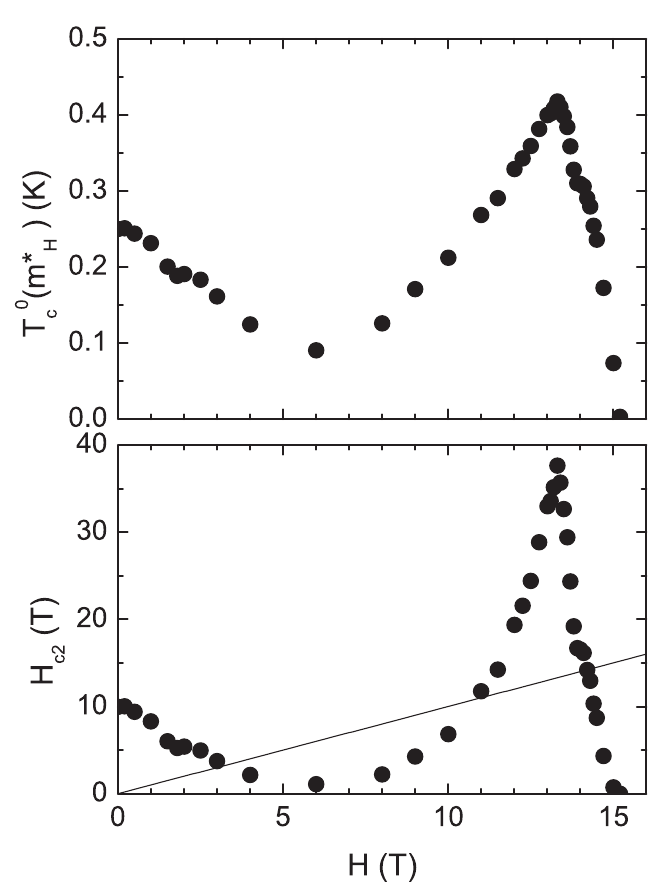}
\end{center}
\caption{(a) Field dependence of the SC transition $T_{\rm SC}^0$($m_H^*$) for URhGe,evaluated at zero field assuming  $m^{**}$($H$) is equal to the $H = 0$ effective mass. 
(b) Calculation of $H_{\rm c2}^2$(0)$ \propto (m_H^\ast T_{\rm SC})^2$ taking into account the variation of $T_{\rm SC}^0$ assuming with the hypothesis of the invariance of $m_{\rm B}$.~\cite{MiyakeJPSJ2008,MiyakeJPSJ2009}}
\label{fig30}
\end{figure}

The feedback on $H_{\rm c2}$ (Fig.~\ref{fig30}) was to boost the reference zero field $T_{\rm SC}$ corresponding to $m_H^{**}$ and also act on the slope of $H_{\rm c2}(T)$ by increasing the total effective mass $m^{*}(H) = m^{**}(H) + m_{\rm B}$.
Thus, $T_{\rm SC}(m_H^{**})$ was estimated to vary with the McMillan-type formula
\begin{equation*}
T_{\rm SC} = T_0 \exp \left(-\frac{\lambda + 1}{\lambda}\right)\\
\hspace{2em} \mbox{with} \hspace{3mm}\lambda \equiv \frac{m^{**}}{m_{\rm B}},
\end{equation*}
where $T_0$ is the renormalized electronic energy related to $m_{\rm B}$.
The orbital limit gives the $H_{\rm c2}$ dependence $H_{\rm c2} (0)\sim (m^\ast T_{\rm SC})^2$. 
From the specific heat measurements, $\lambda =0.5$ at $H = 0$ and $\lambda = 1$ at $H_{\rm R}$.

Another way to evaluate the field dependence of $\lambda$~\cite{WuNatComm2017,WuThesis} is to use the $H_{\rm c2}(T)$ dependence and then verifies its agreement with the field variation of $\gamma$ according to the relation:
\begin{equation*}
\lambda(H) = \frac{\gamma(H)}{\gamma(0)}\left(1+\lambda(0)\right) - 1.
\end{equation*} 
A series of $H_{\rm c2}$ curves with a fixed $\lambda$ are drawn via a conventional treatment for strong coupling SC and are adjusted to extract $\lambda(H)$. 
The modified McMillan formula $T_{\rm SC} \propto \exp{\left(-1/(\lambda-\mu^*)\right)}$ was used with the Coulomb repulsion parameter $\mu^* = 0.1$.
In this analysis, $\lambda(H = 0) = 0.75$ and $\lambda(H_{\rm R}) \sim1.4$.

\subsection{Effects of pressure and uniaxial stress}\label{sec:URhGe_pressure_uniaxial}
\begin{figure}[!tbh]
\begin{center}
\includegraphics[width=0.8\hsize,pagebox=cropbox,clip]{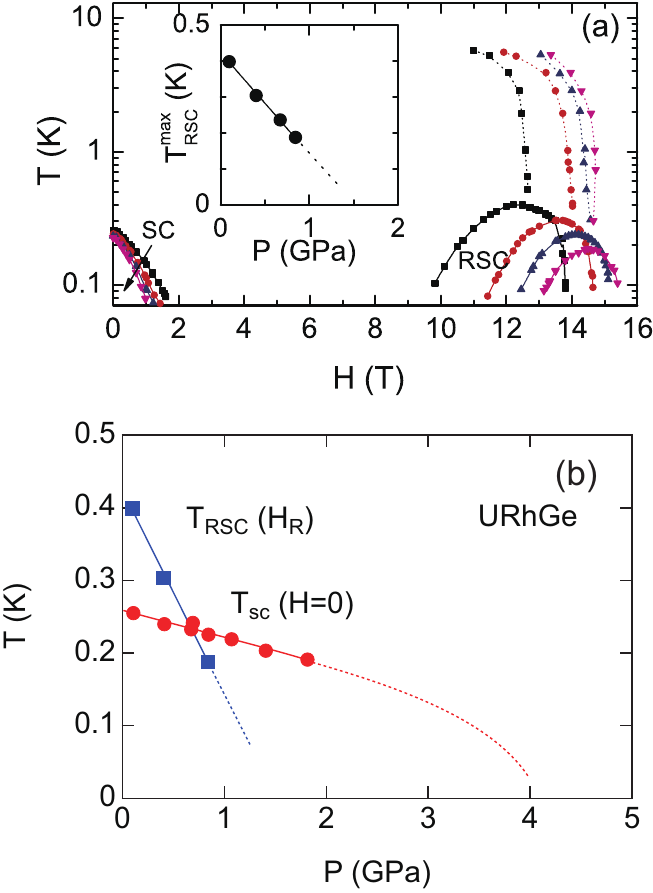}
\end{center}
\caption{(Color online) Pressure ($P$) dependence of the ($T, H$) phase diagram of URhGe for $H\parallel b$  with the shrinkage of RSC. RSC collapses for $P_{\rm RSC}$ = 1.5 GPa, which is roughly $P_s$/2.~\cite{MiyakeJPSJ2008}}
\label{fig31}
\end{figure}

The $P$ dependence of RSC predicted by the first approaches indicates that RSC disappeared at a pressure of $P_{\rm RSC} \sim1.5\,{\rm GPa}$,~\cite{MiyakeJPSJ2009} 
i.e., much lower than the pressure $P_S$ where SC collapses (Fig.~\ref{fig31}). 
Under pressure, $T_{\rm Curie}$ increases with $H_{\rm R}$, while $m^\ast (H_{\rm R})$ decreases.

Note that the quasi-coincidence of the collapse of RSC to $H_{\rm QCEP}$ with increasing $\theta$.
The increase in $H_{\rm R}$ along the $b$-axis with the $H_c$ component is a direct consequence of the wing structure. 
As $\theta$ increases in the ($H_b$, $H_c$) plane, $H_{\rm R}$ increases, whereas the effective mass $m^{**}$  and $\chi_b$ decrease.     
The butterfly SC shape given by $H_{\rm c2}$ (see Fig.~\ref{fig41}) is a direct consequence of not only adding an $H_c$ component with increasing of $H_{\rm R}$ but also decreasing $T_{\rm SC}^0$($m^*$) as $m_{\rm H}^\ast (\theta)$ decreases. 
RSC will collapse for a critical value of $T_{\rm SC}^0 (\theta)$.

\begin{figure}[!tbh]
\begin{center}
\includegraphics[width=0.8\hsize,pagebox=cropbox,clip]{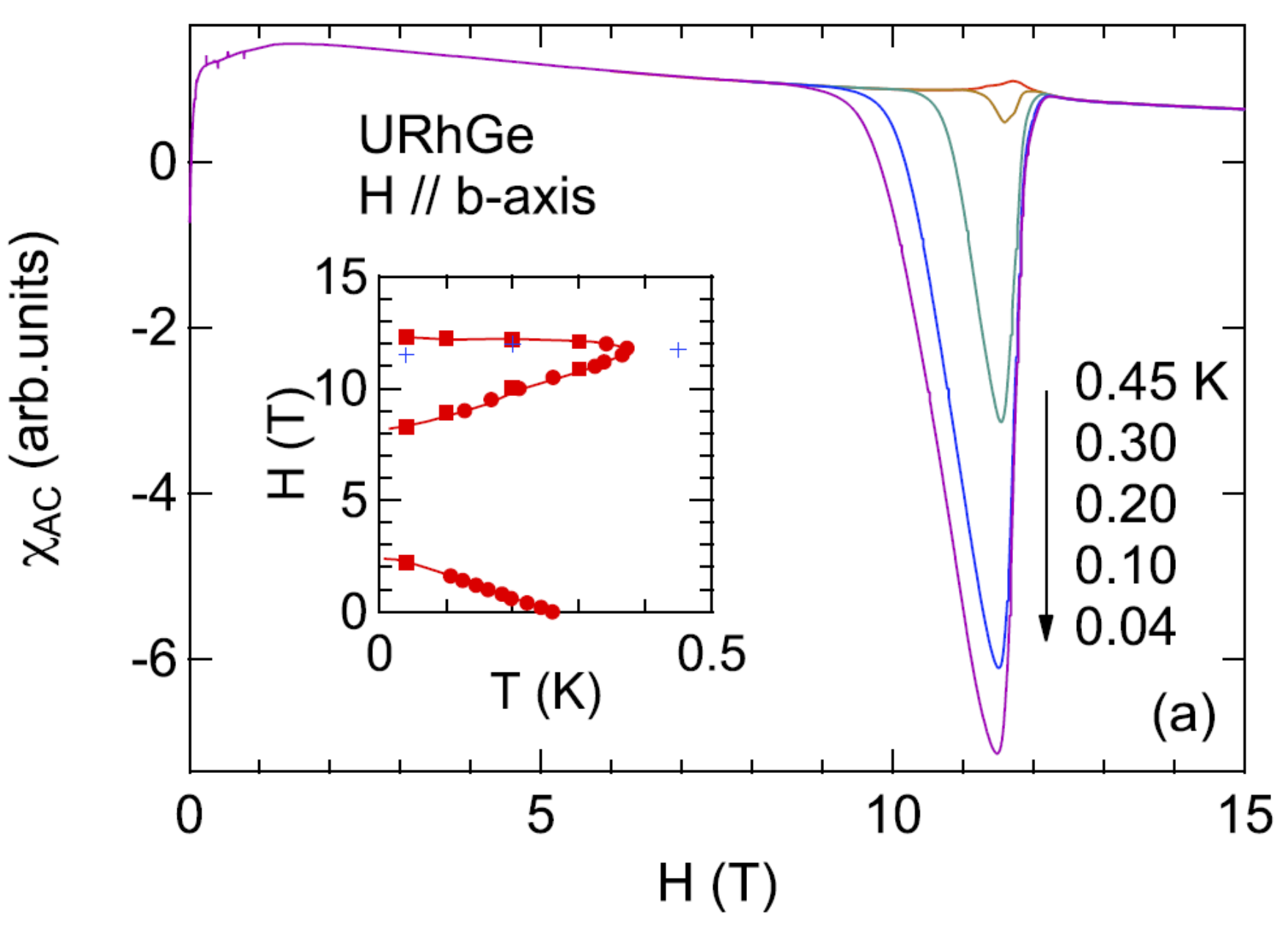}
\end{center}
\caption{(Color online) Diamagnetic shielding response in URhGe detected from AC susceptibility ($\chi_{\rm ac}$) as a function of $H$ for $H \parallel b$.
The inset shows the temperature dependence of $H_{\rm c2}$.}
\label{fig34N}
\end{figure}

The RSC domain appears to be quite robust, at least for $H_{\rm R} < H_{\rm QCEP}$ .
As shown in Fig.~\ref{fig34N} for a moderately clean crystal, diamagnetic shielding of the low-pressure phase is negligible while a clear diamagnetic signal is observed in the RSC domain.
The RSC domain above $H_{\rm R}$ is narrow close to 0.2 T.
This experimental observation provides evidence of a change in the order parameter at $H_{\rm R}$.

An interesting point is that $m^{\ast\ast}(H)/m^{\ast\ast}(0)$ depends only on the ratio $H/H_{\rm R}$.~\cite{MiyakeJPSJ2009}
A decrease in $m^{**}(0)$ leads to a decrease in $m^{**}$($H_{\rm R}$).  
This situation is reminiscent of the case of CeRu$_2$Si$_2$, where a sharp pseudo-metamagnetic crossover at $H_{\rm M}$ occurs from a nearly AFM phase at $H$ = 0 to a polarized PM phase at $H_{\rm M}$ with $M(H_{\rm M})=\chi_0 H_{\rm M}$ 
($\chi_0$: initial-low field susceptibility).~\cite{Flo06_review}
Scaling of $H/H_{\rm M}$ is observed under pressure of $m^*(H)/m^*(0)$. 
Then $P$ motion of $H_{\rm M}$ occurs for the critical value of magnetization, $M(H_{\rm M})$~\cite{AokiJPSJ2011,Flo06_review}.
A change in magnetic correlations is associated with a drastic change in the FS; when the magnetic polarization reaches a critical value,~\cite{Flo06_review} the low-field PM FS becomes unstable.
As we will see in Sect. 6, a Lifshitz transition at $H_{\rm R}$ can strongly enhance $m^*_H$ as it will occur qualitatively through the crossing of the FM instability.
Scaling of $H/H_{\rm R}$ is an additional key signature of the Lifshitz transition.
There is an additional mechanism to FM spin fluctuations with the assumption of invariance of the FS for RSC.

\begin{figure}[!tbh]
\begin{center}
\includegraphics[width=\hsize,pagebox=cropbox,clip]{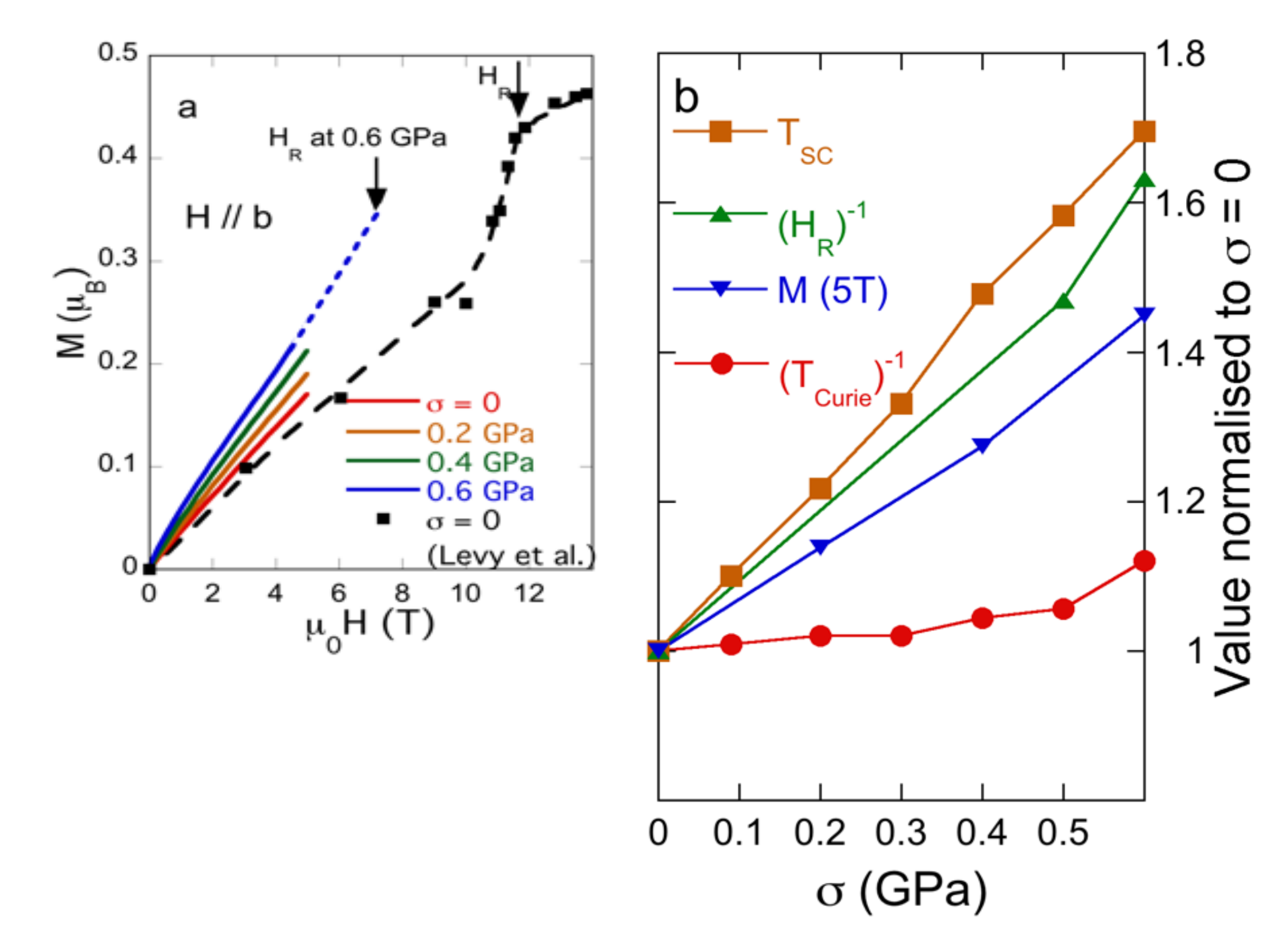}
\end{center}
\caption{(Color online) (a) Effect of uniaxial stress $\sigma$ on the URhGe magnetization curve with $H \parallel b$. 
(b) Uniaxial stress dependence of the zero field $T_{\rm SC}$(0), the inverse $H_{\rm R}$, and the  inverse $T_{\rm Curie}$ at $H = 0$.~\cite{BraithwaitePRL2018}}
\label{fig32}
\end{figure}

A novel possibility for driving the FM instability between the $c$- and $b$-axes is to apply uniaxial stress $\sigma$ along the $b$-axis, as thermal-expansion experiments demonstrate that $T_{\rm Curie}$ strongly decreases with increasing $\sigma$.
The target is to reach the FM quantum criticality along the $c$-axis and even to switch to FM along the $b$-axis by changing the sign of the magnetocrystalline energy\cite{BraithwaitePRL2018}.
At least the tendency to reach the FM instability along the $b$-axis is clear from the increase in the susceptibility $\chi_b$ [Fig.~\ref{fig32}(a)]. 
Furthermore, as $H_{\rm R}$ would occur when $\chi_b H_{\rm R} \approx M_0$, the expected concomitant effect is a decrease in $H_{\rm R}$.
Figure~\ref{fig32}(b) shows the uniaxial stress dependence of $T_{\rm Curie} (H = 0)$, $T_{\rm SC}(H = 0)$, $M_b$ (the magnetization at 5 T) and the increase of $H_{\rm R}^{-1}$. 
Let us emphasize the major boost of $T_{\rm SC}(\sigma)$ at zero field associated with the increasing of $\chi_b$.
At $\sigma = 1.2\,{\rm GPa}$, the maximum $T_{\rm SC}$ reaches $1\,{\rm K}$ at $H_{\rm R}=4\,{\rm T}$, while $T_{\rm SC}$ at zero field increases to $0.5\,{\rm K}$.

\begin{figure}[!tbh]
\begin{center}
\includegraphics[width=0.8\hsize,pagebox=cropbox,clip]{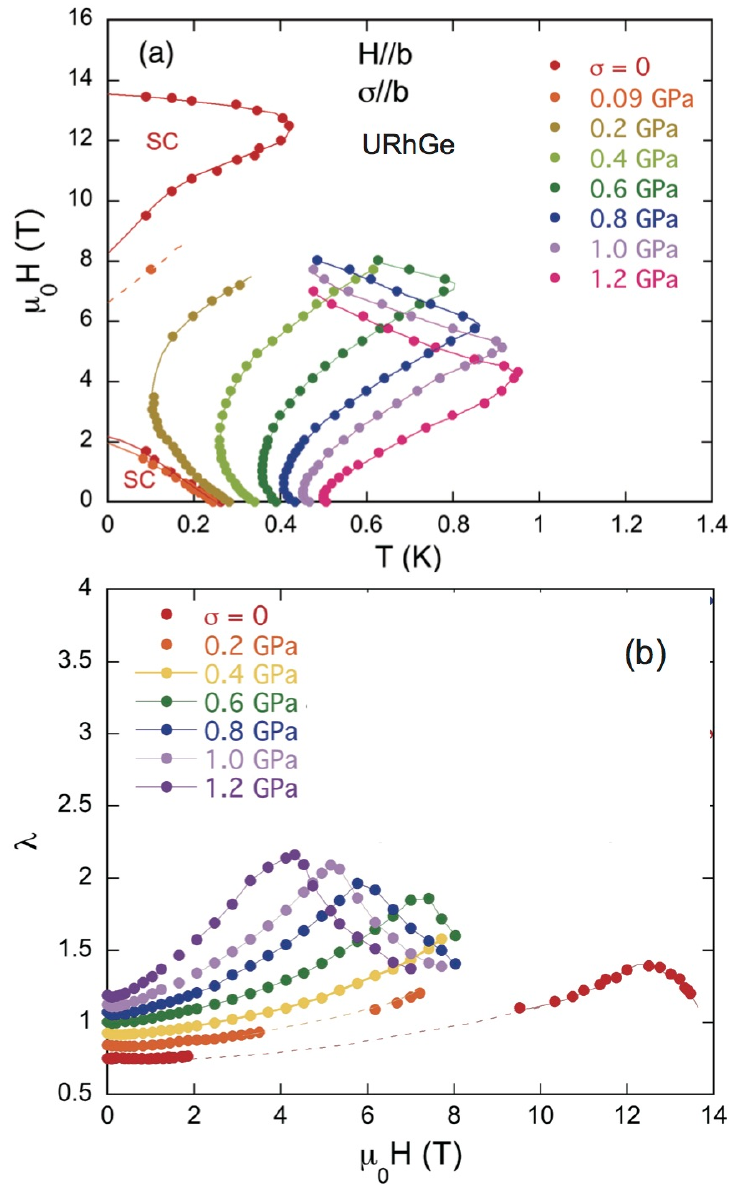}
\end{center}
\caption{(Color online) (a) Phase diagram of URhGe at different $\sigma$ with $H\parallel b$. (b) Dependence of $\lambda$ on field in URhGe at different $\sigma$ with $H \parallel b$}
\label{fig33}
\end{figure}

In the $\sigma$ experiments, it was not possible to quantitatively derive the normal-phase parameter. presumably because of insufficient $\sigma$ homogeneity. 
However, as shown in Fig. \ref{fig33}(a), drastic changes occur in the behavior of $H_{\rm c2}$, RSC is replaced by upward enhancement of SC. 
Analysis of $H_{\rm c2}$ leads to the field dependence of $\lambda(H)$ at different stresses $\sigma$ [Fig.~\ref{fig33}(b)].
Furthermore, scaling of $\lambda(H)/\lambda(0)$ as a function of $H/H_{\rm R}$ is obeyed.
Approaching the FM instability at $H = 0$ under uniaxial stress result in the marked enhancement of SC at $H_{\rm R}$.     

\section{Properties of UCoGe} \label{sec:UCoGe}
UCoGe offers the opportunity to study in more detail the interplay of FM and SC with the decrease in $T_{\rm Curie}$ down to 2.5 K and the increase in  $T_{\rm SC}$ up to 0.6 K at ambient pressure.
Furthermore, in contrast to URhGe, a moderate pressure ($P \sim1$ GPa) drives the FM-PM instability. 
Special attention is given to NQR and NMR results.

\subsection{NQR view of FM and SC transition}
\begin{figure}[t]
\begin{center}
\includegraphics[width=0.8\hsize,pagebox=cropbox,clip]{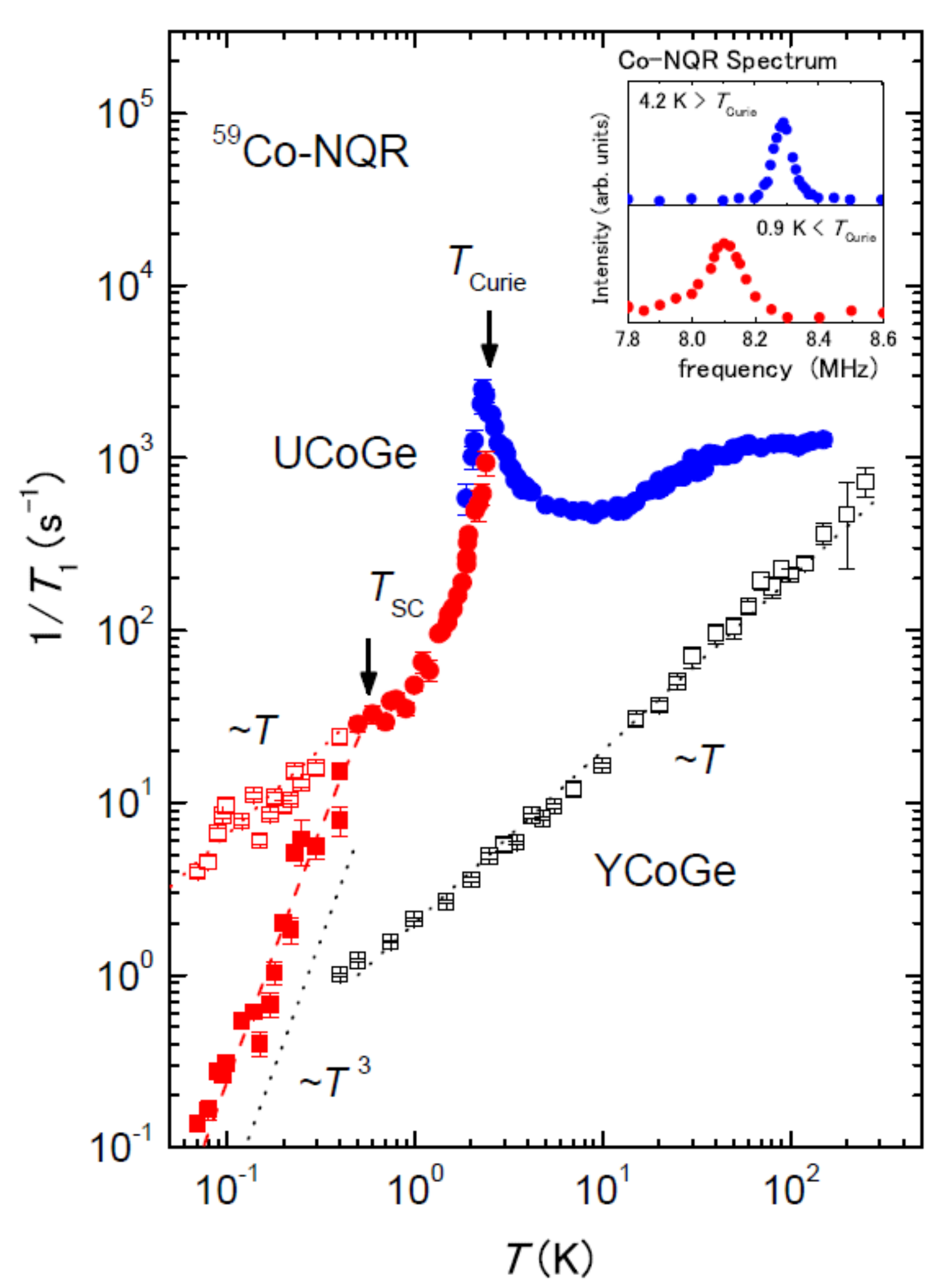}
\caption{(a) (Color online) Temperature dependence of $^{59}$Co NQR $1/T_1$ in single-crystal UCoGe. $1/T_1$ was measured at the PM ($8.3$ MHz) frequency above $2.3$ K, shown by blue circles. Below $2.3$ K, $1/T_1$ was measured at the FM ($8.1$ MHz) frequency. Two $1/T_1$ components were observed in the SC state: the faster (slower) component denoted by red solid (open) squares. The red broken curve below $T_{\rm SC}$ represents the temperature dependence calculated assuming a line-node gap with $\Delta_0 /k_{\rm B} T_{\rm SC} = 2.3$\cite{OhtaJPSJ2010}. The inset shows the Co-NQR spectra corresponding to the $E_{\pm 5/2}\leftrightarrow E_{\pm 7/2} (\nu_3)$ transitions above and below $T_{\rm Curie}$.}
\label{Fig-T1}
\end{center}
\end{figure}

\begin{figure}[t]
\begin{center}
\includegraphics[width=0.9\hsize,pagebox=cropbox,clip]{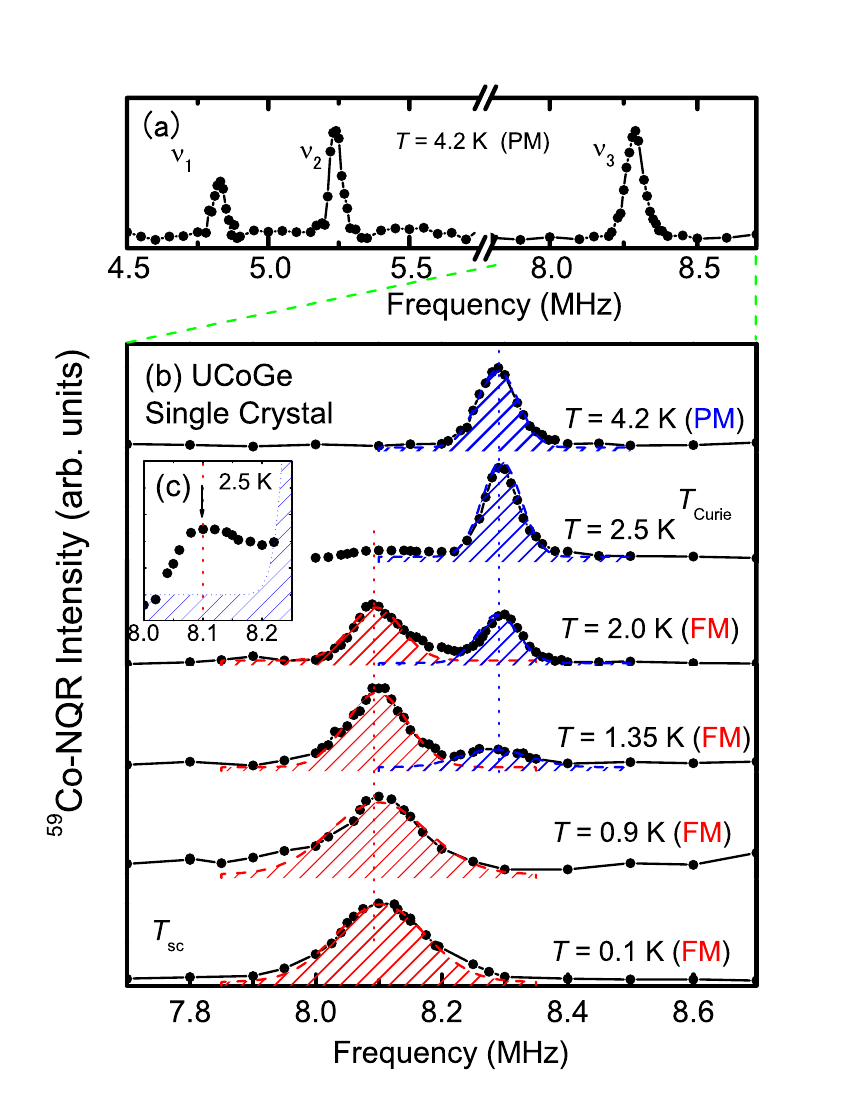}
\vspace*{-15pt}
\caption{(Color online) (a) $^{59}$Co NQR spectrum of single-crystal UCoGe in PM state. (b) Temperature variation of the $^{59}$Co NQR spectrum from the $\pm5/2 \Leftrightarrow \pm 7/2$ transitions ($\nu_3$) in the single-crystal sample\cite{OhtaJPSJ2010}. The inset (c) shows the FM signal at 8.1 MHz at $T$ = 2.5 K.}
\label{CoNQRSpectrum}
\vspace*{-15pt}
\end{center}
\end{figure}

Figure \ref{Fig-T1} shows the temperature dependence of $1/T_1$ measured by $^{59}$Co-NQR in single-crystal UCoGe down to 70 mK\cite{OhtaJPSJ2010}. 
$1/T_1$ in the single crystal remains nearly constant down to $T^* \simeq40$ K and gradually decreases below $T^*$. 
The magnetic susceptibility deviates from the Curie-Weiss behavior and the electrical resistivity along the $c$-axis shows metallic behavior below $T^*$; $T^*$ is regarded as the characteristic temperature below which the U-$5f$ electrons become itinerant with relatively heavy electron mass. 
Below 10 K, $1/T_1$ increases and shows a large peak at $T_{\rm Curie} \simeq2.5$ K owing to the presence of FM critical fluctuations. 
However, the $^{59}$Co-NQR spectrum gives evidence of first-order transition behavior at $T_{\rm Curie}$.
As shown in Fig.~\ref{CoNQRSpectrum}\cite{OhtaJPSJ2010}, with decreasing temperature, the intensity of the 8.3 MHz NQR signal arising from the PM region decreases below $\simeq$3.7 K, while an 8.1 MHz signal from the FM region, which shifts as result of the presence of the internal field ($H_{\rm int}$) at the Co site, appears below 2.7 K. 
The two NQR signals coexist between 1 and 2.7 K, but the PM signal disappears below 0.9 K.
This indicates that although the phase separation between the PM and FM regions occurs at $T_{\rm Curie}$, the single-crystal UCoGe is in the homogeneous FM state, which is proof of the absence of the PM signal below 1 K.
Also note that the frequency of the FM signal, 8.1 MHz, is nearly unchanged from its first appearance. 
The experimental results of the discontinuous appearance of the FM signal and of the coexistence of FM and PM signals around $T = 2$ K show that the FM transition is of the first order.
However, when the temperature variation of the NQR intensity of the PM and FM signals was recorded in cooling and warming processes, no hysteresis behavior was observed, the energy difference between the PM and FM phases was very small.
The results are consistent with the previous discussion on UGe$_2$ that the low-temperature transition in itinerant ferromagnets is generally of the first order (see Sect.~\ref{sec:UGe2} on UGe$_2$)~\cite{BelitzPRL1999,BelitzPRL2005}.   
The FM transition of UCoGe is close to the TCP.

In the SC state, the fast component of $1/T_1$ in the FM signal is roughly proportional to $T$, indicating that it originates from non-superconducting regions. 
In contrast, the slow component in the FM signal decreases rapidly below $T_{\rm SC}$, roughly as $T^3$, suggestive of line nodes on a SC gap. 
The red broken line in Fig.~\ref{Fig-T1} shows a fit using the line-node model $\Delta (\theta) = \Delta_0 \cos \theta$ with $\Delta_0 = 2.3 k_{\rm B} T_{\rm SC}$. 
The detection of the SC gap via the FM signal provides unambiguous evidence for the microscopic coexistence of ferromagnetism and superconductivity. 
In addition, the results below $T_{\rm SC}$ provide some new insight on the nature of the superconductivity in UCoGe. 
From the relaxation in the FM signal, nearly half of the sample volume remains non-SC even at 70 mK, wheseas the sample is in a homogeneous FM state below 1 K.  
The inhomogeneous SC state is expected to be related with the SIV state, as discussed below.

Co-NQR measurements of the reference compound YCoGe were performed\cite{KarubeJPSJ2011}. 
YCoGe has the same TiNiSi crystal structure and similar lattice constants to UCoGe but has no $f$ electrons.
The band calculation suggests that the contribution of Co-3$d$ electrons to the density of states is similar in UCoGe and YCoGe.
As shown in Fig.~\ref{Fig-T1}$, 1/T_1$ of Co follows the $T$-linear relation below 250 K down to 0.5 K; this is a typical metallic behavior, and neither ferromagnetism nor superconductivity was observed down to 100 mK\cite{KarubeJPSJ2011}.
These results prove that the ferromagnetism and unconventional superconductivity in UCoGe originate from U-5$f$ electrons.

\subsection{Self-induced-vortex (SIV) state}    
Since the temperature where the second component of $1/T_1$ emerges coincides with $T_{\rm SC}$ in UCoGe, the two-relaxation behavior of $1/T_1$ is considered to be intrinsic.
Furthermore, from recent pressure NQR measurements on the same single crystal, the non-SC component of $1/T_1$ disappears in the high-pressure SC state, where the FM state is suppressed.
This strongly indicates that the non-SC component is not an extrinsic effect, such as impurities or inhomogeneity of the sample, but an intrinsic effect induced by the presence of the FM moments.   
The plausible origin of the non-SC component is ascribed to the SIV, as pointed out in Sect. 2.   
The SIV state in UCoGe has also been indicated by a muon experiment\cite{VisserPRL2009}, and is now clearly observed in magnetizatiion data.
  
\begin{figure}[!tbh]
\begin{center}
\includegraphics[width=0.6\hsize,pagebox=cropbox,clip]{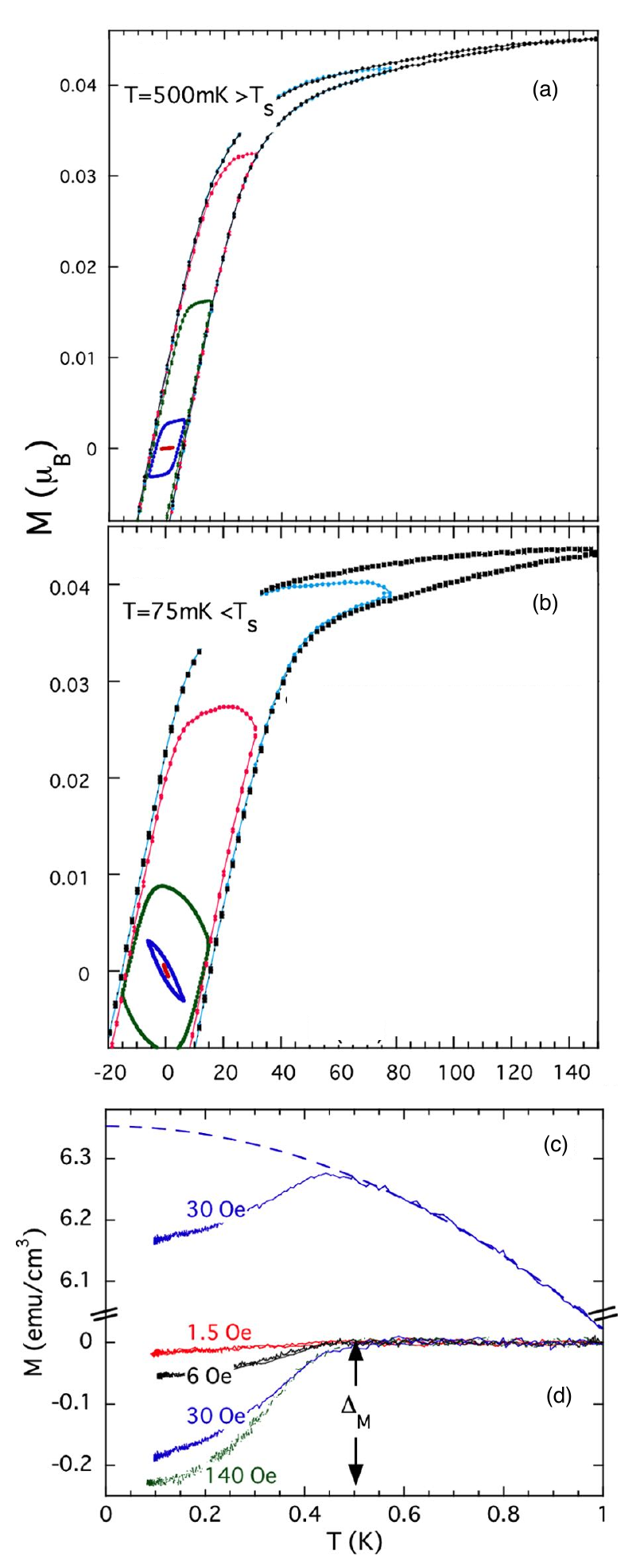}
\end{center}
\caption{(Color online) Magnetization curve of UCoGe in SC and normal phases for $H \parallel c$. (a and b) Hysteresis cycle measured in the PM and SC domains. (c) Field-cooled magnetization measured along the c-axis at 30 Oe, showing flux expulsion below 500 mK. The dashed line is the extrapolation to $T=0\,{\rm K}$ in the case of normal-phase behavior. (d) $c$-axis field-cooled data after deduction of normal-phase contribution.~\cite{PaulsenPRL2012}}
\label{fig44}
\end{figure}

It is worth examining the response of the magnetization of UCoGe to the interplay between FM and SC\cite{DeguchiJPSJ2010, PaulsenPRL2012}. 
A rough estimation of the local critical field $H_{c1}$ from $H_{\rm c2}$ and $H_c$ (thermodynamic critical field estimated from specific heat) gives $H_{c1} \sim3$ G along the $c$-axis and approximately 0.1 G along the $a$- and $b$-axes. 
Clearly, owing to the strength of the internal field near 100 G (see table II), SIVs already exist at $H = 0$. 
In Fig.~\ref{fig44}, it is worth observing the change in the hysteresis cycle above 500 mK and below $T_{\rm SC}$ at 75 mK: the coercive field is 6 G at 500 mK and increases 16 G at 75 mK\cite{HykelPRB2014}. 
Expulsion of the flux is shown in Fig.~\ref{fig44}. 
For $H \parallel c$, the flux expulsion is directly related to the bulk magnetization; it operates on each FM domain. 
Scanning SQUID microscopy~\cite{HykelPRB2014} helps to clarify the macroscopic figures but the vortex lattice has not yet been observed.
No shrinkage of FM domains has been detected, as proposed theoretically\cite{Fau05,Dao11}.
Recent calculations of the magnetization in the FM-SC phase confirm slight magnetization expulsion in the frame of two FM bands with equal spin pairing.~\cite{Min18}
The regime near the vortex core may be the origin of the fast component of $T_1$. 
However, the number of vortices that can be derived from specific heat measurement (Fig. \ref{fig:Cp}) cannot quantitatively explain the large relaxation component detected by NQR.
Further improvement of the crystal purity may help clarify the SIV phase.

\subsection{Ising fluctuation with strong $H$ dependence in longitudinal and transverse $H$ scan NMR studies}
\begin{figure}[!tbh]
\begin{center}
\includegraphics[width=0.8\hsize,pagebox=cropbox,clip]{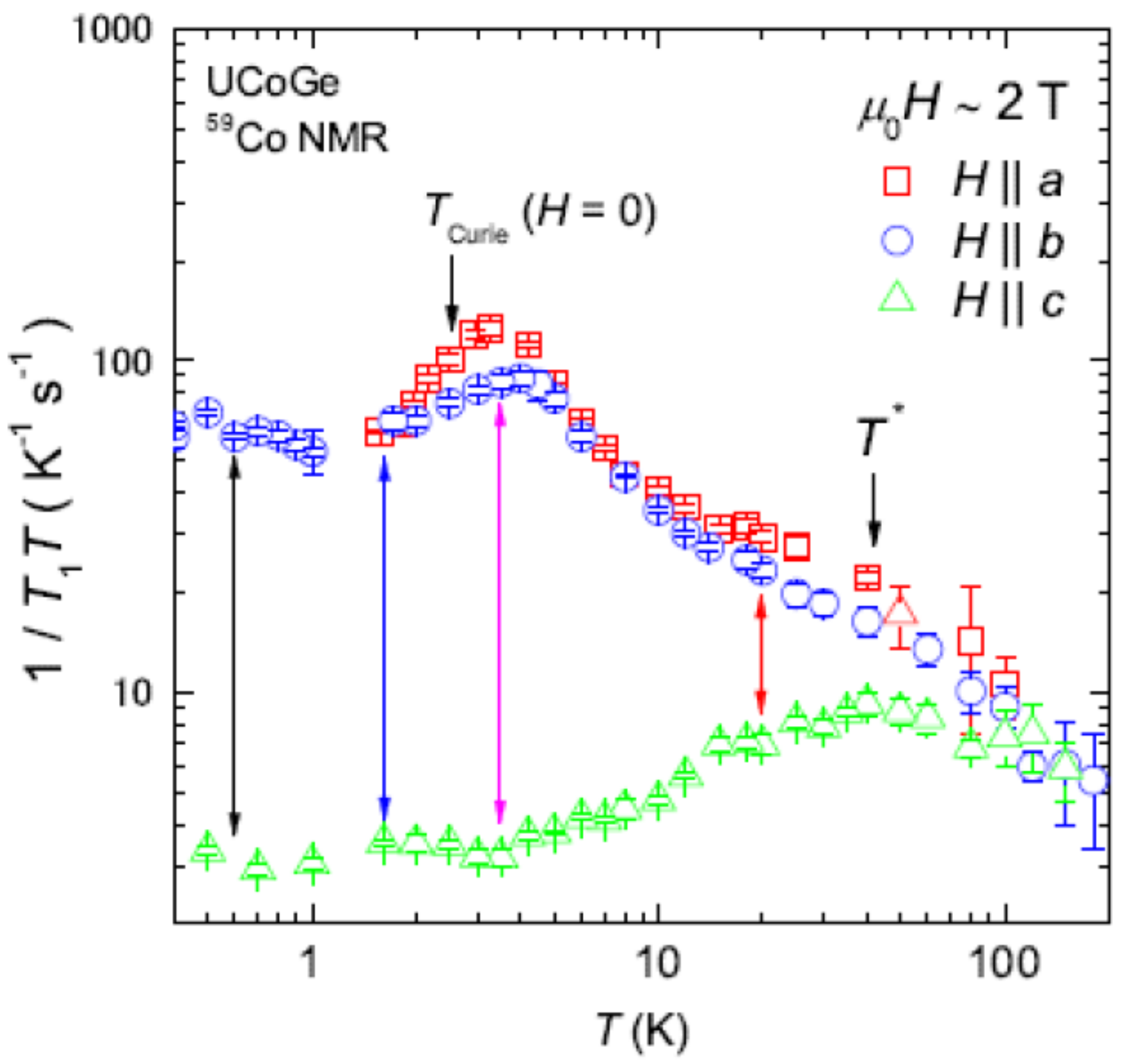}
\end{center}
\caption{(Color online) Temperature dependences of $1/T_1T$ with field along each crystalline axis. Data below 1 K were measured using a $^3$He-$^4$He dilution refrigerator. The Ising FM fluctuation along the $c$-axis grows below 50 K, where the resistivity along the $c$-axis becomes metallic, and remains above $T_{\rm SC} \sim0.6$ K. The arrows indicate the temperature where the angle dependence of $1/T_1T$ in Fig.~\ref{fig35}(a) was measured.  }
\label{fig34}
\end{figure}

\begin{figure}[!tbh]
\begin{center}
\includegraphics[width=\hsize,pagebox=cropbox,clip]{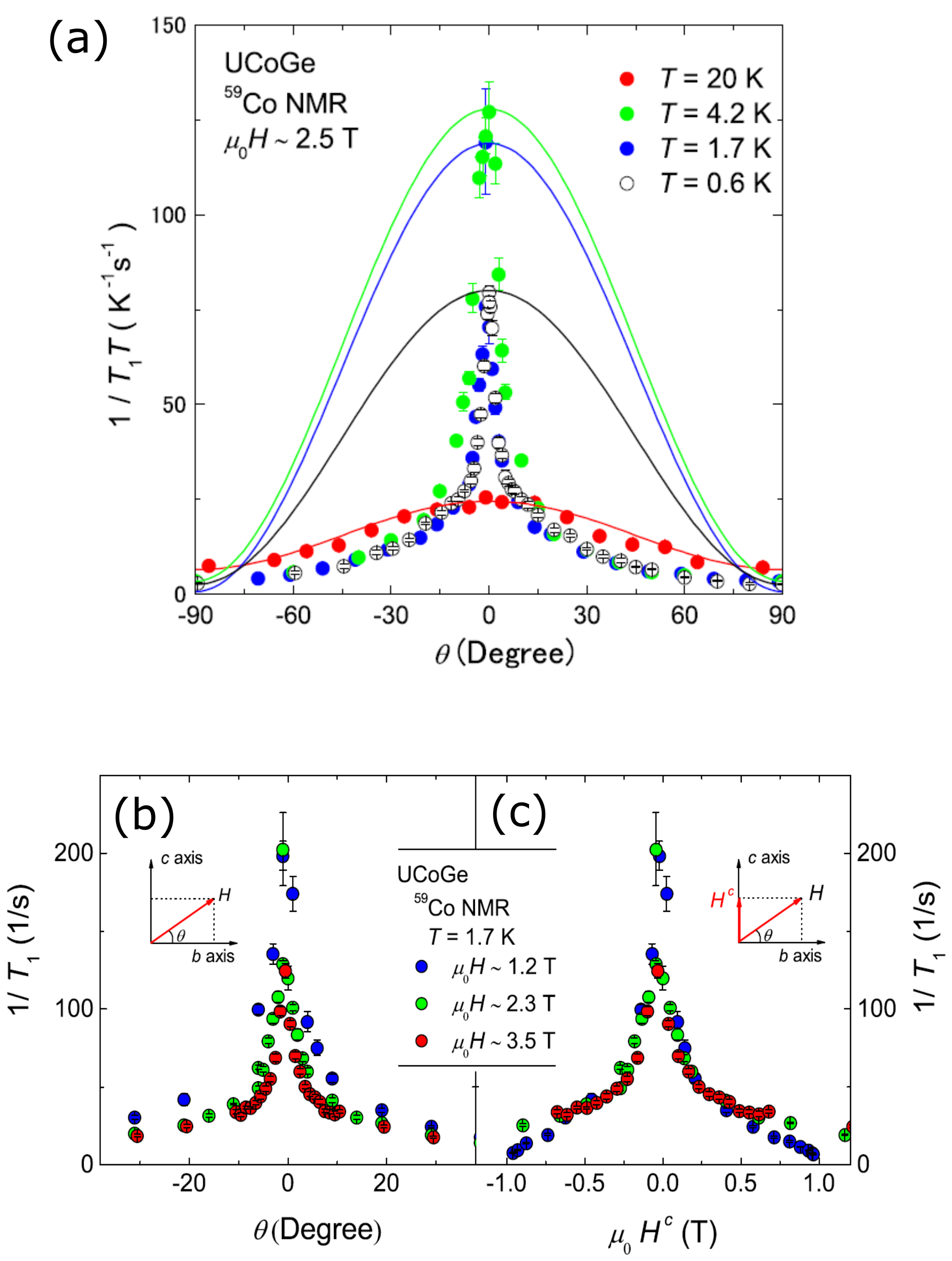}
\end{center}
\caption{(Color online) (a) Angular dependence of $1/T_1$ at various temperatures. $\theta$ is the angle from the $b$-axis in the $bc$-plane. The curves show the equation $1/T_1(\theta) = 1/T_1^b \cos^2{\theta} +1/T_1^c \sin^2{\theta}$, with $1/T_1^{b,c}$ a constant value. This curve can consistently explain the smooth variation at 20 K but not the sharp angle dependence observed below 4.2, 1.7, and 0.6 K, which shows a cusp centered at $\theta = 0^{\circ}$. (b) Angular dependences of $1/T_1$ in the $bc$-plane measured at three different magnetic fields at $T$ = 1.7 K. (c) Plot of $1/T_1$ against the $c$-axis component of the field $H^c = H \sin{\theta}$.}
\label{fig35}
\end{figure}

\begin{figure}[!tbh]
\begin{center}
\includegraphics[width=0.8\hsize,pagebox=cropbox,clip]{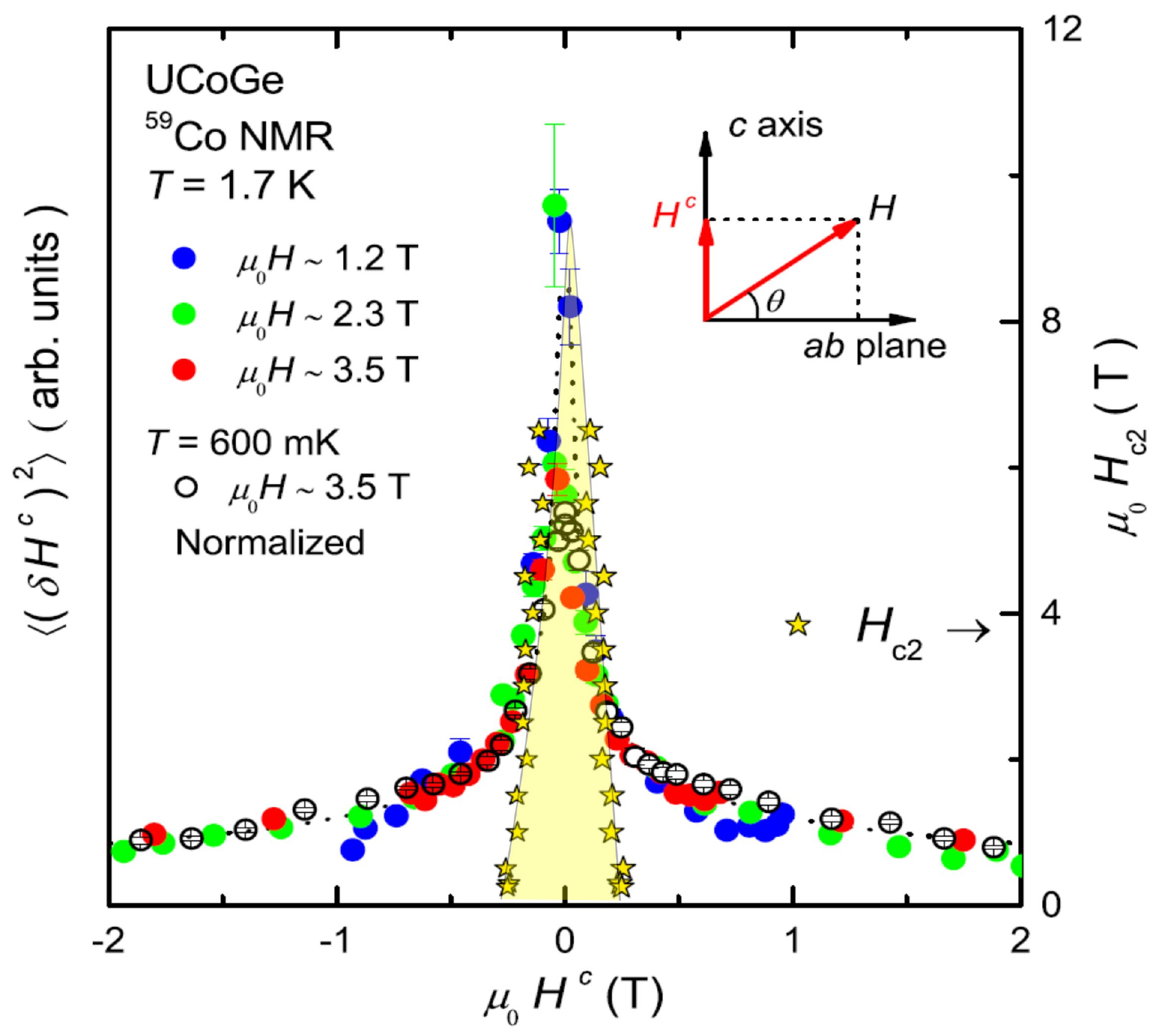}
\end{center}
\caption{(Color online) $H^c$ dependence of magnetic fluctuations along the $c$-axis $\left<(\delta H^c)^2\right>$ at 1.7 and 0.6 K, extracted using Eq.~(\ref{eq:fluctuation}). $H_{\rm c2}$ determined from $\chi_{\rm ac}$ is plotted against $H^c = H_{\rm c2}\sin{\theta}$, and the $H^c$ region where superconductivity is observable is shown by the yellow area. The relation $\left<(\delta Hc)^2\right> \propto 1/\sqrt{H^c}$ is shown by the dotted line  }
\label{fig35a}
\end{figure}

The weakness of $M_0$ and $T_{\rm Curie}$ leads to the new feature that the FM Ising-type interaction in the FM and PM ground states will collapse rapidly under a longitudinal magnetic field $H \parallel c \parallel M_0$\cite{HuyPRL2008, AokiJPSJ2009}.
Further neutron inelastic experiments,\cite{StockPRL2011} as well as measurements of $T_1$\cite{IharaPRL2010, HattoriPRL2012} by NMR with $H \parallel a, \parallel b$, and $\parallel c$, show that the magnetic fluctuations are of the Ising type with fluctuations of the magnetism parallel to $M_0$. 
Figure~\ref{fig34} shows the temperature dependences of $1/T_1T$ in the three main directions. 
This longitudinal fluctuation is strongly affected strongly by the strength of the $H^c$ component of $H$ along the $c$ axis, as shown in Fig.~\ref{fig35}(a), where $\theta$ is the angle from the $b$-axis in the $bc$-plane. 
Although the data follow the classical equation of 
\begin{equation}
\frac{1}{T_1}(\theta) = \frac{1}{T_1^b} \cos^2{\theta} +\frac{1}{T_1^c} \sin^2{\theta} \label{eq:T1vstheta} 
\end{equation}
with constant values of $T_1^{b}$ and $T_1^{c}$ at 20 K, the low-temperature data below 4.2 K do not follow the relation at all, since $T_1^b$ depends on $H^c$. 
As shown in Fig. \ref{fig35}(c), $H^c$ is the key parameter\cite{HattoriPRL2012}.
To investigate how the longitudinal FM fluctuations along the $c$-axis  $\langle (\delta H^{c} )^2 \rangle$ couple to the external field, $\langle (\delta H^{c} )^2 \rangle$ is derived as a function of $1/T_1 (\theta)$ by combining the equation of $1/T_1T$ and eq. (\ref{eq:T1vstheta}) assuming that the magnetic fluctuations in the $ab$-plane are isotropic in the low-field region
($\langle (\delta H^a)^2\rangle \sim \langle (\delta H^b)^2\rangle$):
\begin{equation}
\langle (\delta H^c)^2\rangle \propto \frac{1}{\cos^2{\theta}}\left(\frac{1}{T_1}(\theta)-\frac{(1+\sin^2{\theta})}{2}\frac{1}{T_1^c}\right).\label{eq:fluctuation}
\end{equation}
Figure \ref{fig35a} is a plot of $\left<(\delta H^c)^2\right>$ at 1.7 and 0.6 K against $H^c$.
When $H_{\rm c2}$ along the $b$-axis is drawn in the same figure, superconductivity is observed in this narrow field region where the longitudinal FM spin fluctuations are active.

The longitudinal FM fluctuations  $\left<(\delta H^c)^2\right>$, which are coupled with superconductivity, are different from the ordinary spin-wave excitation observed in the FM ordered state. 
In the conventional FM state, the low-lying spin excitation is a transverse mode corresponding to the Nambu--Goldstone mode, 
but the FM fluctuations observed in UCoGe are an longitudinal mode of the U-5$f$ moment.

\begin{figure}[!tbh]
\begin{center}
\includegraphics[width=0.8\hsize,pagebox=cropbox,clip]{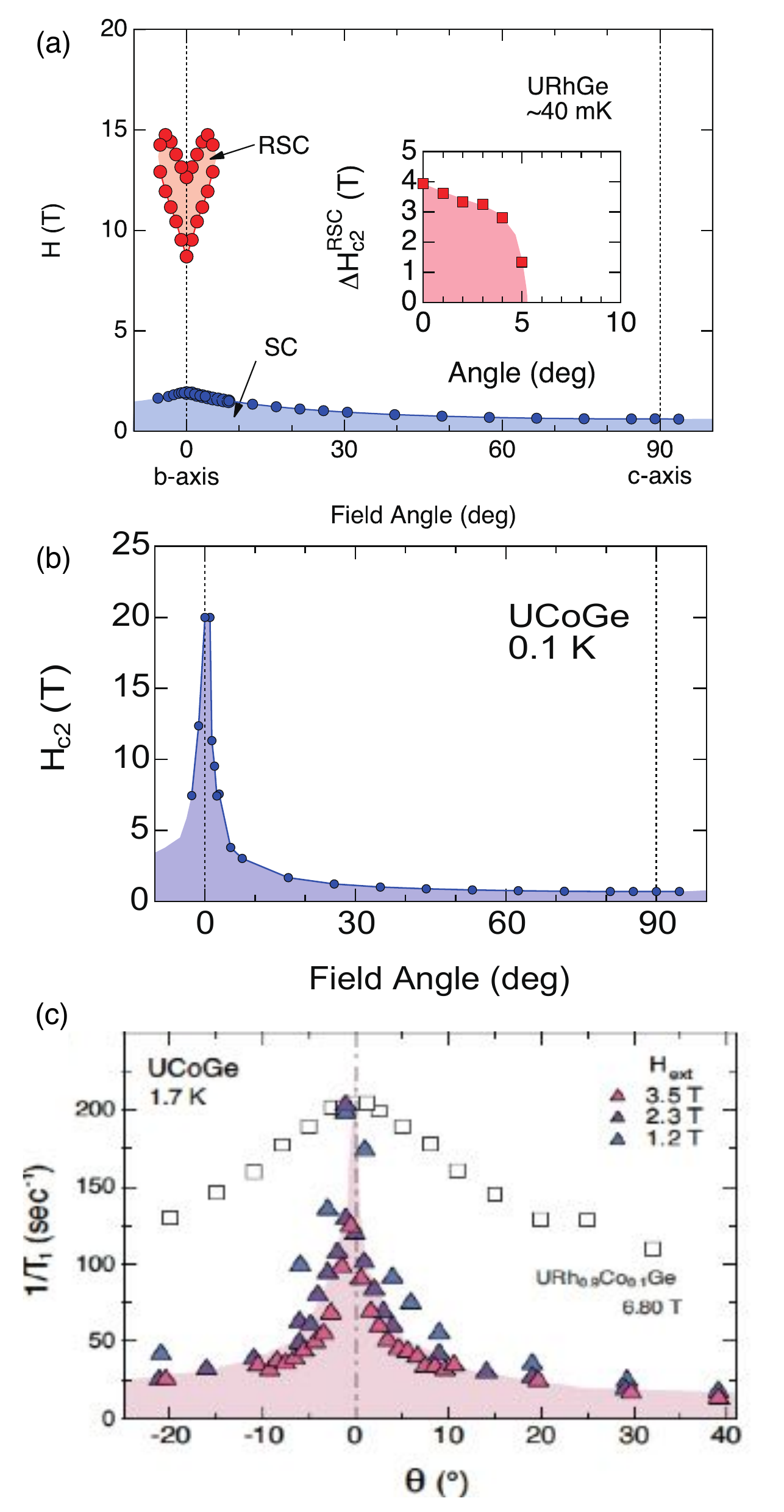}
\end{center}
\caption{(Color online) Angular singularity in $H_{\rm c2}(\theta)$ in (a) URhGe and (b) UCoGe for $H \parallel b \perp M_0$; note the collapse of the RSC of URhGe for $\theta = 5^{\circ}$. (c) Angular dependence of $1/T_1$ measured in UCoGe and URhGe for $H \parallel b$ in low-field scan. }
\label{fig41}
\end{figure}

In addition, to point out the link between the angle dependence of $H_{\rm c2}(\theta)$ and that of the low-field FM fluctuations seen via $T_1(\theta)$, 
Fig.~\ref{fig41} shows the difference in $H_{\rm c2}$ between UCoGe and URhGe  as a function of the angle $\theta$ from the $b$-axis: the link between the Ising character of the fluctuation along the $c$-axis is sensitive in UCoGe, but rather insensitive in URhGe. 
It is noteworthy that there are two SC regions in UCoGe as well as in URhGe: one is in the SC state, which has an extremely large $H_{\rm c2}$ and is sensitive to $\theta$ for $H \parallel b$, 
and the other is in the SC state, which has a small $H_{\rm c2}$ and is weakly sensitive to deviation from $\theta = 90^\circ$ for $H\parallel c$.
The former SC state is considered to be induced by critical FM fluctuations.      


\begin{figure}[!tbh]
\begin{center}
\includegraphics[width=0.8\hsize,pagebox=cropbox,clip]{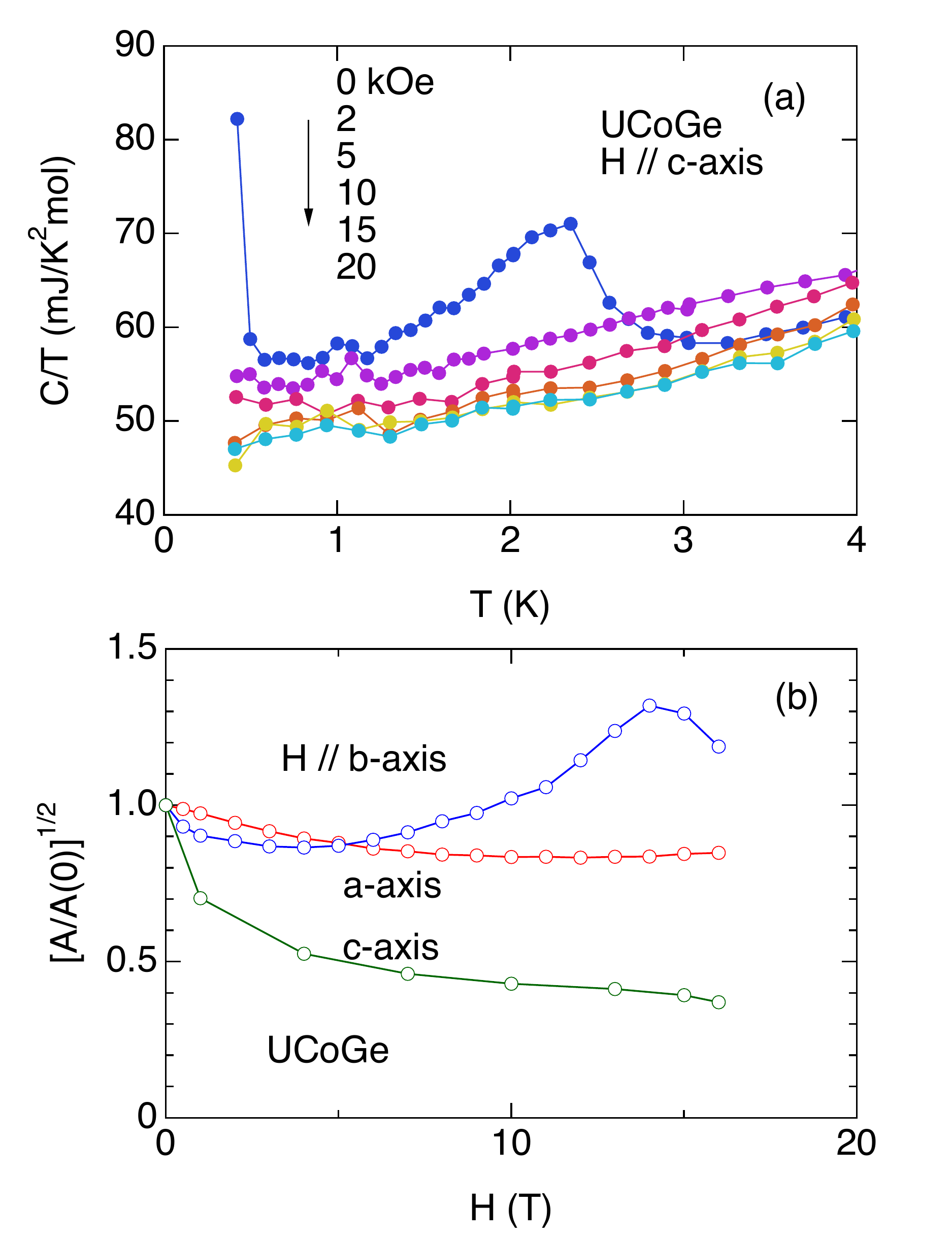}
\end{center}
\caption{(Color online) (a) $T$ dependence of $C/T$ in UCoGe at different fields for $H \parallel c$. 
(b) $H$ decrease of $A$ in longitudinal scan ($H\parallel M_0 \parallel c$) and transverse scan ($H\parallel b$ and $H \parallel a$) }
\label{fig36}
\end{figure}

\begin{figure}[!tbh]
\begin{center}
\includegraphics[width=0.8\hsize,pagebox=cropbox,clip]{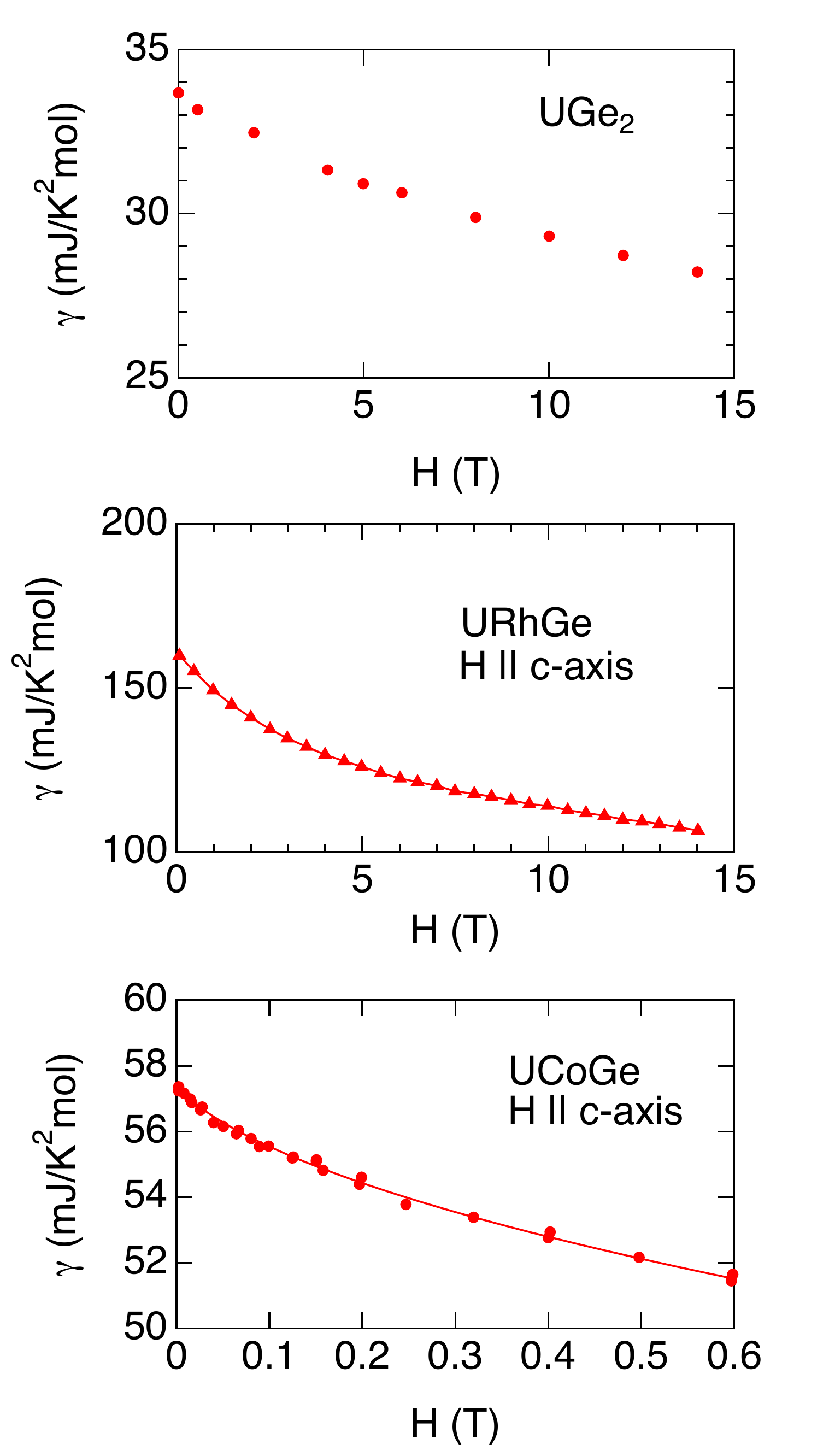}
\end{center}
\caption{(Color online) Decrease in $\gamma(H)$ as a function of $H$ for UGe$_2$, URhGe, and UCoGe for longitudinal field scan ($H \parallel H_0$.~\cite{Har09_UGe2,Har11,WuThesis} The decrease in $\gamma(H)$ in UCoGe will lead to a decrease in $\lambda(H) $ and the universal upward curvature of $H_{\rm c2}$ for $H \parallel b$. }
\label{fig37}
\end{figure}

The strong reduction of the FM fluctuation for $H \parallel c$ is also clearly detected in the specific heat measurements shown in Fig. \ref{fig36}, as well as in the field dependence of $A$. 
Comparing the field dependence of $\gamma$ between UGe$_2$,\cite{Har09_UGe2} URhGe,\cite{AokiJPSJ2014Rev} and UCoGe\cite{AokiJPSJ2014Rev,WuNatComm2017} (Fig.~\ref{fig37}), we see that the relative decrease depends roughly on the ratio $a H/T_{\rm Curie}$. 
For UCoGe, in agreement with the NMR data, the $H$ dependence of $\gamma$ is large for $H < 0.3$ T.

\subsection{Consequence on superconductivity: $H_{\rm c2}$ data and modeling}
\begin{figure}[!tbh]
\begin{center}
\includegraphics[width=\hsize,pagebox=cropbox,clip]{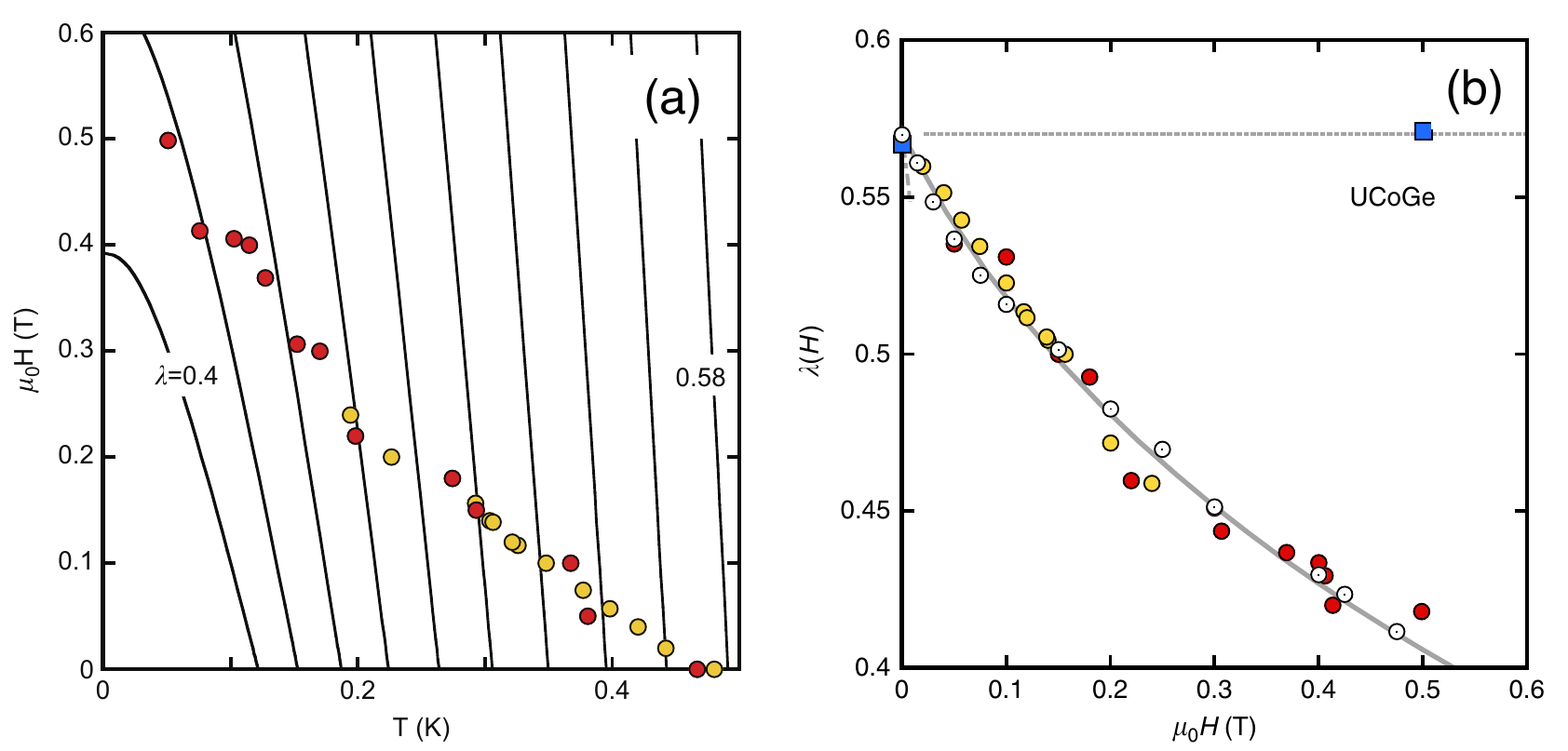}
\end{center}
\caption{(Color online) (a) Universal upward curvature of $H_{\rm c2}$($T$) for $H \parallel M_0 \parallel c$ in UCoGe. The $H_{\rm c2}(T)$ curve is adjusted by selecting $\lambda(H)$ and $T_{\rm SC}$($\lambda (H)$) point by point.~\cite{WuNatComm2017,WuThesis} (b) Analysis of the field dependence of the pairing strength for $H \parallel c$ in UCoGe\cite{WuNatComm2017,WuThesis}. Filled (open) circles were estimated from the experimental $H_{c2}$ curve (from specific-heat measurement). The solid line is a calculation of $\lambda(H)$ derived from the magnetization measurements.~\cite{MineevPUsr2017,HassingerJPSJ2008}
The low-field regime for $H\parallel b$ is presented by blue squares.}
\label{fig38}
\end{figure}

The direct consequence on SC is that the field dependence of $\gamma(H_c)$ will also lead to a drastic decrease in $\lambda(H_c)$, driving the usual upward curvature of $H_{\rm c2}(H \parallel c)$ due to the term $(dT_{\rm SC}/d\lambda)(d\lambda/dH)$ in the expression for $H_{\rm c2}$\cite{WuNatComm2017}
\begin{equation*}
\left( \frac{dT_{\rm SC}}{dH_{\rm c2}} \right) 
= -\frac{1}{\alpha_0 T_{\rm SC} {m^\ast}^2} + \frac{dT_{\rm SC}}{d\lambda}~\frac{d\lambda}{dH}
\end{equation*}
Figure~\ref{fig38}(a) shows how the $H_{\rm c2}$ data for $H \parallel c$ can be parameterized taking into account the field dependence of $\lambda (H)$ on $T_{\rm SC}(m_H^*)$ and $H_{\rm c2}$ caused by the decrease in $m^*_H$. 
Figure~\ref{fig38}(b) shows $\lambda$ ($H \parallel c$) derived from the analysis of $H_{\rm c2}$, from the direct determination of $\gamma$, and from the phenomenological model described later in Sect. 6. 
From NMR measurements above $T_{\rm SC}$ at 1.5 K, it was proposed that the FM fluctuation $\left<\delta H_c\right>^2$ decreased as $\sqrt{H_c}$; this $H_c$ dependence will lead to an infinite derivative of $\lambda$ for $H \rightarrow 0$ from the relation  $\left<(\delta Hc)^2\right> \propto 1/\sqrt{H^c}$\cite{HattoriPRL2012}. 
Measurements of the specific heat down to $T_{\rm SC} \sim0.7$ K show that $\gamma$ and thus $\lambda$ initially decrease linearly with increasing $H$.
Thus, the $H$ dependence of $1/T_1$ at intermediate temperatures is markedly enhanced, probably owing to the strong coupling with the critical FM fluctuations.

\begin{figure}[!tbh]
\begin{center}
\includegraphics[width=0.8\hsize,pagebox=cropbox,clip]{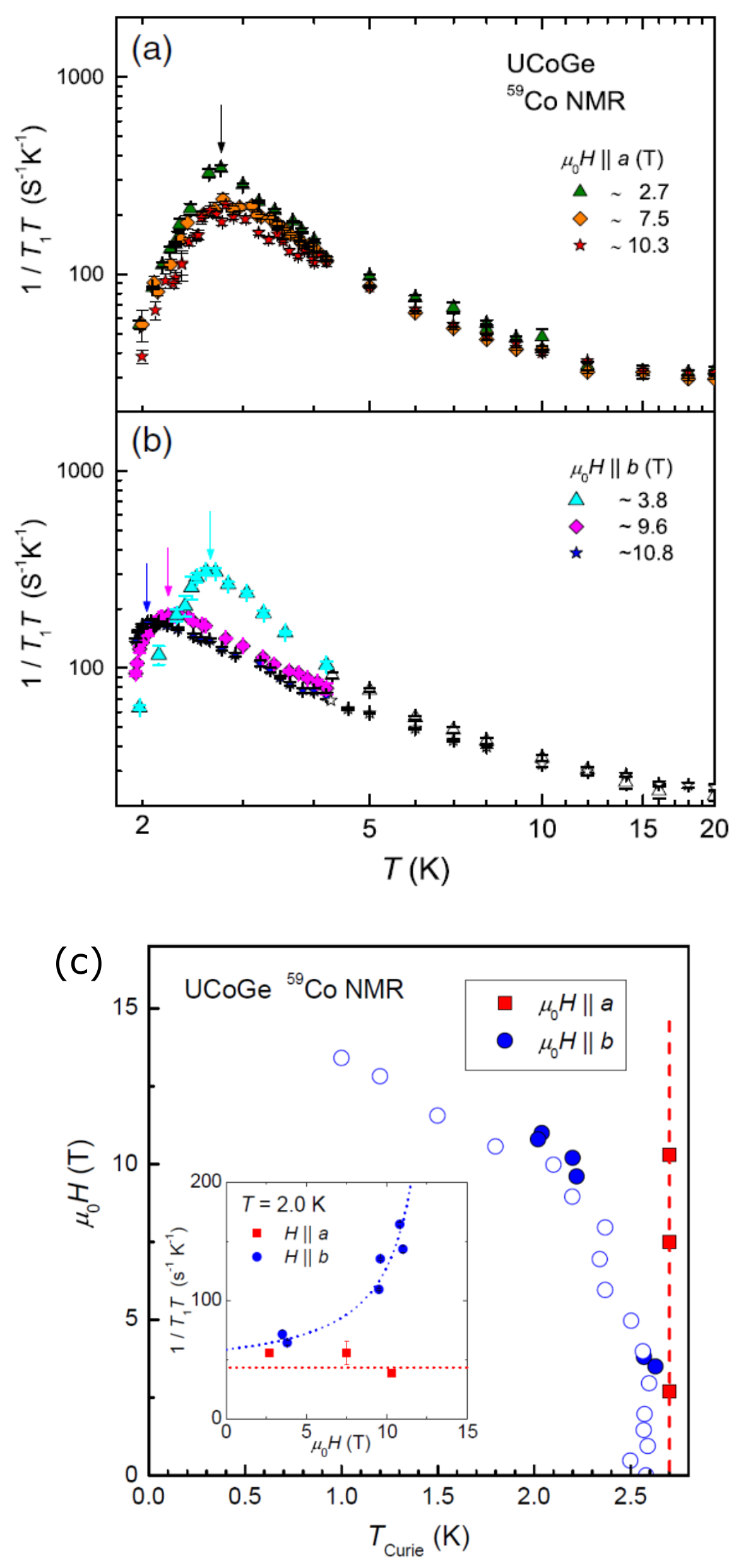}
\end{center}
\caption{(Color online) Temperature dependence of $^{59}$Co-NMR $1/T_1T$ in various fields along the (a) $a$-axis  and (b) $b$-axis. (c) Field dependence of $T_{\rm Curie}$ determined by the peak of $1/T_1T$ against temperature in the fields along the $a$- and $b$-axes.
The inset shows the field dependences of $1/T_1T$ measured by $^{59}$Co-NMR at $T =$ 2.0 K ($< T_{\rm Curie}$). The dotted lines in the inset are guides for the eye. While the field along the $a$-axis does not change the magnetic properties, the magnetic field along the $b$-axis enhances the magnetic fluctuations.  }
\label{fig40}
\end{figure}

\begin{figure}[!tbh]
\begin{center}
\includegraphics[width=0.8\hsize,pagebox=cropbox,clip]{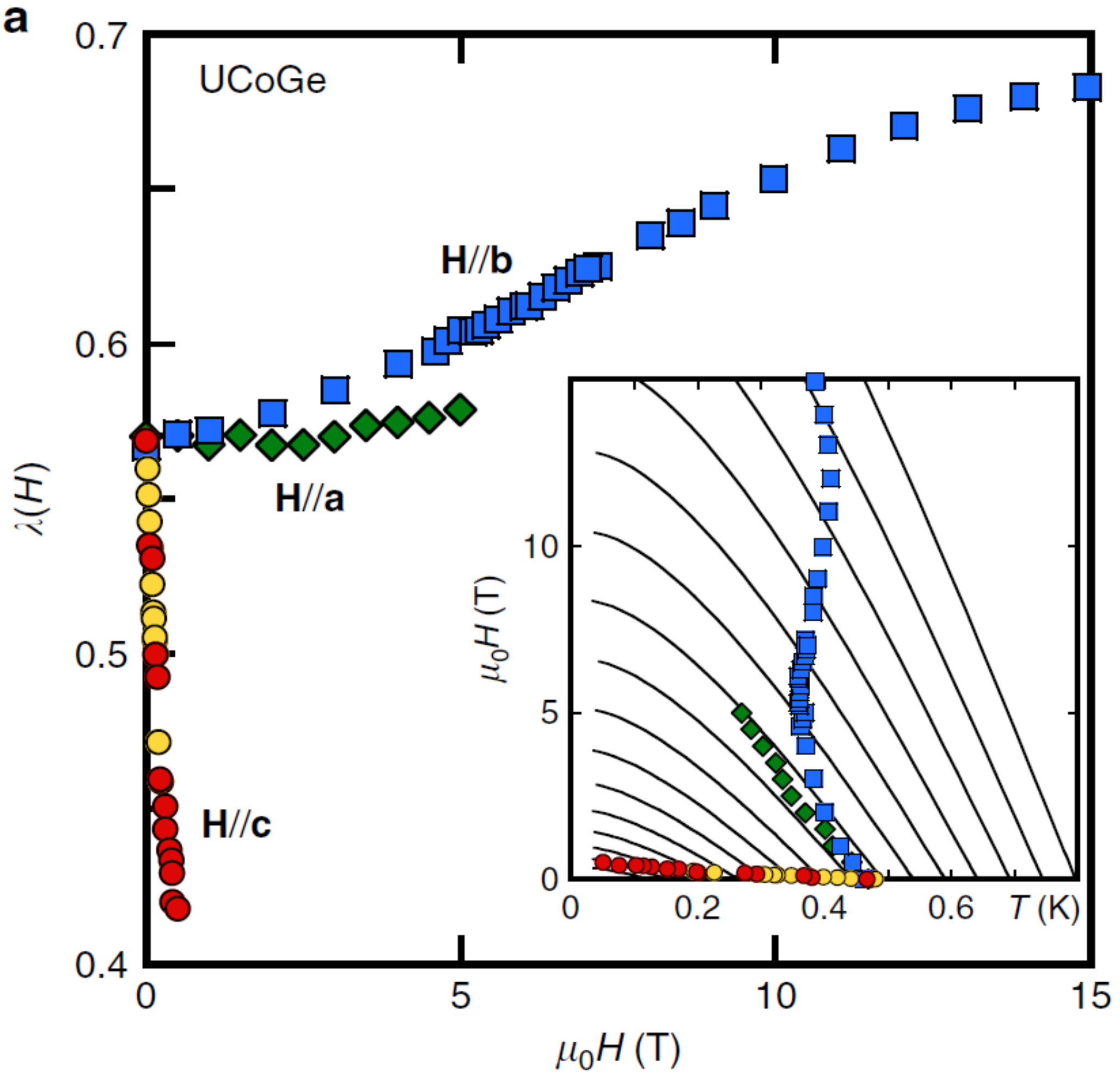}
\end{center}
\caption{(Color online) Field dependence of $\lambda(H)$ in UCoGe, for $H \parallel c$ longitudinal mode and $H \parallel b$ and $a$ transverse modes derived from the $H_{\rm c2}$ analysis with the variation of $\lambda$ as shown in the inset. See  in Refs.~\citen{WuNatComm2017,WuThesis}.}
\label{fig39}
\end{figure}

As already underlined, for $H \parallel b$ the consequence is that the long-range FM along the $c$-axis will collapse for $H^*_b \sim12\,{\rm T}$. 
This estimation is in excellent agreement with the strong $H$ shift of the maximum of $1/T_1T$ on approaching 12 T, which is shown in Fig.~\ref{fig40}. 
Furthermore, the strong increase in $1/T_1T$ at $T$ = 2 K by a factor of 2 between $H = 0$ and $12$ T [the inset of Fig. \ref{fig40}(c)]\cite{HattoriJPSJ2014} is also in good agreement with the increase in $A$ by a factor of 1.8 (in a crude electronic model with $1/T_1T \sim\gamma^2 \sim A$).
Parameterization of $\lambda$ via $H_{\rm c2}$ ($H \parallel b$) (see Fig.~\ref{fig39}) leads to an increase from $\lambda \simeq0.57$ at $H = 0$ to $\lambda \simeq0.68$ at $H^*_b$; the estimation of $\gamma(H^*_b)/\gamma(0) \sim1.06$ is much lower than the value of 1.4 expected from $1/T_1T$ or $A$.~\cite{WuNatComm2017,WuThesis}
The strength of the ratio $T_{\rm Curie}/T_{\rm SC}$ must be related to the switch from the weak to the strong coupling condition.
Between URhGe and UCoGe, this ratio differs by one order of magnitude (40 in URhGe compared with 3.8 in UCoGe).
As the magnitude of $\lambda$ must be connected to the proximity of the FM-SC instability ($P_{\rm c}$),
one may expect at $H=0$ that $\lambda$ of UCoGe is greater than that of URhGe.
The derivation of $\lambda$ via $H_{\rm c2}$ gives the opposite result
($\lambda = 0.75$ in URhGe and $\lambda=0.57$ in UCoGe).
We remark that the quasi-invariance of $T_{\rm SC}$ against pressure in UCoGe through $P_{\rm c}$ is not compatible with the expected variation of $\lambda (P \to P_{\rm c})$ or a McMillan-type dependence.
We again discuss the origin of this discrepancy in Sect. 6.


\subsection{Attempt to determine the order parameter }
\begin{figure}[!tbh]
\begin{center}
	\includegraphics[width=0.8\hsize,pagebox=cropbox,clip]{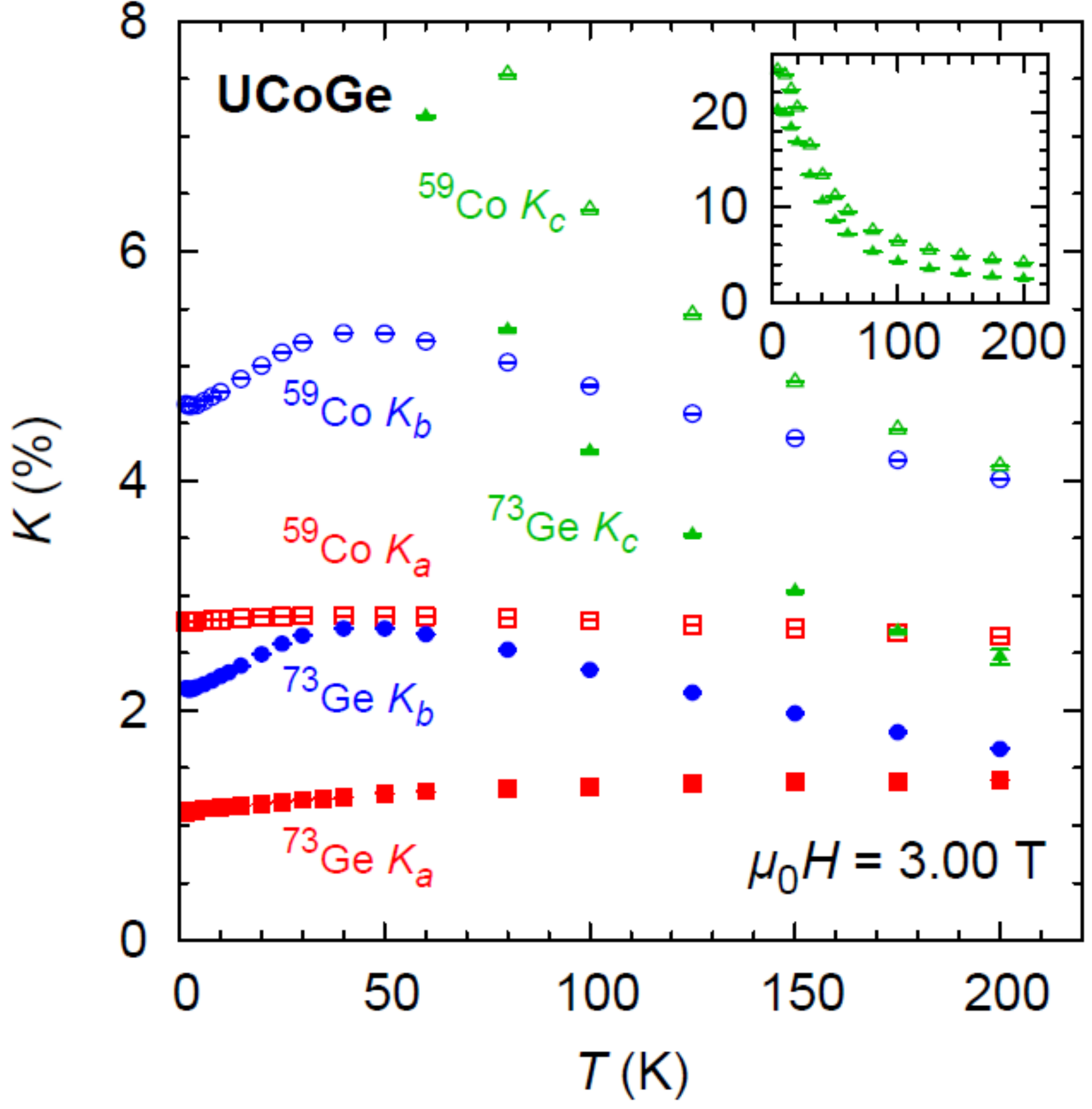}
\end{center}
	\caption{(Color online) %
	$^{73}$Ge (closed symbols) and $^{59}$Co (open symbols)
	Knight shifts measured at a central line ($1/2 \leftrightarrow -1/2$)
	with the field of 3 T
	parallel to the $a$ (squares)-, $b$ (circles)-, and $c$ (triangles)-axes.
	The inset shows the result along the $c$ direction on a different scale.
	}
\label{fig40A}
\end{figure}

One of the reliable methods of determining the order parameter is to measure the measurement of Knight shift in the SC state.   
To estimate the spin susceptibility related to superconductivity, the Knight shift at the Co and Ge sites was measured.  
Figure~\ref{fig40A} shows the $^{59}$Co and $^{73}$Ge Knight shift along three directions in normal-state UCoGe\cite{ManagoPRB2018}.
The Knight shift along the $i$ direction ($i$ = $a$, $b$, and $c$) at the Co and Ge sites is described as\\
\begin{eqnarray*} \label{eq:knightshift}
^{m}K_{i} = {}^{m}A_{i} \chi_{\text{spin},i} + {}^{m}K_{\text{orb},i}      \\
                   \mbox{($m$ = 59 for $^{59}$Co and 73 for $^{73}$Ge)},
\end{eqnarray*}
where $^{m}A_{i}$ is the hyperfine coupling constant, $\chi_{\text{spin},i}$ is the spin susceptibility, and $^{m}K_{\text{orb},i}$ is the orbital part of the Knight shift.
The latter part is usually independent of temperature, and $\chi_{\text{spin}}$ is no longer pure spin in an $f$ electron system because of the strong spin-orbit interaction, but we use the term ``spin susceptibility'' for simplicity.   

When the field is parallel to the $b$- or $c$- axis, the Knight shift at two sites shows the same behavior in a wide temperature range.
This indicates that the dominant temperature dependence of the Knight shift can be attributed to the single component of the spin susceptibility from the U-$5f$ electrons, and that the simple treatment of the Knight shift described above is valid even in a $5f$ electron system since the system has a large spin susceptibility and the temperature dependence of $K_\text{orb}$ is relatively small.
The hyperfine coupling constants of $^{73}$Ge are estimated from the linear relations and are $\sim0.9$ times those at $^{59}$Co, suggesting that the U-$5f$ electrons couple to the $^{59}$Co and $^{73}$Ge nuclei almost equally.
When the field is parallel to the $a$-axis, the temperature dependence of the Knight shift at both sites is relatively small.
This result suggests that the spin susceptibility along the $a$-axis is much smaller than those along the $b$- and $c$- axes since $^{m}A_i$ is considered to be isotropic in this system. 
Note that the magnitude of the $^{59}$Co Knight shift along the $a$-direction in URh$_{0.9}$Co$_{0.1}$Ge at low temperatures [$^{59}K_a \sim3.5~(2.8) $\% in URh$_{0.9}$Co$_{0.1}$Ge (UCoGe)]  is a similar value to that of UCoGe, although the difference in $^{59}K_b$ is huge [$^{59}K_b \sim18~(4.1) $\% in URh$_{0.9}$Co$_{0.1}$Ge (UCoGe)].  
This suggests that the spin susceptibility along the $a$ axis in URhGe is also negligibly smaller than those along the $b$- and $c$-axes, in good agreement with the susceptibility and $M(H)$ data shown in Figs. \ref{fig7} and \ref{fig8}.

\begin{figure}[!tbh]
\begin{center}
\includegraphics[width=\hsize,pagebox=cropbox,clip]{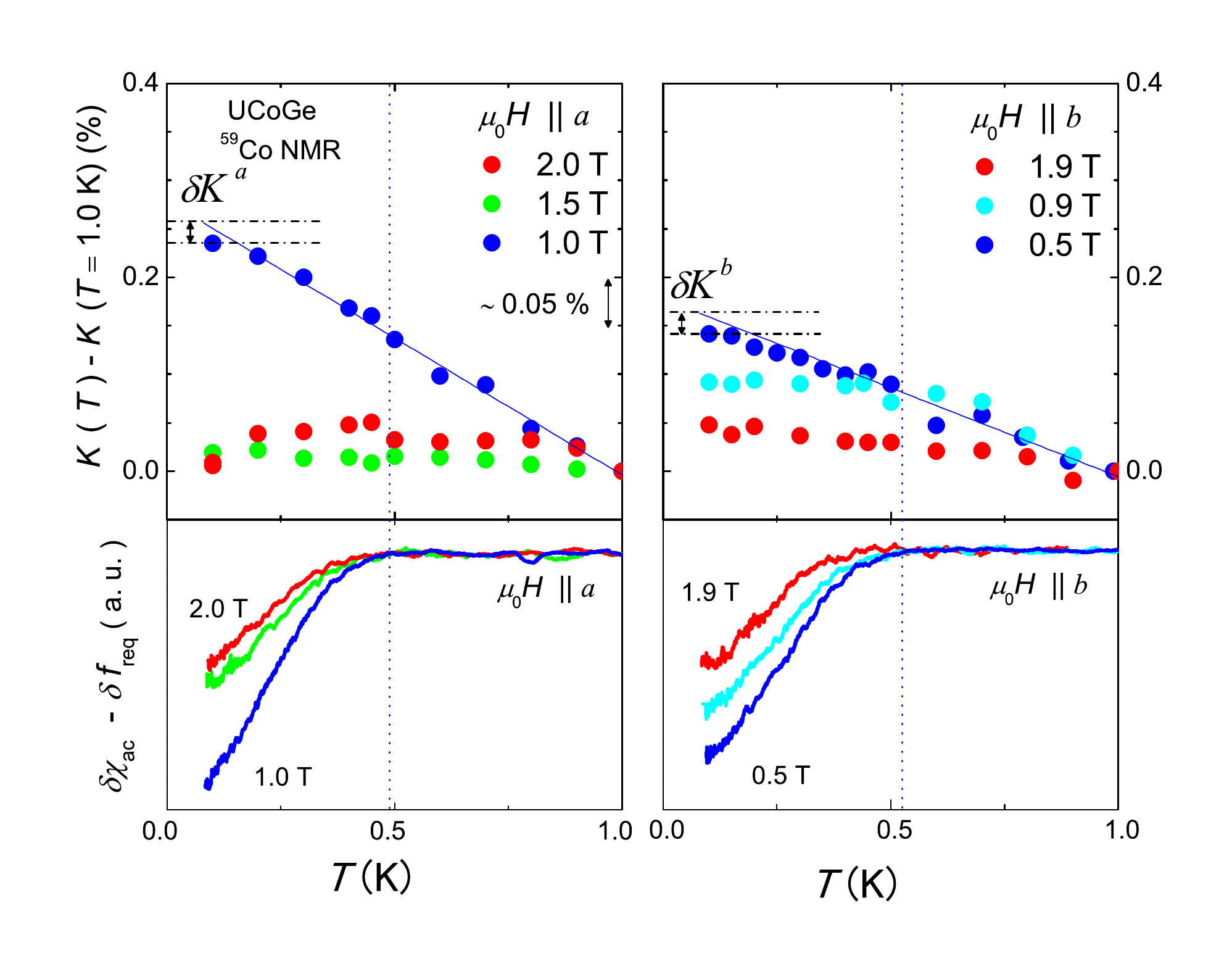}
\end{center}
\caption{(Color online) Temperature dependences of $^{59}$Co NMR Knight shift and Meissner signal for $H \parallel a, b$ below 1 K. The change from the value at 1 K $\left[\Delta K \equiv K - K(1~{\rm K}) \right]$ is shown. The blue lines in the top two figures are the extrapolations of the linear fit between $T_{\rm SC}$ and 1 K. The vertical dotted lines indicate the onset of the superconductivity probed by the measurements of the ac susceptibility $\chi_{\rm ac}$.}
\label{fig40M}
\end{figure}

The Knight shift in the SC state was measured in various fields along the $a$- and $b$-axes. 
Figure \ref{fig40M} shows the temperature dependence of $^{59}K$ and the Meissner signal below 1 K. 
The deviation from $^{59}K$ at $T = 1$ K $[\Delta K \equiv K - K(1~{\rm K})]$ is plotted since, as pointed out above, the NQR measurement shows that the whole region of the same single-crystal sample is in the FM state below 1 K\cite{OhtaJPSJ2010}. 
In the normal state, $\Delta K$ increases with decreasing temperature, following the development of the FM moments. 
At $\mu_0 H = 1$ T for $H \parallel a$ and $\mu_0 H = 0.5$ T for $H \parallel b$, the increase in $^{59}K$ looks small or saturates around the SC transition temperature, below which the diamagnetic signal appears (vertical dotted lines in Fig.~\ref{fig40M}).
The extrapolation of $^{59}K$ is determined from the linear fit of $^{59}K$ from 1 K to $T_{\rm SC}$, which is thought to give the upper limit of $^{59}K$ at $T = 0$ K. 
Therefore, the derivation from the linear extrapolation of $^{59}K$, $\delta K^{a,b}$, which is the maximum value of the suppression of $^{59}K$ due to the occurrence of superconductivity, is estimated to be less than 0.05\%.
The tiny amount or absence of $^{59}K$ suppression below $T_{\rm SC}$ excludes the spin-singlet pairing state, since an appreciable decrease in the Knight shift, which is of the order of $10^{-1} \sim2$\%, is expected in $K_b$ when the spin-singlet pairing is formed.
Actually, a clear decrease in the Knight shift was reported in the U-based superconductor UPd$_2$Al$_3$, which is a spin-singlet superconductor coexisting with the antiferromagnetism.~\cite{Kyogaku93,Kitagawa18}
For a spin-triplet superconductor, the spin component of the Cooper pair is expressed by the SC $\mbox{\boldmath $d$}$ vector, which is defined to be perpendicular to the spin component of the Cooper pair, and the Knight shift decreases below $T_{\rm SC}$ when $H$ is applied parallel to the $\mbox{\boldmath $d$}$ vector fixed along a certain crystal axis.   
However, the situation is not so simple in the present FM superconductors.
It was pointed out that the decrease in the Knight shift will be reduced when there is spontaneous magnetization ($M_c$), which splits the up-spin and down-spin bands significantly~\cite{MineevPRB2010}.
Thus, Knight shift measurements in the SC state without the FM ordering, which is achieved under pressure, are crucial to determine the SC $\mbox{\boldmath $d$}$ vector.
These measurements are now in progress.

\begin{figure}[!tbh]
\begin{center}
\includegraphics[width=0.8\hsize,pagebox=cropbox,clip]{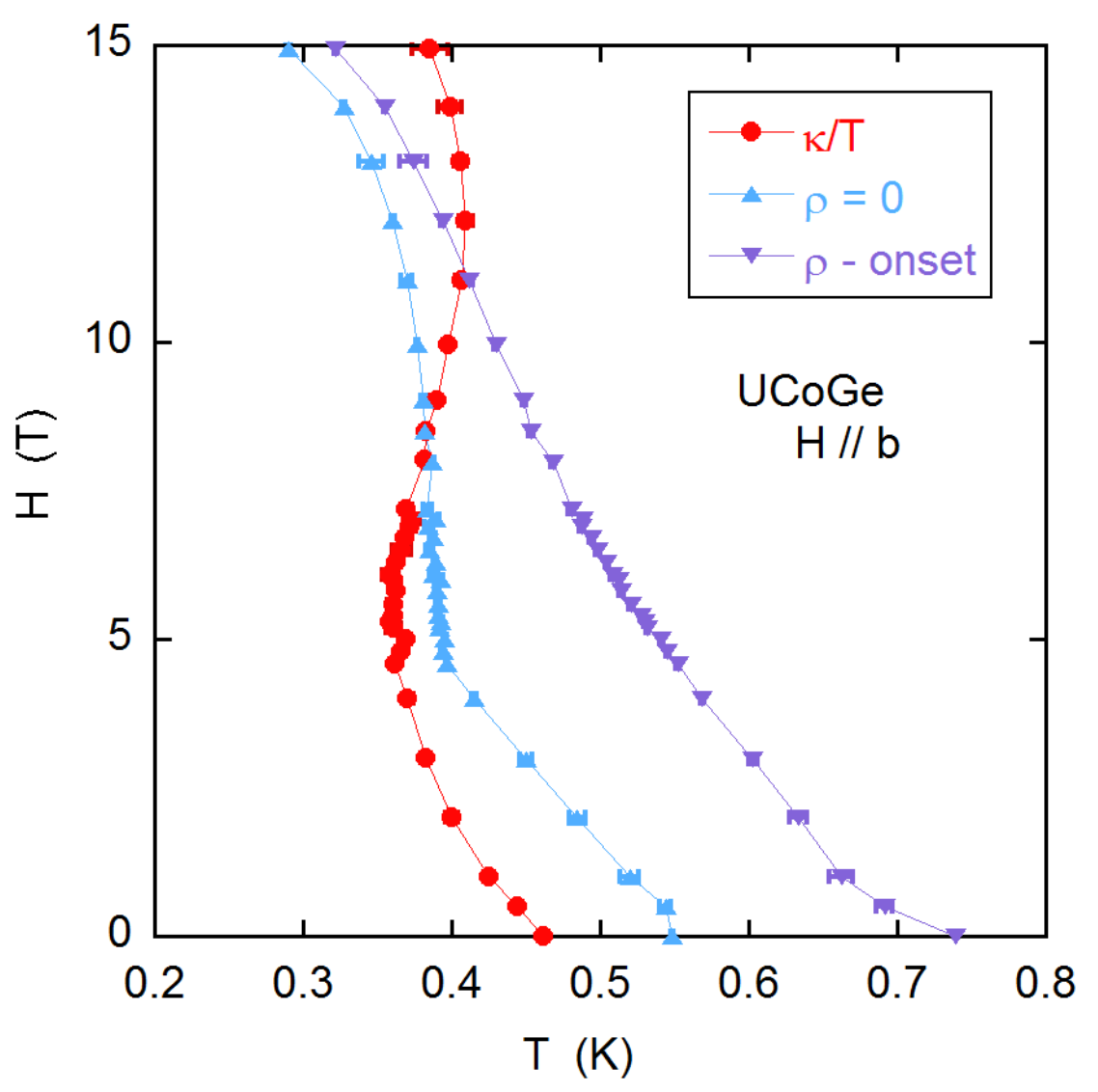}
\end{center}
\caption{(Color online) $H_{\rm c2}$ in $H\parallel b$ scan for UCoGe probed by thermal conductivity ($\kappa$) (red circles) and resistivity (blue triangles for $T_{\rho=0}$ and purple triangles for the onset of SC).~\cite{Wu18}
The results indicate a possible change in the vortex liquid phase for $H >7$ T associated with a rotation of the SC order parameter. Note also that the broadening of the resistive transition decreases by a factor of approximately 3 after the crossing of $\kappa$ and $\rho$ curves. }
\label{fig:UCoGe_Hc2_rho_kappa}
\end{figure}

Thermal conductivity in the normal and SC phases of FM UCoGe~\cite{TaupinPRB2014} was carefully measured in a situation far from very clean material conditions.
In the normal phase, the strong anisotropy of $\kappa$ is in excellent agreement with the Ising nature of the magnetic fluctuations and thus provides extra proof of the strong itinerant character of the magnetism of UCoGe. 
In the SC phase,  rather isotropic behavior of $\kappa /T$ was observed, presumably governed by the dominant isotropic impurity effect. 
More fascinating behavior~\cite{Wu18} was recently reported in the determination of $H_{\rm c2}$ detected by from the resistivity ($\rho$) and thermal conductivity in a transverse field $H_b$ scan.
As shown in Fig.~\ref{fig:UCoGe_Hc2_rho_kappa}, close to the field $H^*_b$ where $T_{\rm Curie}$ collapses, the anomalous difference between $\kappa$ and $\rho$ is associated with a drastic decrease in the SC resistivity broadening detected from a gap.
The concomitant features have led to the proposal of a field-induced vortex liquid phase coupled to a change in the SC order parameter.               

\subsection{Entering the PM regime under pressure}
\begin{figure}[!tbh]
\begin{center}
\includegraphics[width=0.8\hsize,pagebox=cropbox,clip]{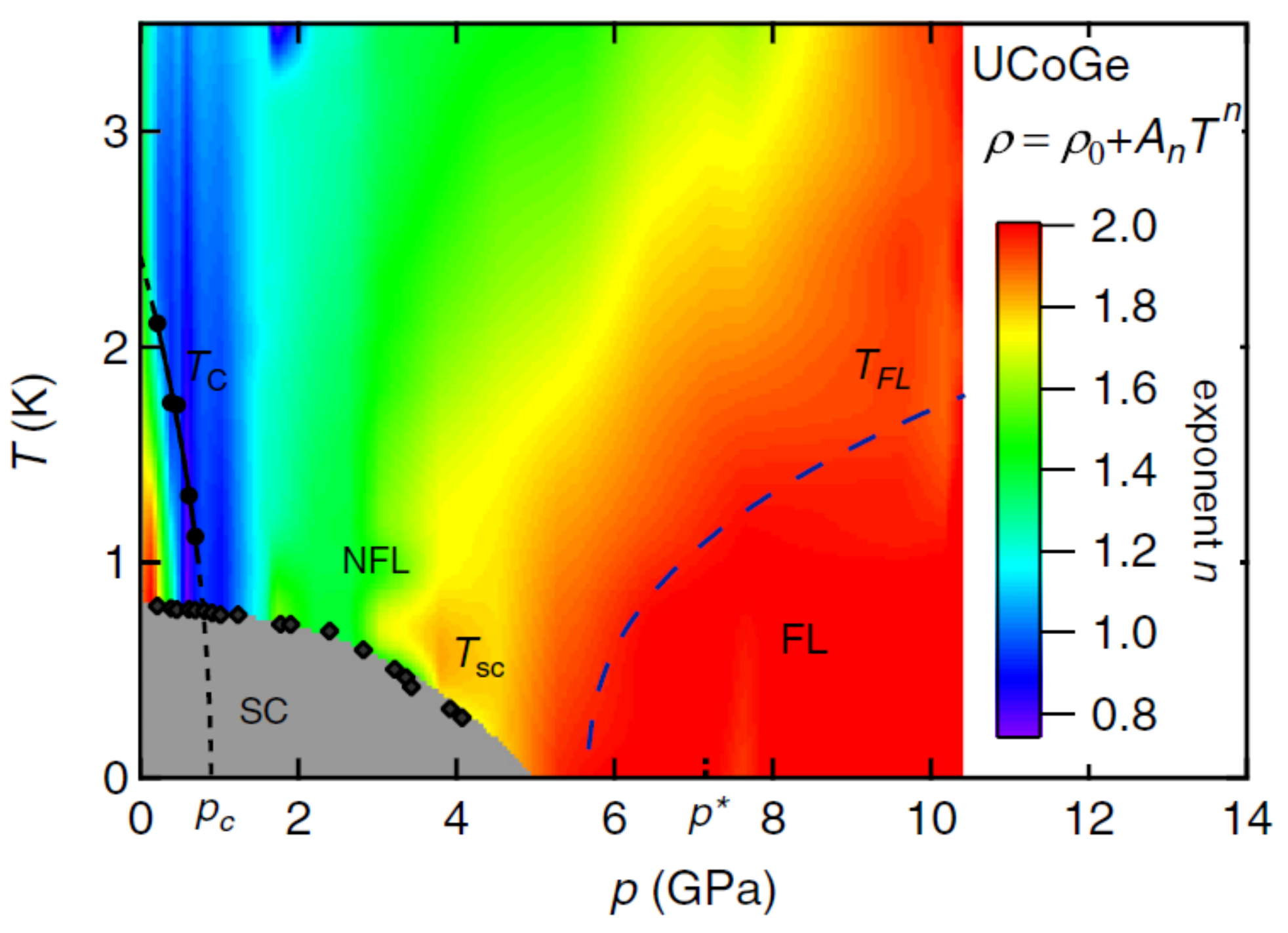}
\end{center}
\caption{(Color online) Contour plot of the resistivity exponent $n$ of UCoGe as a function of $T$ and $P$ in the SC and FM domains. When SC collapses, Fermi liquid behavior ($n \sim2$) is recovered below $T_{\rm FL}$.~\cite{BastienPRB2016,Wu18}}
\label{fig42}
\end{figure}

The application of pressure puts the UCoGe system in the PM region when $T_{\rm Curie} = T_{\rm SC}$\cite{HassingerJPSJ2008, SlootenPRL2009, BastienPRB2016, BastienPhD}. 
SC survives far above $P_c \sim$1 GPa up to $P_s \sim$4 GPa.~\cite{BastienPRB2016} 
In a large $P$ window close to $P_c$ up to $P_s$, the Fermi liquid regime ($AT^2$ in resistivity) is masked by the occurrence of SC (Fig.~\ref{fig42}); the characteristic spin fluctuation energy appears to remain low. 
Analysis of $H_{\rm c2}$ under pressure shows that the strong enhancement of $H_{\rm c2}$ for $H \parallel b$ collapses at $P_c$.
The unusual field dependence of $\lambda$ for $H \parallel c$ decreases slowly with increasing $P$ (Fig.~\ref{fig43}). 
The upward curvature of $H_{\rm c2}$ for $H\parallel c$ survives very close to $P_s$ and far from $P_c$. 
The corresponding variation of $\lambda (H)$ with $p$ for $H \parallel c$ is shown in Fig.~\ref{fig43}.~\cite{BastienPRB2016,BastienPhD}
The variation of $H_{\rm c2}$ through $P_{\rm c}$ does not support the proposal~\cite{Kus13} that in the vicinity of $P_{\rm c}$, a switch from type II SC to type I SC may occur.
The collapse of $M_0$, and thus of $H_{\rm int}$, leads to the disappearance of the SIV state, as observed already in NQR experiment.
Note that no Pauli depairing effect was observed even in the pressure region above $P_c$ along the $b$ axis (Fig.~\ref{fig43}).
This implies that the SC $\mbox{\boldmath $d$}$ vector is perpendicular to the $b$-axis, and thus along the $a$-axis with high possibility, at least when the magnetic field is near $H_{\rm c2}$ along the $b$-axis.
To confirm this and determine the SC $\mbox{\boldmath $d$}$ vector thoroughly, it is important to measure $H_{\rm c2}$ along the $a$-axis above $P_c$ and the Knight shift under pressure.

\begin{fullfigure}[!tbh]
\begin{center}
\includegraphics[width=0.8\hsize,pagebox=cropbox,clip]{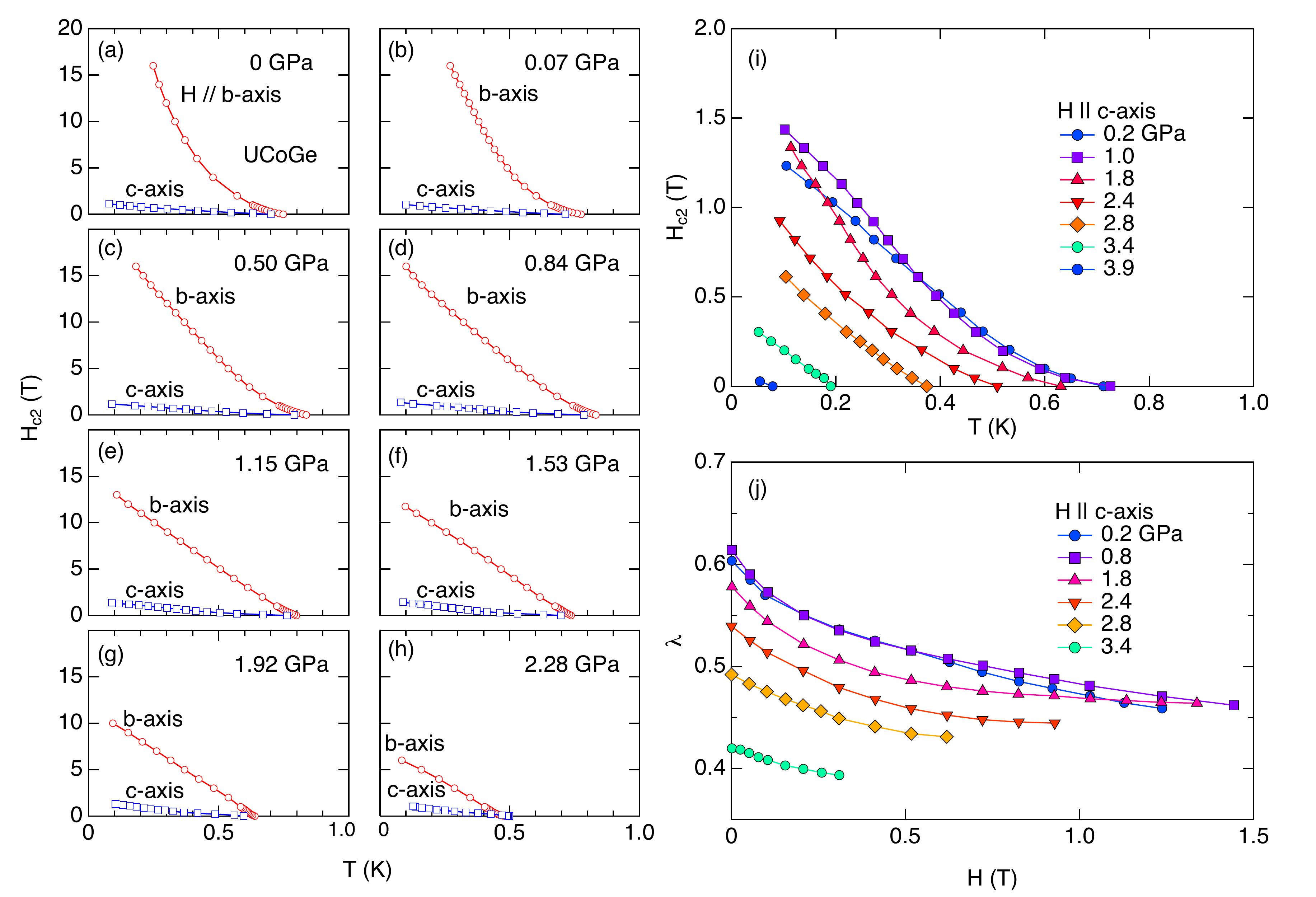}
\end{center}
\caption{(Color online) (a)-(h) $H_{\rm c2}$ of UCoGe for $H \parallel c$ and $H \parallel b$ below $P_{\rm C} \sim$ 1GPa and above $P_{\rm C}$. Note the collapse of the upward curvature of $H_{\rm c2} (H \parallel b)$ above $P_{\rm C}$. 
(i) Enhancement of $H_{\rm c2}$ for $H \parallel c$ up to $3.9\,{\rm GPa}$.
(j) $\lambda(H)$ derived for $H \parallel c$  from $H_{\rm c2}$ calculation at different pressures.~\cite{BastienPRB2016,Wu18}}
\label{fig43}
\end{fullfigure}

\section{Theory}
Progress has been realized in clarifying the link between FM fluctuations, the order of quantum criticality, and SC pairing. 
The difficult issue remaining is the description of the strong-coupling case in UCoGe when $T_{\rm Curie}$ becomes comparable to $T_{\rm SC}$.
Special focus is given to different theoretical approaches to explaining the RSC of URhGe. 
The main approaches are based on the field dependence of the magnetic coupling. 
A recent attempt involved determining how the shift of the electronic sub-bands with $H$ modifies the SC pairing. 
Finally, particular phases may enter the class of topological superconductors.  
\subsection{Quantum criticality: TCP, QCEP}
Different theoretical approaches to studying AFM and FM quantum critical points (QCPs) assuming a second-order transition at $P_c$ ($M_0$ will collapse continuously at $P_c$) can be found in Refs.~\citen{HertzPRB1976, MillisPRB1993, MoriyaJPSJ1995}. 
However, taking into account the nonanalytic term in the free energy, the compressibility, and the interaction with acoustic phonons, a phase transition can switch the QCP to the TCP\cite{MillisPRB1993,LarkinJETP1969}. 
A generalization of the idea given by Larkin and Pikin has been proposed recently,\cite{Cha18} with the conclusion that for a first-order transition, quantum criticality will be restored by zero-point fluctuations.

\begin{figure}[!tbh]
\begin{center}
\includegraphics[width=0.8\hsize,pagebox=cropbox,clip]{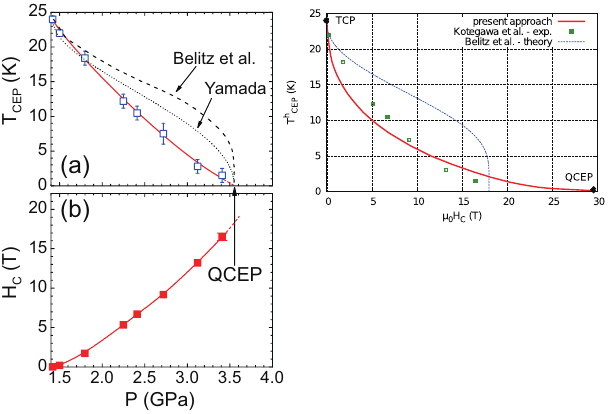}
\end{center}
\caption{(Color online) Measured and theoretically calculated $T_{\rm CEP}$ in UGe$_2$.~\cite{KotegawaJPSJ2011,WysokinskiPRB2015,BelitzPRL2005,Kad18}.}
\label{fig45}
\end{figure}

It was shown for UGe$_2$ that under pressure\cite{MineevCRP2011}, a TCP exists, in agreement with the experimental results. 
This is also the case for URhGe as $H$ approaches $H_{\rm R}$ in an $H \parallel b$ scan, and it has already occurred for UCoGe at $P = 0$. 
At least, the theory for a clean three-dimensional itinerant ferromagnet predicts $P_{\rm TCP} < P_c$\cite{MillisPRL2002}.
The possibility of an intermediate state below the TCP~\cite{Chu09,Ped13} has been stressed if a drastic FS reconstruction drives a change in the nature of the interaction itself.  
Figure~\ref{fig45} shows the data with the prediction~\cite{BelitzPRL2005} based on long-wavelength correlation effects.
Poor agreement with the predictions~\cite{MillisPRL2002} has been previously obtained. 
Recently, a microscopic approach to UGe$_2$\cite{WysokinskiPRB2015} led to the main features of the CEP, $H_x$, $H_c$, TCP and QCEP being obtained even a good evaluation of the QCEP.~\cite{WysokinskiPRB2015}. 
A key ingredient to obtain two FM2 and FM1 phases is to start with a two-band mode; quantum critical fluctuations are coupled to the instability of the FS topology.  
Changes in the topology of the FS via the Lifshifz transition were considered first in Ref.~\citen{SandemanPRL2003}.   
   
The feedback between the localized and itinerant duality of $4f$ or $5f$ electrons leads to spin dynamics different from those in  $3d$ itinerant systems.~\cite{HuxleyPRL2003,StockPRL2011}
It was demonstrated that the residual damping detected in an inelastic neutron scattering experiment on UGe$_2$ is a mark of the duality.~\cite{Chu14,Min13}

\subsection{Superconductivity}
\begin{figure}[!tbh]
\begin{center}
\includegraphics[width=\hsize,pagebox=cropbox,clip]{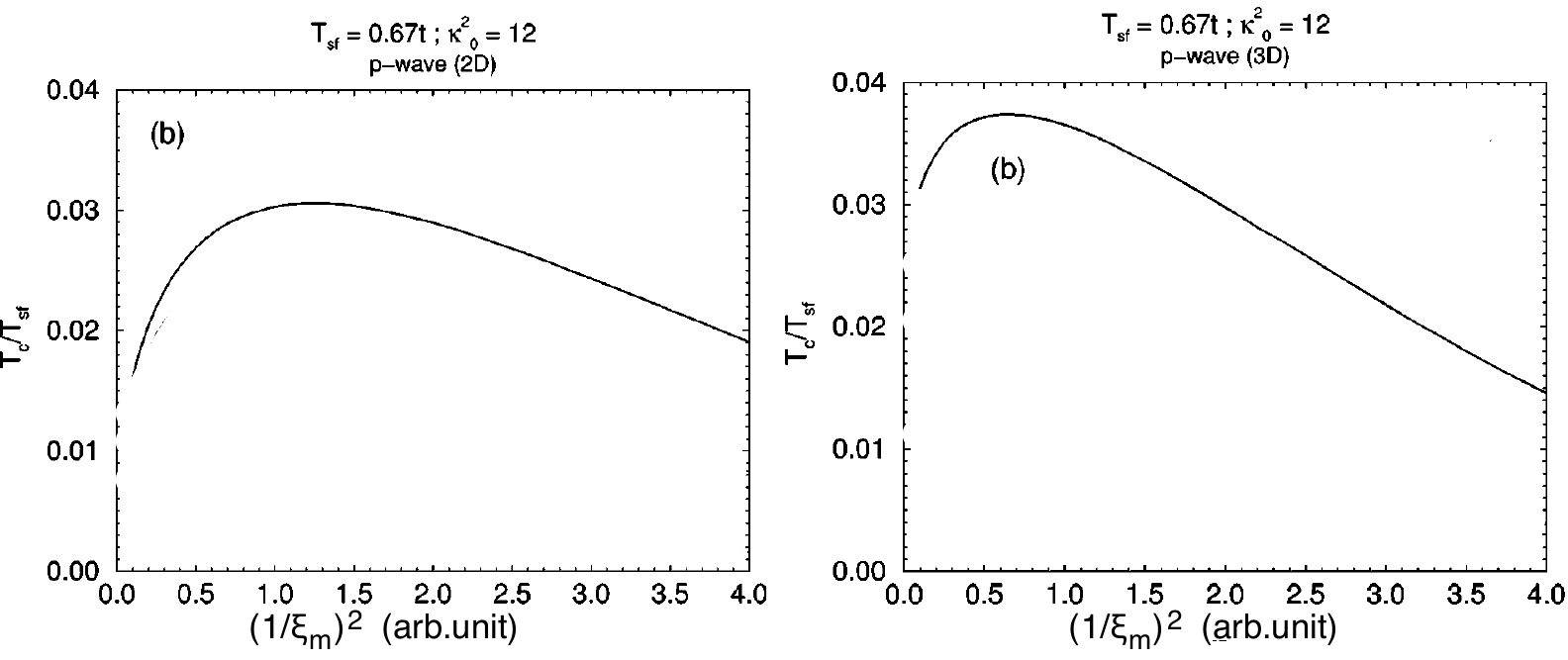}
\end{center}
\caption{Variation of $T_{\rm SC}/T_{\rm sf}$ as a function of $\xi_m$ for two and three dimensional models for $p$-wave pairing provided by FM fluctuation ($T_{\rm sf}$: spin-fluctuation temperature) }
\label{fig46}
\end{figure}

In a magnetic medium, an attractive pairing between electrons at sites $\mbox{\boldmath$r$}$ and $\mbox{\boldmath$r'$}$\cite{RoussevPRB2001} can be mediated by FM interactions.
In the case of a triplet state, the interaction is
\begin{equation*} 
V(\mbox{\boldmath$r$}-\mbox{\boldmath$r'$}) = -C \sum_{\alpha, \beta} \mbox{\boldmath$S$}_{j, \alpha} \chi_{\alpha, \beta}(r - r') \mbox{\boldmath$S$}_{i, \beta},
\end{equation*}
where $\chi_{\alpha, \beta}$ is the static susceptibility, and $\mbox{\boldmath$S$}_i$ and $\mbox{\boldmath$S$}_j$ are the spins at sites $\mbox{\boldmath$r$}$ and $\mbox{\boldmath$r'$}$, respectively~\cite{deGennes}. 
The coexistence of $p$-state superconductivity with itinerant ferromagnetism\cite{FayPRB1980} was discussed in the framework of a Hubbard-type exchange interaction $I$ corresponding to a Stoner enhancement factor of $S = (1 - I)^{-1}$; a maximum of $T_{\rm SC}$ was found on both sides of $I = 1$ (at $P_c$), but $T_{\rm SC}$ decreases to zero at $P_c$. 
It was pointed out that this collapse is an artificial consequence of the approximation; $T_{\rm SC}(P_c)$ is determined by low- but fixed-energy spin fluctuations.~\cite{Shen03}
Breakdown of the coexistence of singlet superconductivity and itinerant ferromagnetism for the same electrons was emphasized in Ref.~\citen{Shen03}.
The possible existence of triplet superconductivity in an almost localized Fermi liquid frame was pointed out in Ref.~\citen{Val84}. 
The possibility of maintaining a high $C/T$ term below $T_{\rm SC}$ was stressed in Ref.~\citen{Kar01}.
Pairing gaps near an FM QCP were discussed in Ref.~\citen{EinenkelPRB2015} for two-dimensional itinerant FM systems: a superconducting quasi-long-range order is possible according to an Ising-like hypothesis but will be destroyed in Heisenberg ferromagnets (the case of ZrZn$_2$ is now accepted to be non superconductivity~\cite{Yel05}).
To justify that superconductivity in UGe$_2$ occurs only in the FM phase, it was proposed that a coupling with FM magnons will boost $T_{\rm SC}$ in the FM state;~\cite{Kirkpatrick2001PRL,Kir03,Kar03} however, experiments failed to detect any magnons.
Magnetically mediated superconductivity with AFM and FM coupling near their magnetic instability was calculated in the quasi-two and three dimensional models\cite{MonthouxPRB2001}. 
The results are shown in Fig.~\ref{fig46}, where $\xi_m$ is the magnetic coherence length, which diverges at $P_c$. 
For large $\xi_m$, we are very far from the McMillan formula used in the previous analysis of the experimental results.
However, the collapse of FM through a first-order transition leads to a finite value of $\xi_m$ at $P_c$, and thus a good description of $T_{\rm SC}$ is recovered with the McMillan formula $T_{\rm SC} \propto \exp{(-1/\lambda)}$ when $\xi_m$ approaches a distance of a few atomic distances.

Just after the discovery of SC in UGe$_2$, a model where localized spins are the source of singlet pairing for the quasiparticles was presented.~\cite{Suh01,Abr01}
However, two decades of experiments rule out this possibility.

After the discovery of superconductivity in UGe$_2$ and URhGe, the symmetry of the triplet SC states in these orthorhombic materials was proposed. 
If the $S_z$ = 0 component is negligible (equal spin pairing), the two superconducting states are the A state with a point node along the $z$ easy magnetization $M_0$ axis and the B state with a line node in the ($x, y$) plane.~\cite{MineevPUsr2017}
Experimentally, there is no convincing proof of the A and B states.
One of the difficulties is the already discovered SIV state.
In addition, the purity of crystals may be too low to avoid dominant impurity effects at very low temperatures.

The main aspects of FM superconductivity are well described in the framework of BCS weak-coupling theory~\cite{MineevPUsr2017} with the pairing interaction expressed through the static magnetic susceptibility of an FM medium with orthorhombic anisotropy.
Such an interaction in anisotropic ferromagnet was discussed in Ref.~\citen{Min13}.
The key results are:
\begin{enumerate}
\item the magnetic field dependence of the pairing interaction\cite{MineevPRB2011}, 
\item the prediction that $T_{\rm Curie}$ decreases as $H^2$ in $H$ transverse scan~\cite{MineevPRB2014}, 
\item RSC near TCP with $T_{\rm SC}(H_{\rm R}) \sim (1/2) T_{\rm TCP}$\cite{MineevPRB2015}, 
\item the drop of TCP with $\sigma$ associated with that of $H_{\rm R}$\cite{MineevPRB2017_96}. 
\end{enumerate}
The proposal of a change of the order parameter around $P_c$ (see Ref.~\citen{MineevPUsr2017}) is actually very difficult to verify experimentally.

\subsection{Special focus on $H_{\rm c2}$ and $H$ dependence of the pairing strength}
We will present the different models developed to explain the unusual shape of $H_{\rm c2}$.
The first three proposal deal with the magnetism: the SC pairing is given by spin interactions.
The last one shows the effect of FS reconstruction on the pairing.

As indicated in the sections on URhGe and UCoGe, comparisons have been made between the $\lambda$ values derived from $H_{\rm c2}$ analysis~\cite{WuNatComm2017,WuThesis} and those predicted theoretically in the previous Landau approach~\cite{MineevPRB2011}. 
The calculated variations of $\lambda(H_c)$ and $\lambda(H_b)$ are 
\[
\lambda(H_c) = \lambda(0) \frac{\left[1+(\xi_{\rm mag} k_{\rm F})^2\right]^2}{\left[\frac{3M_z^2}{2M_0^2}-\frac{1}{2}+(\xi_{\rm mag} k_{\rm F})^2\right]^2}
\]
and
\[
\lambda(H_b) = \lambda(0) \frac{\left[1+(\xi_{\rm mag} k_{\rm F})^2\right]^2}{\left[\frac{T_{\rm SC}-T_{\rm Curie}(H)}{T_{\rm SC}-T_{\rm Curie}(0)}+(\xi_{\rm mag} k_{\rm F})^2\right]^2},
\]
respectively, where $\xi_{\rm mag}$ is the magnetic coherence length and $k_{\rm F}$ is the Fermi wave vector.
When the $5f$ itinerant character dominates, $\xi_{mag} k_F > 1$. 
Excellent agreement is found between the field dependence of the magnetization (see Fig.~\ref{fig38}) at low temperatures and $\lambda(H \parallel c)$ determined by comparison with the experiments.
The respective values of $\xi_{\rm mag} k_{\rm F}$ for URhGe and UCoGe are linked to the size of $M_0$.
For $H \perp M_0$ in UCoGe, the derivation of $\lambda$ (Fig.~\ref{fig39}) does not agree with the data derived in the normal phase from NMR and transport measurements (see Sect. 5).~\cite{WuNatComm2017,WuThesis} 
When $T_{\rm Curie} \leq T_{\rm SC}$, a strong-coupling approach is necessary and, a McMillan representation of $T_{\rm SC}$ is certainly not correct.
The agreement between experiments and theory for $H\parallel c$ indicates that in longitudinal $H$ scan, no drastic change in the coupling conditions occurs, that is, the perturbation is weak.

\begin{figure}[!tbh]
\begin{center}
\includegraphics[width=0.8\hsize,pagebox=cropbox,clip]{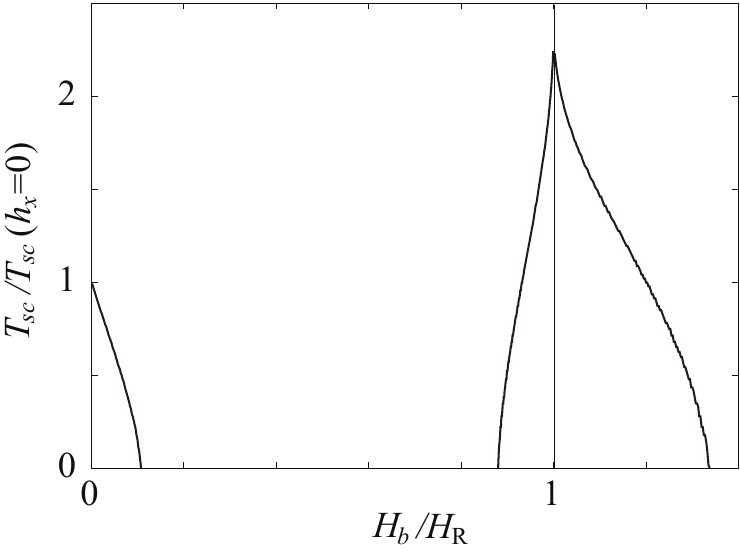}
\end{center}
\caption{Appearance of two SC domains in the duality between $c$ and $b$ easy magnetization axes described in the framework of soft magnons, which generate a strong attractive interaction on approaching $H_{\rm R}$~\cite{HattoriPRB2013URhGe}. However, inconsistency is observed in the field region above $H_{\rm R}$.}
\label{fig47}
\end{figure}

The RSC in URhGe was explained by a microscopic model,~\cite{HattoriPRB2013URhGe} where in a transverse magnetic field, soft magnons generate a strong attractive interaction on approaching $H_{\rm R}$ with the change in the spin components of the Cooper pair\cite{HattoriPRB2013URhGe}. 
The FM XXZ model describes the coupling of localized moments, and the interaction of magnons and conduction electrons is mediated via anisotropic AFM coupling.
The lowest-order fluctuations yield a magnon quasiparticle interaction. 
In agreement with the experiments, two domes are found for the SC domain.
Figure~\ref{fig47} shows the prediction of $T_{\rm SC}$ with the pairing interaction expressed through the static susceptibility of the FM medium with the orthorhombic anisotropy
(see Chapter IV in Ref.~\citen{deGennes} for isotropic ferromagnets).
As shown in Fig.~\ref{fig31}(a), the RSC domain above $H_{\rm R}$ is quite narrow in experiments. 
This is due to the suppression of FM fluctuations by $H_b$.
This effect is not taken into account in the model.

\begin{figure}[!tbh]
\begin{center}
\includegraphics[width=0.8\hsize,pagebox=cropbox,clip]{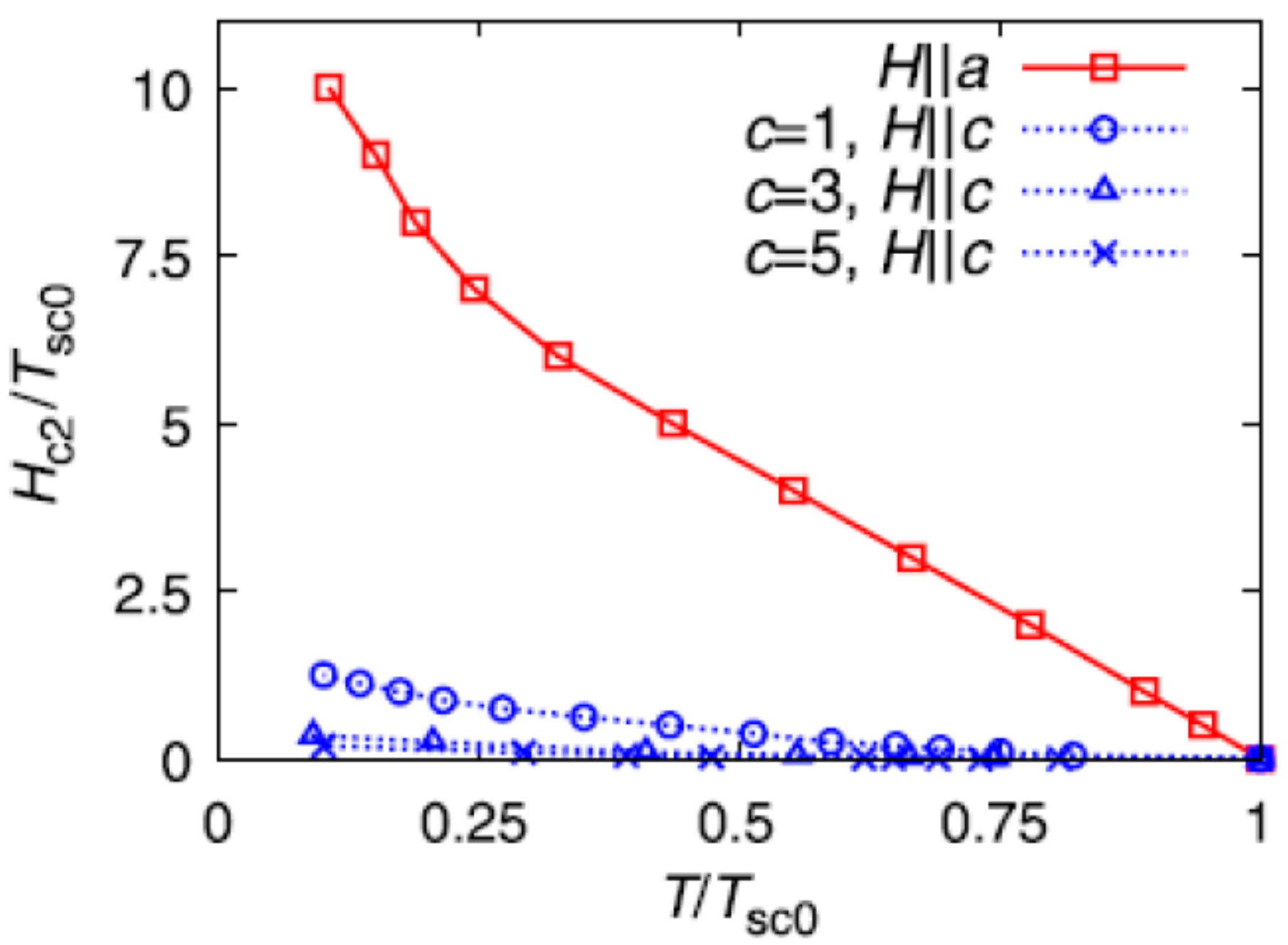}
\end{center}
\caption{(Color online) Calculation of $H_{\rm c2}$ in UCoGe for $H \parallel a$ and $H \parallel c$ using NMR results.~\cite{TadaJPCS2013}}
\label{fig48}
\end{figure}

On the basis of mainly on NMR results, the superconductivity in UCoGe was analyzed in a model calculation when longitudinal FM fluctuations induce the spin-triplet pairing\cite{HattoriPRL2012,TadaJPCS2013}.
The Ising FM fluctuations are described by the susceptibility
\begin{equation*}
\chi_z=\left(\delta(H_c) +q^2+\frac{\omega}{\gamma_q}\right)^{-1}, 
\end{equation*}
where $\gamma$ is the Fermi velocity and $\delta(H_c)$ indicates the fluctuations taken at this time, given as $1+c\sqrt{H_c/H}$, from the NMR data\cite{HattoriPRL2012}. 
Good agreement was found for the A state, as shown in Fig. \ref{fig48}.
Experimentally, we have already pointed out that the $\sqrt{H_c}$ dependence is not obeyed at low temperatures.

\begin{figure}[!tbh]
\begin{center}
\includegraphics[width=0.8\hsize,pagebox=cropbox,clip]{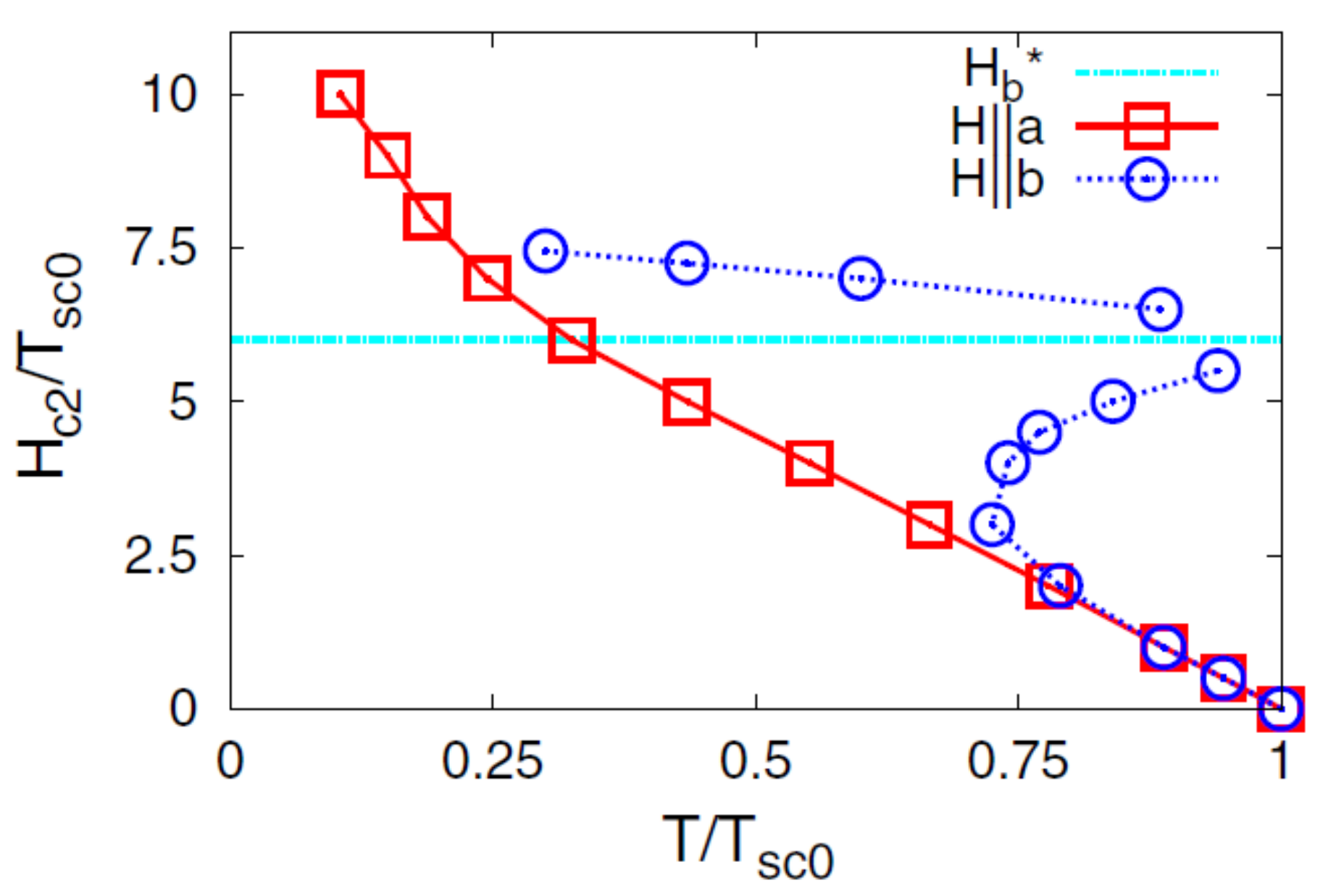}
\end{center}
\caption{(Color online) Calculation of $H_{\rm c2}$ in UCoGe for $H \parallel b$ and $H \parallel a$,  taking into account the Dyaloshinski--Moriya interaction created by the zigzag chain structure.~\cite{TadaPRB2016}}
\label{fig:UCoGe_Hc2_calc}
\end{figure}

Magnetism and superconductivity in UCoGe were recently revisited with aim of explaining the $a$-$b$ anisotropy of $H_{\rm c2}$ with special focus on the Dyaloshinskii-Moriya interaction produced by the zigzag chain structure of UCoGe\cite{TadaPRB2016}. 
In agreement with the phenomenological model, the unusual $S$ shape of $H_{\rm c2}$ is linked with the enhancement of FM fluctuations owing to the decrease in $T_{\rm Curie}$. 
Rotation of the $\mbox{\boldmath $d$}$ vector of the order parameter is a noteworthy feature. 
Figure~\ref{fig:UCoGe_Hc2_calc} shows the $H_{\rm c2}$ predictions for $H \parallel a$ and $H \parallel b$.    

\begin{figure}[!tbh]
\begin{center}
\includegraphics[width=0.8\hsize,pagebox=cropbox,clip]{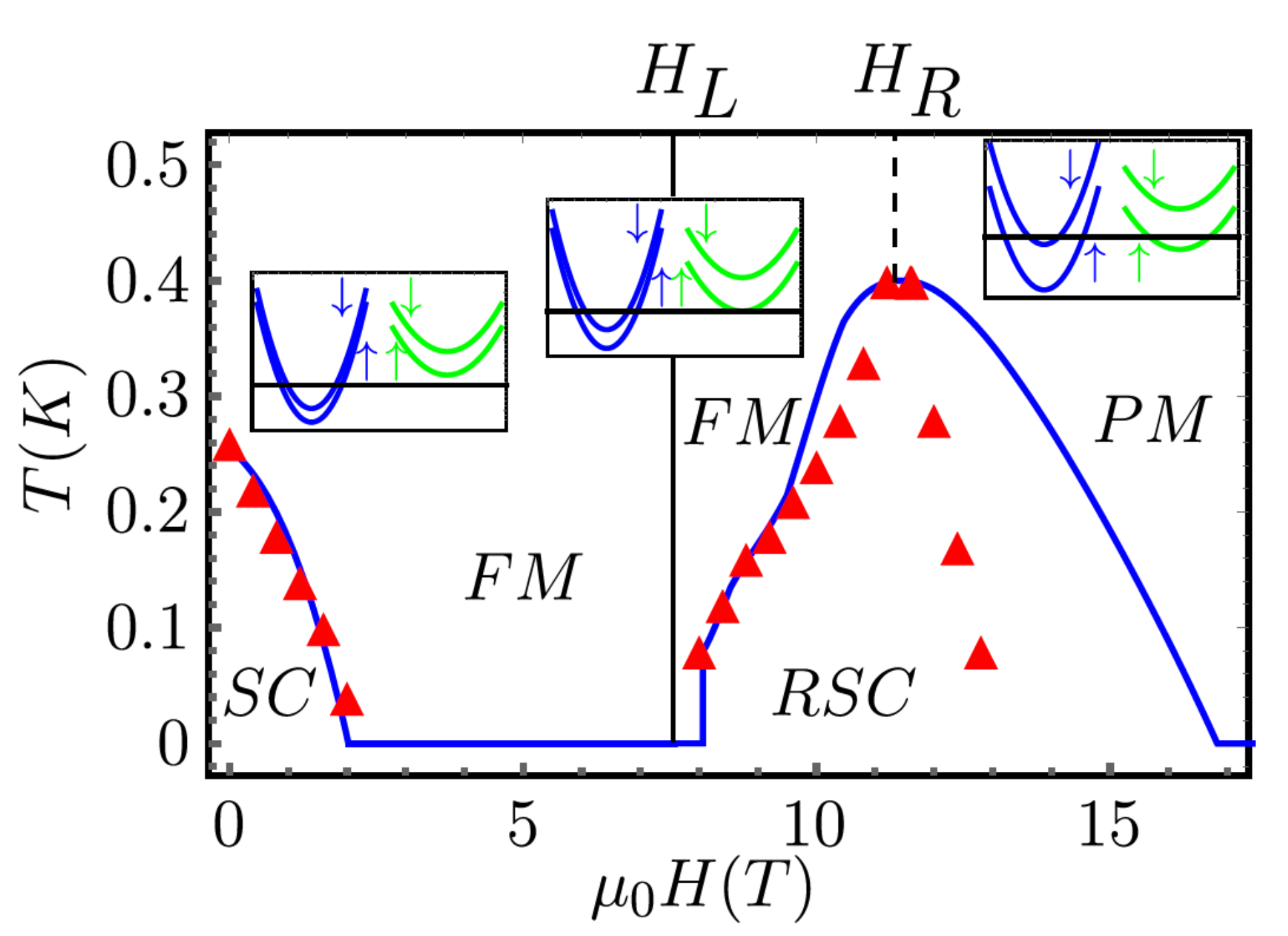}
\end{center}
\caption{(Color online) Phase diagram of URhGe for $H \parallel b$. The experimental data are shown by red triangles. The solid line is the fitting by the model. The inset shows the schematic description of two electron bands separated by $K_0$, which are shifted up increasing the Zeeman energy.~\cite{She18}}
\label{fig58}
\end{figure}

To see the effect of FS instability on phenomena such as RSC, the first step is to develop a two-band model with a field-induced instability for RSC in URhGe.
Quite recently, such an approach was made to derive the effect of a Lifshitz transition on SC.~\cite{She18}
Attention was given to the $H$-induced Lifshitz transition inside a minimum frame of two bands with unequal dispersions and band minima.
The band shifts with the field, as shown in Fig.~\ref{fig58}.
A Lifshitz transition occurs at $H_{\rm L}$ when the spin-up branch of band 2 reaches the chemical potential; the relative motions of bands 1 and 2 lead to the maximum of $\chi_0$ and $m^*$ being located at $H_{\rm R} = 1.5 H_{\rm L}$. 
Eliashberg treatment for the superconductivity shows that the $H$ enhancement of $m^*_H$ drives the increase in $T_{\rm SC}(m^*_H)$ without an orbital effect; upon adding pair-orbital breaking at $H_{\rm c2}$, an RSC domain appears as shown in Fig.~\ref{fig58} 
(see Fig.~\ref{fig30} in Sect.~\ref{sec:URhGe}).
This inconsistency was seen in the field range above $H_{\rm R}$, as in the case shown in Fig.~\ref{fig47}

From comparison with the experiments, the FS topology should be known.
A few quantum oscillation experiments have been reported for $H \parallel b$.
In the first one, a misalignment of $\theta \sim12^{\circ}$ allows us to escape from RSC;~\cite{YellandNatPhys2011} a frequency of 600 T was observed, which collapses exactly at $H_{\rm R}$.
This disappearance was the first evidence of a possible Lifshitz transition.
It was hypothesized that it leads to reduction of the SC coherence length as the Fermi velocity vanishes, and thus to the recovery of SC; no proof was given that this selected orbit plays a key role in the SC pairing.
In a rigid frame, the contribution of this orbit is only 2\% of the total FS volume.
However, in this highly correlated multiband system, the reconstruction of one subband reacts to that of the other. 
In a second experiment with the crystal perfectly aligned along the $b$ axis, one frequency was detected at around 200 T below $H_{\rm R}$ and two frequencies were detected, at 600 and 1200 T above $H_{\rm R}$.
A detailed experimental detection of the FS below and above $H_{\rm R}$ is missing.
TEP measurements (see Fig.~\ref{fig28}) indicate that a Lifshitz transition occurs in a sharp window through $H_{\rm R}$ (not in a large domain). 
Specific heat measurements show that the increase in $m^*$ starts near $H_{\rm R}/1.5$.~\cite{Har11}
The interplay between FS and FM instability is obviously stronger than that described in the model.
Clearly, this frame is valid in the sense that for a given magnetic polarization (reaching a critical value of the magnetization), a Lifshitz transition drives the enhancement of $m^*$, in good agreement with the scaling observed in pressure and uniaxial experiments (see Sect..~\ref{sec:URhGe_pressure_uniaxial}).

\subsection{Possible topological cases}
The symmetrical and topological properties of FM superconductors were stressed in Refs.~\citen{Che16,Nom17,Kob16} for the SC-PM phase of UCoGe with the remark that the superconductivity here will be the electronic analog of the B phase of $^3$He. 
It was emphasized that UCoGe is an excellent candidate for unconventional SC with hidden protected line nodes.~\cite{Nom17}
It was proposed that the SC-PM phase is a promising candidate for Z$_4$ nontrivial topological nonsymmorphic crystalline SC.~\cite{Dai18}
A comparison with $^3$He was also made in Ref.~\citen{HuxleyPhysicaC2015}.

\section{Fermi Surface and Band Structure}
An open question is why FM triplet superconductivity has so far been restricted to U compounds. 
A key point of interest is whether future experiments and band-structure calculation will converge towards a clear view of the interplay between FS topology and SC pairing.

There are similarities among Ce, Yb, and U intermetallic compounds. 
However, major differences exist in the Coulomb repulsion ($U$), spin orbit ($H_{\rm SO}$), crystal field (CF), and hybridization directly linked to the localization of the $f$ electrons.~\cite{Fournier85}
Typical values for these parameters for a lanthanide (La) and actinide (Ac) are listed in Table~\ref{table2} with the unit of eV.
$H_{\rm ex}$ is the order of magnitude of exchange coupling. 
In the case of $U$, as the hybridization is strong like Ce and Yb,  Kondo coupling pushes the system close to magnetic instability.
Up to now, no SC has been found in FM Ce or Yb compounds.
The specificity of U compounds as well as the role of the zigzag configuration in these three compounds must to be clarified.
From the rather high values of $T_{\rm SC}$ of the actinide heavy-fermion compounds [PuCoGa$_5$ ($T_{\rm SC}= 18.5\,{\rm K}$)~\cite{Sar02} and NpPd$_5$Al$_2$ ($T_{\rm SC}= 4.9\,{\rm K}$)~\cite{Aok07_NpPd5Al2}],
the specificity of the U band structure may favor the observation of SC at a moderate temperature.   

\begin{table}[!tbh]
\caption{Typical values for Coulomb repulsion ($U$), spin orbit ($H_{\rm SO}$), crystal as above field (CF), and exchange coupling ($H_{\rm ex}$) for lanthanide (La) and actinide (Ac).
The unit of these energies is eV.}
\begin{tabular}{ccc} \hline
             & La                   &  Ac  \\ \hline 
$U$           & 20                   &  10                       \\
$H_{\rm SO}$   & 0.1                   & 0.3                  \\
CF           &  $\leq 0.01$                   & 0.01--0.1           \\
$H_{\rm ex}$  & 0.001--0.01         &   0.01                 \\  \hline
\end{tabular}
\label{table2}
\end{table}

\begin{figure}[!tbh]
\begin{center}
\includegraphics[width=0.5\hsize,pagebox=cropbox,clip]{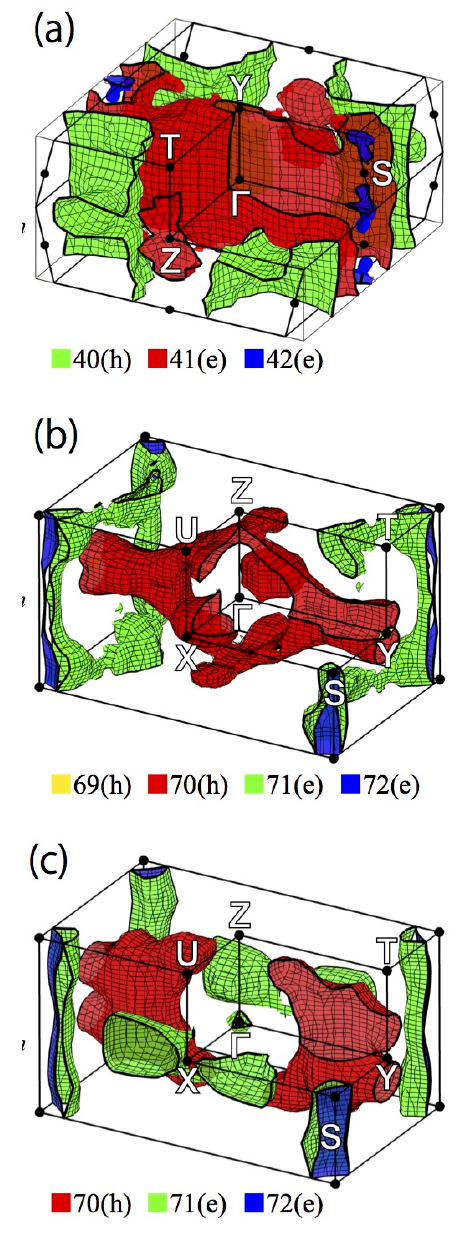}
\end{center}
\caption{(Color online) Fermi surfaces (FSs) of (a) UGe$_2$, (b) URhGe, and (c) UCoGe in their PM phase according to Ref.~\citen{FujimoriJPSJ2016}}
\label{fig50}
\end{figure}

Up to now, the main features of the electronic structure have been revealed by soft X-ray photoelectron spectroscopy\cite{FujimoriJPSJ2016}. 
Despite the debate between the localized and itinerant description of the 5$f$ electrons\cite{TrocPRB2012}, the itinerant treatment of the 5$f$-U state is required to analyze the experiments focusing at the Fermi level. 
Similarities and differences between ARPES experiments and LDA band calculations are discussed in Ref.~\citen{FujimoriJPSJ2016}.
Because of the low symmetry of the orthorhombic crystals and the different sites of the U atoms in the lattice, the calculation of the FS topology is notoriously difficult. 
Figure~\ref{fig50} shows the FS topology obtained with relativistic linear augmented plane waves within the LDA. 
UGe$_2$ is a model with a large carrier concentration, while URhGe and UCoGe can be considered semimetals. 

As already pointed out, the FS of UGe$_2$ has been investigated via elegant quantum oscillation experiments, notably for the FM2 and PM phases. 
The agreement with band structure calculations remains poor\cite{ShickPRL2001, MoralesPRB2016}. 
Measurements of the magnetic form factor\cite{KuwaharaPhysicaB2002, HuxleyJPCM2003} conclude that the ratio $m_{\rm L} /m_{\rm S}$ of the orbital moments to the spin moments increases on switching from FM2 to FM1, whereas the theory predicts a decrease.
At least in UGe$_2$, a sufficient numbers of frequencies of the orbits on the FS have been determined to improve the calculations.

The band structure in the FM phase of URhGe was first determined by the local spin density approximation (LSDA)\cite{ShickPRB2002} with the proposal of an additional AFM component.
However, neutron scattering experiments ruled out this possibility\cite{LevyScience2005}.
ARPES measurements above and below $T_{\rm Curie}$ suggest a small change in the band structure between the PM and FM phases\cite{FujimoriPRB2014}. 
Clearly, experimental progress must be made in this area to give a sound basis for future calculations.

In the case of UCoGe, the first study was on combining density functional theory (DFT) and the Kohn--Sham equation with fully relativistic self-consistent resolution, \cite{DivisPhysicaB2008} but the proposal of a large moment carried by the Co sites disagrees with the NMR results\cite{KarubeJPSJ2011} and with other measurements\cite{VisserPRL2009,TaupinPRB2014}.
A similar DFT approach\cite{MoraJPCS2009} gives an FM moment of 1.35 $\mu_{\rm B}$/U, far above the experimental value of 0.06 $\mu_{\rm B}$/U.
Another calculation was done by resolving the Kohn--Sham--Dirac equation\cite{CzekalaJPCM2010}.
The predicted FM moment (0.47 $\mu_{\rm B}$/U ) is basically one order of magnitude higher than the measured one; an interesting feature is the large difference between FM and PM FSs.

An interesting feature in UCoGe is the cascade of field-induced Lifshitz transitions for $H \parallel c$ much higher than $H_{\rm c2}$.~\cite{Bas16}
For $H \parallel b$ at $\mu_0 H_b^\ast \sim12\,{\rm T}$, preliminary TEP measurements also suggest also a Lifshitz transition in this configuration\cite{MalonePRB2012}; 
confirmation is required using a high-quality crystal.
Note that dichroism measurement confirms that in UCoGe\cite{TaupinPRB2014}, in agreement with ARPES, the U 5$f$ electron count is close to 3; the Co 3$d$ moment induced by the U moment is only 0.1 $\mu_{\rm B}$ at 17 T, 
In URhGe, no evidence has been found for a change in orbital and spin components across $H_{\rm R}$\cite{WilhelmPRB2017}. 

\section{Conclusion}
In the U compounds discussed in this review, equal spin-triplet pairing has been established via a large variety of experiments. 
The important difference between the three materials is the magnitude of ordered moments, leading to different $(T,P)$ phase diagrams.
All materials show strong Ising anisotropy and a switch from a second- to first-order transition at the TCP.
In spite of the first order transition, the relatively low temperature of the TCP preserves the FM fluctuation and the interplay with the FS instability.
This may play an important role in the realization of superconductivity.
Note that URhGe is a special case of an ``Ising'' ferromagnet because of the weak anisotropy between the $b$- and $c$-axes, retaining only hard magnetization along $a$-axis.
Taking into account the fact that the combined FS topology and anisotropy of the local contribution of the magnetism governs the Ising character, for URhGe at a low field, a specific subband may play a dominant role in the Ising character a SC pairing, and another subband may dominate the physical properties around $H_{\rm R}$ (a key role of the Lifshitz transition).

In UCoGe, owing to the low energies involved in the renormalization of the quasiparticle, the magnetic field modifies the strength of the FM fluctuation with a strong decrease for $H \parallel M_0 \parallel c$ and an increase for $H = H_b^* \perp M_0$.
In both UCoGe and URhGe, a transverse field scan ($H \perp M_0$) leads to the collapse of $T_{\rm Curie}$ for $H \sim H_b^*$.

URhGe exhibits a singular situation realized by the switch of the easy-magnetization $c$- axis to the $b$-axis at $H = H_b^* =H_{\rm R}$. 
RSC is a remarkable phenomenon directly linked to the TCP and to the wing structure in the ($H_b, H_c$) plane.
RSC disappears exactly when the QCEP is reached. 
Quantum oscillation and TEP experiments indicate that Lifshitz transition is coupled to the variation of the FM fluctuation. 
Scaling in $m^\ast (H_{\rm R})/m^\ast (0)$ as a function of $H/H_{\rm R}$ is a signature that the $H$-induced magnetization along the $b$-axis reaches a critical value at $H_{\rm R}$.
URhGe belongs to a large class of compounds with an $H$-induced transition driven by a Lifshitz transition such as CeRu$_2$Si$_2$, for example.
Further experiments are required to confirm that the Lifshitz transition occurs close to $H_b^*$ in UCoGe.         

UGe$_2$ is an example of a clean superconductor  with simultaneous FS, FM, and PM instabilities. 
An experimental paradox is that despite the realization of very clean crystals, the intrinsic boundary of SC in the FM domain remains poorly determined. 

In pressure experiments, the details of the various instabilities when $T_{\rm Curie} < T_{\rm SC}$ are yet unresolved owing to the strong decrease in $T_{\rm Curie}$ towards $P_c$. 
It is a major experimental challenge to select an accurate tool at pressure in the vicinity of $P_{\rm c}$ to monitor changes in $T_{\rm Curie}$, $M_0$ and $T_{\rm sc}$.
A transverse field scan is an elegant way to reach this boundary.

The theory has been focused on explaining the experiments.
Many properties are explained by a phenomenological weak-coupling approach without consideration of the FS reconstruction even the FM criticality.
Up to now, there has been no global consideration combining the FM wing structure, the additional Lifshitz transition, and the consequence on superconductivity, notably above the QCEP.
A sound viewpoint may be given by remarks on unconventional quantum criticality (see Ref.~\citen{Ima10}).
Recently for URhGe, a crude two-band model has shown the key role of a field induced-Lifshitz transition in the enhancement of $m^\ast$ and in RSC. 
Combined progress in band structure calculation and in FS determination should be realized,
Here, we have not discussed the selection of a given SC order parameter (between A and B) since a new generation of experiments must be performed to resolve the choice.
Experimental observation suggests a change in the SC order parameter with $H$ in the transverse field scan of URhGe as well as in a similar scan of UCoGe at $H_{\rm R}$ and $H_b^\ast$.

A major interest in these three FM-SC uranium compounds is that the origin of the pairing (FM fluctuations) is well established and that its strength can be tuned easily in $H$, $P$ and $\sigma$ scans.
In comparison with other SCs in the class of strongly correlated electronic systems (cuprates, pnictides,, etc.), a negative point is the complexity of their multiband structures. 
High-$T_c$ cuprates with their initial single-band calculations appear to have a simple band structure.~\cite{Dam03}
However, it took two decades to show that the observed frequency at 530 T occupies 2\% of the first Brillouin zone in the underdoped regime~\cite{Seb15} and four years of active research to establish that this nodal electron Fermi pocket is created by a charge order.~\cite{WuT2011,Ghi12}
In iron-based superconductors, as three orbitals, $d_{xy}$, $d_{yz}$, and $d_{zx}$, contribute to the electronic states at the FS~\cite{Chu15}, the multiband character plays an important role in the pairing.
Often quantum oscillation studies give a full view of the FS.~\cite{Ghi12}
An illustrative result is the report of orbital-selective Cooper pairing in FeSe.
An interesting development is the creation of spin-triplet Cooper pairs at SC interfaces with FM materials and their use in superconductivity spintronics.~\cite{Esc15}

Despite the fact that studies on FM-SC have been restricted to a few groups, key results have been discovered that impact on considerably different materials including high-$T_{\rm c}$, pnictide, organic superconductors and a quantum liquid ($^3$He). 
A basis has been given for the interplay between FS instability and FM fluctuations.           

The microscopic coexistence of FM and SC has so far only been found in uranium compounds (UGe$_2$, URhGe, UCoGe, UIr~\cite{AkazawaJPCM2004}) despite a long search for new materials.
This is probably due to the unique characteristics of the 5$f$ electron in the U atom.
It is known that
the 5$f$ electron has an intermediate nature between 3$d$ and 4$f$ electrons, 
in other words, duality of the itinerant/localized character, leading to strong electronic correlations. 
Furthermore, the strong spin-orbit coupling favors the Ising anisotropy of FM.
These two factors might be quite important for FM superconductors.
If one can tune the strong correlation and the strong spin-orbit coupling by, for example, the crystal structure, 5$d$ electrons, or other artificial controls, a new FM superconductor with novel properties might be discovered.

Looking at the number of publications, the subject of FM superconductivity in U compounds appears to be marginal (350 publications) compared with cuprate superconductors (65000 publications) and Fe-based superconductors (already 5000 publications). 
In the class of heavy-fermion superconductors, the numbers of publications are similar for CeCoIn$_5$, UPt$_3$, and CeCu$_2$Si$_2$, which have, respectively 321, 576, and 230 articles with the compound name plus superconductivity in the title.
We hope that the reported data will clarify key issues in the general understanding of unconventional SC in strongly correlated electronic systems, which should be further examined by investigating the physical properties of other materials. 
The case of FM triplet superconductivity has often omitted and AF coupling as a source of SC pairing has mostly been discussed (see, for example, Ref.~\citen{Dav13}).
Research should meet the interests of researchers.
The complexity of 5-$f$ electron behavior commonly observed in low-symmetry crystals may repel researches who only like pure initial conditions. 
Fortunately, using few renormalized parameters, the complexity vanishes and new territory is reached.

The hope is that an ideal material that is easy to grow, easy to purify, easy to cleave, and easy to model with a unique U atom and a high-symmetry structure will be discussed..


\section*{Acknowledgements}
D. A. and J. F. thank their colleagues in Grenoble: 
J. P. Brison, D. Braithwaite, A. Huxley, G. Knebel, V. Mineev, and A. Pourret. 
Many results were obtained in the Ph. D. work carried out in Grenoble by I. Sheikin, F. Hardy, F. Levy, V. Taufour, E. Hassinger, L. Howald, M. Taupin, G. Bastien, A. Gourgout, and B. Wu. 
We also thank A. Miyake, H. Kotegawa, and Y. Tokunaga for their works on high pressure and NMR experiments.
Comments on the draft by G. Knebel, K. Miyake, S. Kambe were a great help in improving the manuscript. 
NMR/NQR studies on UCoGe and YCoGe were done in collaboration with the group of N. K. Sato in Nagoya university. 
K. I. thanks N. K. Sato, K. Deguchi, Y. Tada, S. Fujimoto, Y. Yanase, H. Ikeda, Y. Ihara, N. Nakai, T. Hattori, S. Kitagawa, M. Manago, for the collaborations, and A. de Visser, S. Yonezawa, and Y. Maeno for valuable discussions.
The initial discovery of SC in UGe$_2$ by G. Lonzarich and S. Saxena opened the field and their continuous feedback was very stimulating.
We would like to express our gratitude to K. Asayama, Y. Kitaoka and Y. \={O}nuki for their interest in our studies.
This work was supported by ERC (NewHeavyFermion) and KAKENHI (JP15H05882, JP15H05884, JP15K21732, JP15H05745, JP16H04006, JP15K05156).

\subsection*{Note added in proof}
Just during reading of our final proof, a stimulating discovery of the superconductivity in the dichalcogenide UTe$_2$ with $T_{\rm SC} \sim 1.5\,{\rm K}$ at ambient pressure was reported~\cite{Ran2018}, which is located on the PM verge of the FM instability. 
A large residual Sommerfeld coefficient far below $T_{\rm SC}$, approximately half of the normal-state value, was observed, which pushes to the proposal of a spontaneous half gapped superconductivity. 
This statement deserves to carry out careful measurements on the dependence of the specific heat with the material purity. 
It is clear that this discovery open a large variety of studies in the already known rich family of uranium dichalcogenide~\cite{Ike06,Ike09,Sus72,Tro94,Shl99}. 
UTe$_2$ is an excellent promising compound to clarify the interplay of ferromagnetism and SC in the PM side near the FM quantum instability.


\end{document}